\documentclass[11pt,a4paper]{article}
\pdfoutput=1
\usepackage{jheppub}
\usepackage[utf8]{inputenc} 

\usepackage{xcolor}   
\usepackage{graphicx}
\usepackage{amsmath}
\usepackage{amssymb}
\usepackage{braket} 
\usepackage{hyperref}
\usepackage{frontespizio}
\usepackage{comment}  
\usepackage{epsfig}
\usepackage{float}
\usepackage{color}
\usepackage[title]{appendix}
\usepackage{multirow} 
\usepackage{tikz}
\usetikzlibrary{decorations.pathmorphing}
\usepackage{varwidth}
\usepackage{tikz}
\usetikzlibrary{backgrounds,patterns,calc}
\usepackage{xparse}  
\usepackage{subfigure}

 \newcommand{\be}{\begin{equation}}
\newcommand{\bee}{\begin{equation}}
\newcommand{\ee}{\end{equation}}
\newcommand{\beea}{\begin{eqnarray}}
\newcommand{\eea}{\end{eqnarray}}
\newcommand{\bea}{\begin{eqnarray}}

 \def\cB{{\cal B}}
 
 \def\cM{{\cal M}}

 \title{Vacuum Transitions in Two-Dimensions  and their Holographic Interpretation}

\author[]{{Veronica Pasquarella}{\ and Fernando Quevedo}}
 
\affiliation[]{Department of Applied Mathematics and Theoretical Physics (DAMTP)} 
\affiliation[]{University of Cambridge, Wilberforce Road, CB3 0WA, Cambridge, UK}

\emailAdd{vp360@damtp.cam.ac.uk, fq201@damtp.cam.ac.uk}

 \abstract{\small{We calculate amplitudes for 2D vacuum transitions by means of the Euclidean methods of Coleman-De Luccia (CDL)  and Brown-Teitelboim (BT), as well as the Hamiltonian formalism of Fischler, Morgan and Polchinski (FMP). The resulting similarities and differences in between the three approaches are compared with their respective 4D realisations. For CDL, the total bounce can be expressed as the product of relative entropies, whereas, for the case of BT and FMP, the transition rate can be written as the difference of two generalised entropies.  
By means of holographic arguments, we show that the Euclidean methods, as well as the Lorentzian cases without non-extremal black holes, provide examples of an AdS$_2$/CFT$_1 \subset $  AdS$_3$/CFT$_2$ correspondence. Such embedding is not possible in the  presence of islands for which  the setup corresponds to AdS$_2$/CFT$_1 \not\subset $  AdS$_3$/CFT$_2$. We find that whenever an island is present, up-tunnelling is possible.
}}

 
\begin{document} 
\maketitle

 
\section{Introduction}   




Together with the quantum study of black hole physics, the study of vacuum transitions provides a rich arena to explore quantum aspects of gravity, and may shed some light towards a proper understanding of quantum gravity. Recent progress in the study of black hole information (for a review see \cite{JM2}) has been achieved, in great part, studying concrete examples in 2D, and it is natural to explore if the concepts used there, such as quantum extremal surfaces, generalised entropies and islands  may also play a role in addressing questions regarding other physical systems involving quantum aspects of gravity, namely, early universe cosmology and vacuum transitions.

The present work aims at providing a general perspective towards understanding  vacuum transitions in 2D. The reason for doing it is multi-fold. Among the most important, we highlight the following motivations:  

\begin{enumerate} 

\item To understand, from the simple 2D set-up, the stability of vacua with different values and signs of the vacuum energy, in particular having in mind the string theory landscape.

\item To study the possible emergence of unitarity-violating behaviour, such as encountered in the black hole information paradox \cite{SWH, Banks:1983by,Giddings:1995gd}, in a different physical system.   

\item To identify potential holographic interpretations of vacuum transitions, \cite{Maldacena:2010un, JM}, and their generalisations to higher-dimensional holographic embeddings and defect field theories. 

\item To explore the possibility of assigning an effective entropy to transition amplitudes for spacetimes with arbitrary sign of the cosmological constant. The main motivation for this comes from recent developments towards studying the emergence of spacetime from entanglement \cite{RT, BB14, VanRaamsdonk:2016exw, VanRaamsdonk:2020ydg}.

\end{enumerate}

Throughout the last four decades, the study of quantum transitions in 4D has been addressed by means of three main approaches, differing in terms of the way the cosmological constants are being defined, and the use of Euclidean or Lorentzian techniques. In order of appearance in the literature, they are:

\begin{itemize} 

\item Coleman-de Luccia (CDL) \cite{Coleman:1980aw}, describing transitions between different  local minima in a scalar field potential, following a Euclidean approach.

\item Brown-Teitelboim (BT) \cite{Brown:1988kg}, Euclidean vacuum transitions mediated by brane nucleation. 

\item Fischler-Morgan-Polchinski (FMP) \cite{Fischler:1990pk}, transitions between two spacetimes with different cosmological constants,  by means of the Hamiltonian formalism in Lorentzian signature. 

\end{itemize}   
In each case, the quantity of interest is the transition amplitude, $\Gamma$. In the Euclidean methods, this is obtained from the \emph{bounce}, $B$, defined as the difference  between the Euclidean action evaluated on the instanton $S_{E}|_{inst} $ and on the background  $S_{E}|_{bckgr} $: 
\begin{equation}   
\Gamma\ \sim\   \exp\left(-B\right)\ \ \ , \ \ \ B\overset{def.}{=}S_{E}|_{inst}-S_{E}|_{bckgr}.
\label{eq:gamma}   
\end{equation}
The main motivation for this approach follows from the relation between the WKB approximation and  vacuum transitions, with the latter being viewed as the quantum mechanical process of a particle crossing a potential barrier. 

The Lorentzian method of FMP, instead, proposes an alternative definition for $\Gamma$; the transition amplitude from a spacetime $\cM_1$ to another $\cM_2$, is defined as the relative ratio of two probabilities, 
\be
\Gamma_{1\rightarrow 2}
\ 
\overset{\text{def.}}{=}    
\ 
\frac{P_{{\text{nothing}}\rightarrow {\cM_1}{\text{/Wall}}/\cM_2}}{P_{\text{nothing}\rightarrow {\cM_1}}} 
\ 
= 
\ 
\frac{|\Psi_{\text{nothing}\rightarrow {\cM_1}{\text{/Wall/}}\cM_2}|^{2}}{|\Psi_{\text{nothing}\rightarrow {\cM_1}}|^{2}} 
\ \ \ , 
\label{eq:Gamma0}
\ee
with both nucleations out of nothing being identified with Hartle-Hawking (HH)-like states \cite{Hartle:1983ai}. This is a conditional probability, with the numerator and denominator corresponding to the end point of the transition and the original background spacetime, respectively.   
In an earlier paper, \cite{DeAlwis:2019rxg} (see also \cite{Bachlechner:2016mtp, BBOC}), we recovered the original BT result for dS$_{4}\rightarrow$dS$_{4}$ \footnote{The outer and inner vacua are denoted by $o,I$, respectively.}, by means of the FMP method, with total action given by

\be       
\log \Gamma_{\text{dS}\rightarrow \text{dS}}  
= 
\frac{\eta\pi}{2G}\left[\frac{[(H_{o}^{2}-H_{I}^{2})^{2}+\kappa^{2}(H_{o}^{2}+H_{I}^{2})]R_{o}}{4\kappa H_{o}^{2}H_{I}^{2}} + \frac{1}{2H_{o}^{2}} - \frac{1}{2H_{I}^{2}}\right]. \ \ \ 
\label{eq:bck} 
\ee 
Here $H_o, H_I$ refer to the Hubble parameter for the spacetimes outside and inside the bubble, $\kappa$ is the tension of the bubble, $R_o$ its radius at nucleation and $\eta=\pm 1$.
Under mutual exchange of the 2 vacua, the first term in (\ref{eq:bck}), coming from the brane, is symmetric, whereas the other two terms just flip their sign. From this follows that, the ratio between the direct and reverse transition reads

\be   
\frac{\Gamma_{o\rightarrow I}}{\Gamma_{I\rightarrow o}}= e^{\frac{\eta\pi}{2G}\left(\frac{1}{H_o^2}-\frac{1}{H_I^2}\right)}=e^{\eta(S_o-S_I)}.
\label{eq:db1}    
\ee     
For $\eta=1$, since  $S_o$ and $S_I$ are the entropies of the corresponding de Sitter spacetimes, this is a statement of \emph{detailed balance}. According to \cite{LW}, such entropic argument cannot be extended to spacetimes other than dS\footnote{The definition of the entropy for the static dS patch is a longstanding matter of debate. Motivated by our findings, we will be arguing that the entropy definition arising from the study of vacuum transitions is only meaningful when considering a finite portion of the dS patch. Our findings are therefore in agreement with those of others, such as, e.g. \cite{Banihashemi:2022htw}, where the authors outline the need to introduce a boundary to dS for assigning a notion of entropy to it.}, due to the topological change the spacetime would undergo as a consequence of the change in sign of its curvature. Furthermore, upon taking a vanishing background cosmological constant, \eqref{eq:bck} diverges, thereby forbidding up-tunnelling from a flat spacetime. In such case, \eqref{eq:db1} would thereby suggest that the background entropy is divergent. However, this is clearly in contrast with the common lore that the Minkowski vacuum should have vanishing entropy instead.

Following the steps of \cite{Fischler:1990pk}, whose work was in turn motivated by \cite{Farhi:1989yr}\footnote{The configuration described by Farhi, Guth and Guven, \cite{Farhi:1989yr}, has recently been addressed in the literature, \cite{Susskind:2021yvs}, and claimed to be unsuitable for the application of detailed balance. However, we will be able to prove that the claims made in \cite{Susskind:2021yvs} are not applicable in our setup given the locality of the processes being described.}, we showed in \cite{DeAlwis:2019rxg} that this issue can be addressed by  introducing a Schwarzschild black hole in the background, and realising Minkowski as the limit of the vanishing mass of the black hole itself. Indeed, the total action, in such case, reads  
\be  
S_{\text{Sch}\rightarrow \text{dS}}  
= 
\frac{\eta\pi}{2G}\left[S_{bounce} - S_{bckg} \right] 
= 
\frac{\eta\pi}{2G}\left[S_{bounce} - 4G^{2}M^{2} \right].  
\label{eq:bhds}  
\ee  
and clearly remains finite in the formal limit \footnote{Note that this limit is only approximate since the lower bound on a black hole mass is the Planck scale. } $\underset{M\rightarrow0}{\lim}$.
The question of whether, and how, we can possibly reconcile these apparently contrasting behaviours partly  motivated the present work, with the possibility to correctly assign an entropy to the spacetimes involved. Before embarking in the analysis outlined in the present work, our original interpretation was that transitions were taking place in \emph{local} regions of spacetime; as such, topologically-inspired arguments for the inapplicability of detailed balance could be dropped. But still, a probably more interesting question emerged: can we assign an entropic interpretation to the direct amplitude itself and not only to the ratio of the amplitudes, independently of detailed balance? As we shall see, the answer will turn out to be affirmative in 2D. 

In achieving this result, holography will turn out to play a key role. Since its first formulation, \cite{JM}, there have been many generalisations: AdS/BCFT, \cite{Fujita:2011fp}, wedge holography (codim-2), \cite{BB-1}, $T\bar T$, \cite{BB901, BB902, BB70, BB48, BB47, Coleman:2021nor}, the dS/CFT , \cite{BB20}, and dS$_{_{D+1}}$/dS$_{_{D}}$ correspondence, \cite{BB22}, etc. Also, the introduction of \emph{islands}, \cite{JM1, Penington:2019npb, JS}, has pushed the holographic duality to its currently most extreme formulation, since it involves the  entanglement of spatially-disconnected regions. Throughout our treatment we will encounter applications\footnote{For other recent applications of entanglement and islands within a cosmological setup see, e.g. \cite{Antonini:2022xzo, Antonini:2022blk, VanRaamsdonk:2020tlr, MVRHFC, iic, BBTH, Geng:2021iyq, Geng:2021wcq, Geng2:2021wcq,Geng:2022slq,  Langhoff:2021uct, Betzios:2019rds, Betzios:2021fnm, Bousso:2022gth, dadg,daeh,Chen:2020tes}.} of all such cases within the context of 2D vacuum transitions. 
   
Our main result in this article will be to find  expressions like \eqref{eq:bck} for vacuum transition rates in the 2-dimensional case. We will perform the calculation in all 3 formalisms outlined above (CDL, BT and FMP), providing their corresponding holographic interpretation. Given that the BT case has already been addressed by the authors in their original work, \cite{Brown:1988kg}, we will be using it as a ``standard ruler'' to check the results obtained by other means. In doing so, however, we will not simply take it for granted, but, rather, derive it from the suitably adjusted setup of CDL in 2D, which, to the best of our knowledge, has not been dealt with explicitly in the literature. As outlined in section \ref{sec:4}, the common starting point for both Euclidean methods will be the Almheiri-Polchinski action, \cite{AP},

\be   
S_{2D}   
= 
\int\ d^2 x\ \sqrt{-g\ }\ \left[\ \phi^2\  {\cal R} + \lambda\ \phi^{\prime2}-U(\phi),\ 
\right]  
\ee 
which  resembles the starting point of \cite{Coleman:1980aw} for decays in the presence of gravity. 

Due to the specific features of the 2D setup, though, the Ricci scalar or the scalar kinetic term can be eliminated by a suitable rescaling of the metric and of the potential. As we shall see, removing the $\phi^{\prime\ 2}$-term we can calculate the transition amplitude either with or without the \emph{thin-wall} approximation, proper to CDL. On the other hand, upon removing the Ricci scalar term, and adding suitable boundary terms ensuring the action satisfies a variational principle, we recover the gravitational action of BT. The importance of the addition of the boundary terms will be highlighted in due course, but it is worth stressing that, for the case in which the kinetic term is suppressed, they trivially vanish in either vacuum. Correspondingly, this is mapped to the fact that the calculation by means of the CDL method in 2D  describes transitions in \emph{absence} of gravity. From the holographic interpretation, we will see that, indeed, the total bounce is reminiscent of the entropy of an \emph{internal} CFT$_2$, which geometrises the RG-flow interpolating between different values of the dilaton.

\begin{figure}[h!]    
\begin{center}    
\includegraphics[scale=1]{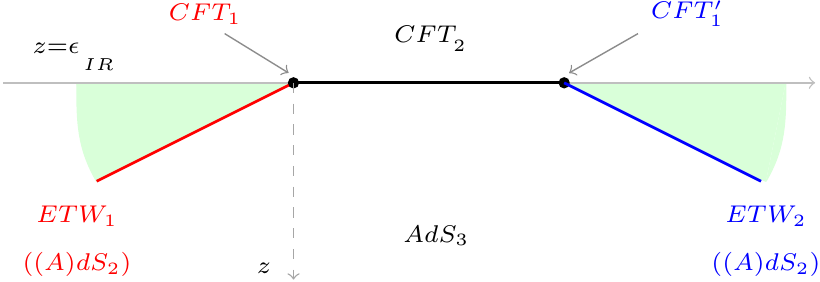}    
\caption{\footnotesize The CFT$_2$ geometrises the RG-flow defining the wall interpolating between the vacua, and in this sense is \emph{internal}. The CFT$_1$s are the duals of the AdS$_2$ spacetimes involved in the transition living on the two ETW branes. Their tensions, hence their opening angles w.r.t. the $z$ direction, correspond to the value of the cosmological constant on the two sides of the wall. }    
\label{fig:plot2f1} 
\end{center}      
\end{figure}

For the case of BT, instead, the addition of the boundary terms, accounting for the presence of gravity, is encoded in boundary entropies of defect CFT$_1$ s placed at the endpoints of the \emph{bulk} CFT$_2$, which in turn are dual to the end-of-the-world (ETW) branes, \cite{ATC1, ATC}, where the spacetimes undergoing the transition live (cf. figure \ref{fig:plot2f1}). This is illustrated in figure \ref{fig:plot2f1}. A more detailed explanation will be provided in section \ref{sec:4}.

This work is structured as follows: in section \ref{sec:4} we extend the CDL formalism to 2D, which, to the best of our knowledge, has not been explicitly done in the literature. In doing so, we use the JT-gravity theory analysed by Almheiri and Polchinski \cite{AP}. Thanks to the nature of the 2D setup, under  suitable  Weyl transformation, we can trade the kinetic term for the dilaton with the Ricci scalar. We thereby perform the calculation in both ways, showing that the case without the kinetic term describes transitions in absence of gravity. Upon removing the Ricci term, instead, the action can be brought back to the same form as in BT, up to a boundary term whose role is essential to account for the gravitational interaction. The second part of the section provides a brief overview of the 2D treatment of BT, as already addressed by the authors in their original work, \cite{Brown:1988kg}.

In section \ref{sec:2}, we turn to the Hamiltonian formalism of \cite{Fischler:1990pk} applied to 2D JT-gravity \cite{Jackiw:1984je,Teitelboim:1983ux, BB21}, and determine the transition rates in absence of black holes, finding that up-tunneling from and down-tunneling to Minkowski spacetimes are not allowed. 

In section \ref{sec:3}, we generalise the results of section \ref{sec:2} by adding a constant term in the action that allows for non-extremal black holes to emerge, as long as their mass lies within a certain range to be specified in due course. The range emerges from requiring the existence of two distinct physical turning points. We conclude that, upon adding a black hole of suitable mass on either side, the flat spacetime limit does not violate unitarity, and we are still left with a well defined transition amplitude. 

In section \ref{sec:4.4} we provide a possible holographic interpretation of the total bounces and actions calculated throughout our work.
Altogether, this leads to an exhaustive explanation of figure \ref{fig:plot2f1}. In particular, we find that the corresponding expression for the transition rate in presence of gravity, and in absence of black holes, is given by the difference of entropies of $T\bar T$-deformed CFTs, hence illustrating the locality of the nucleation process. Upon adding black holes, within a suitable mass range, instead, the total action can be expressed as the difference of generalised entropies, with an island emerging beyond a critical value of the black hole mass. In particular, we show that, whenever an island is present, up-tunnelling is always possible. Furthermore, the results obtained by means of the FMP method are found to agree with the expression provided by \cite{MVR} for describing mutual approximation of boundary states belonging to different CFTs under suitable parametric redefinition. The BT results, which in section \ref{sec:3} were proved to be equivalent to the FMP results in absence of black holes, can be expressed in terms of entropies of BCFT$_2$s with two nontrivial boundary conditions dual to ETW branes. Furthermore, the CDL result can be recast in the form of an entropy product of a CFT$_2$, thereby showing agreement with the expectations following from the analytic behaviour encountered in section \ref{sec:4}. 

One of the key results of our work can therefore be synthesised by saying that the total action (or bounce) associated to the direct vacuum transition process carries an \emph{internal} entropic interpretation. In particular, for the BT and FMP cases, they can always be expressed as the difference of generalised entropies. However, only the formalism of \cite{Fischler:1990pk} provides the right setup for an island to emerge. Following a brief summary of our findings, at the end of section \ref{sec:4.4}, we relate our work to other recent developments. 

\section{Euclidean transitions in 2D }\label{sec:4}

In this first section, we analyse 2D vacuum transitions by means of the Euclidean methods of CDL and BT, with the latter being essentially already known from the original work of \cite{Brown:1988kg}. Due to the specifics of the 2D setup, both results can be obtained from the same formalism, namely that of Almheiri and Polchinski, \cite{AP}. However, along the way, we encounter significant differences, resulting in different final expressions for the bounces. A physical explanation for this will be outlined throughout the treatment, leaving a more detailed justification to the final core section of the work, namely \ref{sec:4.4}, where, by means of holographic arguments, we will be showing the complementarity of the methods applied.

\subsection{Vacuum transitions in 2D without gravity}

\begin{figure}[ht!]
\begin{minipage}[c]{0.45\textwidth}    
\caption{\footnotesize  The transition between the minima of the potential is independent of the specific profile of $V$, and is only a function of $\Lambda_{I,o}$ and the tension of the wall. }  
\label{fig:cdl}  
\end{minipage}\hfill   
\begin{minipage}[c]{0.45\textwidth}      
\includegraphics[scale=0.9]{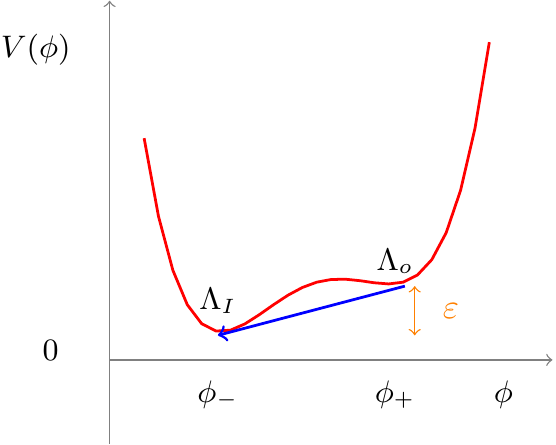} 
\end{minipage} 
\end{figure}

Let us start considering the simplest case of vacuum transitions in 2D without including gravity. As already argued by the authors of \cite{Coleman:1980aw}, the final amplitude does not depend on the particular profile of the potential. However, their calculation still relies upon the potential exhibiting specific features. In particular, for the prurpose of describing the process, it must have at least two minima, corresponding to the vacua involved in the transition. The simplest starting point is therefore to assume a double well potential, such as the one depicted in figure \ref{fig:cdl}, whose analytic expression can be split into two parts as follows, 

\begin{equation} 
V(\phi)\ \overset{def.}{=} \ V_o + V_{\varepsilon},  \label{eq:V}
\end{equation} 
where $V_{_{o}}$ is the ordinary degenerate double well potential satisfying the following properties

\begin{equation}    
V_o(\phi_+)\equiv V_o(\phi_-)\ \ \ ,\ \ \ \frac{d V_o}{d\phi}\bigg|_{\phi_{\pm}}=0, 
\end{equation}
with an additional symmetry breaking term proportional to the energy difference between the minima, 
\begin{equation} 
\varepsilon\overset{def.}{=} V(\phi_+)-V(\phi_-)\ \equiv\ \Lambda_{o}-\Lambda_{I}.
\end{equation}
The interpolation between the two minima is mediated by the scalar field, as depicted in figure \ref{fig:cdl}, since the latter is characterised by a nontrivial spatial profile

According to the Euclidean procedure, the decay rate, $\Gamma$, is defined from the bounce, $B$, as follows 
\be 
\Gamma\sim e^{-B } 
\ \ \ 
, 
\ \ \ 
B\overset{def.}{=}S_{E}|_{inst}-S_{E}|_{bckgr},
\ee    
where $S_{E}$ denotes the Euclidean action associated to a given theory. Specifically, there are 3 main contributions,
\bea 
B_{tot}^{CDL}
&=& 
B_{out}+B_{wall}+B_{inside}. 
\label{eq:BTOT1}    
\eea        
For a scalar field admitting an O(2)-symmetric instanton solution   
$ds^2    
=    
d\rho^2+\rho^2d\theta^2,   $
the corresponding Euclidean action reads:

\begin{equation}   
S_E    
=
\int\ dx^2\ \left[\frac{1}{2}(\partial_{\mu}\phi)^2+V(\phi)\right]       
=
\pi\int\ d\rho\ \rho\ \left[\frac{1}{2}(\phi^{\prime})^2+V(\phi)\right].
\end{equation}  
where $^{\prime}\equiv\frac{d}{d\rho}$. The e.o.m. for $\phi$, in the thin-wall approximation, reduces to:    

\begin{equation}   
\phi^{\prime}\phi^{\prime\prime}    
-V_o^{\prime}=\left(\frac{1}{2}(\phi^{\prime})^2-V_o\right)^{\prime}=0,
\label{eq:eomnew}
\end{equation}
which is exactly integrable once having chosen suitable boundary conditions, $\phi(\rho=\infty)=\phi_+$.

\begin{equation}    
\phi^{\prime} 
=    
\sqrt{2(V_o(\phi)-V_o(\phi_{\pm}))\ }\ \ \ \Rightarrow\ \ \ \rho-\bar\rho 
=    
\int_{\bar\phi}^{\phi}\frac{d\phi}{\sqrt{2(V_o(\phi)-V_o(\phi_{\pm}))\ } }  
\label{eq:eomnew1}
\end{equation}   
with $\bar\phi\overset{def.}{=}\frac{\phi_++\phi_-}{2}$. Equation \eqref{eq:eomnew1} defines the turning point $(\bar\rho,\bar\phi)$.  
The total bounce is simply given by the interior and the wall

\begin{equation}  
\begin{aligned}
B_{tot}   
=&    
B_{in}+B_{wall}  
=   
-\pi\bar\rho^2\ \varepsilon+2\pi\bar\rho S_1,   
\label{eq:btotcdlnog}
\end{aligned}
\end{equation}
where, making use of \eqref{eq:eomnew1}, the tension of the wall is defined as

\begin{equation}    
S_1    
=     
\int d\rho\ \left[2(V_o(\phi)-V_o(\phi_+))\right] .  
\end{equation}
Equation \eqref{eq:btotcdlnog} is extremised at $\bar\rho   
=    
\frac{S_1}{\varepsilon} $, at which the total extremised bounce reads:

\begin{equation}   
\boxed{\ \ B_{tot}^{\ \ extr} 
=     
\frac{\pi\ S_1^2}{\varepsilon}    \ \ }
\label{eq:Schw}   
\end{equation}   
The presence of a single spatial dimension allows for a direct identification of the nucleation process described by \eqref{eq:Schw}    with the Schwinger pair production, \cite{Schwinger:1951nm}, whose bounce reads    

\begin{equation}   
B_{Schw}  
=   
\frac{\pi\ m^2}{|eE_{on}|} 
=\pi m\bar\rho_{_{BT}},
\label{eq:schpr}       
\end{equation}
with $\bar\rho_{_{BT}}\overset{def.}{=}\frac{m}{|eE_{on}|}$ being the value of the turning point extremising the bounce of the corresponding process. Equation \eqref{eq:schpr} suggests the following parametric identification:

\begin{equation}   
m\ \ \leftrightarrow\ \ S_1\ \ \ ,\ \ \ \varepsilon\ \ \leftrightarrow\ \ |e E_{on}|, 
\end{equation}  
where $m$ is the mass of the particles, whereas $|eE_{on}|$ is the difference in energy provided by the electric field prior and after pair creation, and can therefore be associated to the energy difference between the minima in $V$. 

\begin{figure}[ht!]    
\begin{minipage}[c]{0.6\textwidth}    
\caption{\footnotesize Brane nucleation in one spatial dimension, with $\Lambda_{\pm}\overset{def.}{=}V(\phi_{\pm})$ is analogous to the Schwinger process upon interpreting the brane with tension $\sigma$ as the particle/antiparticle pair (denoted by the two red dots).}    
\label{fig:Schw}      
\end{minipage}\hfill  
\begin{minipage}[c]{0.3\textwidth}       
\includegraphics[scale=1]{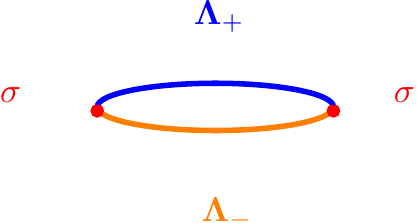}    
\end{minipage}    
\end{figure}

\subsection{CDL from the Almheiri-Polchinski setup }\label{sec:4.1}

We will now turn to the case with gravity. The Euclidean action for gravity coupled to a scalar field should contain a potential and a kinetic term. However, in 2D gravity, the Ricci scalar needs to be coupled to the dilaton to ensure that the theory is nontrivial, as first formulated by Jackiw and Teitelboim in \cite{Jackiw:1984je}. This is a first major difference w.r.t. the 4D treatment, from which a dynamical coupling between $\phi$ and $\rho$ is expected to arise at the level of the equations of motion.

\subsubsection*{Light-cone and cartesian coordinates}

Given the conformal metric 

\be   
ds^2 
= 
-e^{2\omega}\ dx^+dx^-    
= 
4\ e^{2\omega}\left(-dt^2+dz^2\right)      
\label{eq:lineelemlineelem}    
\ee 
with $x^{\pm}\overset{def.}{=}t\pm z$ and $2\ \partial_{\pm}=\partial_t\pm\partial_z$, the Ricci scalar in 2D is given by the following expression:
\bea    
{\cal R}    
&= &       
-4\  e^{-2\omega}\left(\ \partial_{+}\partial_{-}\omega\ -  2\ \partial_+\omega\ \partial_-\omega\ \right)     \nonumber   \\
&=&       
-2 \left[\ \partial_t\left(e^{-2\omega}\  \partial_t\ \omega   \right)- \partial_z\left(e^{-2\omega}\  \partial_z\ \omega   \right)\ \right].
\label{eq:Ricci}               
\eea  
In Lorentzian signature, the 2D action in terms of $x^{\pm}$ reads

\be   
S_{2D}   
= 
\int\ d^2 x\ \sqrt{-g\ }\ \left[\ \phi^2\  {\cal R} + \lambda\  \partial_{+}\phi\ \partial_{-}\phi-U(\phi)\ 
\right].
\label{eq:Se}    
\ee
Given the particular nature of the 2D setup, we can proceed in two ways: either removing the kinetic term or performing a Weyl rescaling of the metric and setting ${\cal R}=0$. Both procedures will be explored in turn in the present section.


\subsubsection*{Removing the kinetic term}

The kinetic term can be removed under suitable rescalings

\be   
\rho\rightarrow\phi^{-\lambda/4}\rho   
\ \ \ 
,    
\ \ \ 
U\rightarrow\phi^{-\lambda/2}U\overset{def.}{=}U_o,   \ \ \    
\label{eq:fredef}    
\ee   
such that the action resembles that of JT-gravity. Once integrated by parts, equation \eqref{eq:Se} is simply

\bea  
S_{2D}       
&=&         
-8\ \left(\phi^2\partial_-\omega\right)\bigg|_{x^+=\text{const.}}-\int d^2 x \ \left[8\left(\partial_+\phi^2\right)\left(\partial_{-}\omega\right)     +U_o e^{2\omega}\right],
\label{eq:newacr}   
\eea         
from which the e.o.m. for $\phi^2$ and $\omega$ can be extracted

\be    
4\partial_+\partial_-\omega    
= 
\frac{e^{2\omega}}{2}\partial_{\phi^2}U_o \equiv   T_{+-},
\ \ \ \ \ \ \ \ \ \ \ \    \ \ \ \ \ \ \ \ \ \ \ \ 
4\partial_-\partial_+\phi^2    
=   
U_oe^{2\omega}   \equiv T_{-+},
\label{eq:SE1}   
\ee   
corresponding to the off-diagonal components of the energy momentum tensor $T_{\mu\nu}$. The remaining components, instead, read:

\be   
T_{\pm\pm}    
= 
0 
\ \ \ 
\Rightarrow    
\ \ \ 
\partial_{\pm}\left(e^{-2\omega}\ \partial_{\pm}\phi^2\right) =0.  
\label{eq:thc1}
\ee    
In cartesian coordinates, the Lorentzian action reads:  
\bea   
S_{2D}         
&=&         
S_{bdy}-\int\ dt\ dz\ \left[\ 2\phi^2\ \partial_z^2\ \omega   +U_o e^{2\omega}\ \right], 
\label{eq:newact2}   
\eea  
where $\omega=\omega(z), \phi^2=\phi^2(z)$. \eqref{eq:newact2} corresponds to the starting point for the extension of CDL to 2D, and the constraints \eqref{eq:thc1} reduce to \cite{AP}, 
\be  
\partial_z\left(e^{-2\omega}\ \partial_z\left(\phi^2\right)\right)    
=     
0.   
\label{eq:3rdcinz}    
\ee

\subsubsection*{Wick rotation and polar coordinates}

Following \cite{Coleman:1980aw}, we turn to Wick-rotated and  polar coordinates:
\be   
ds^2 
= 
e^{2\omega}\ \left(dt^2+dz^2\right) 
= 
\rho^2\ \left(dr^2+r^2d\theta^2\right),    
\label{eq:lineelem}    
\ee 
 where,
\be   
\begin{cases}   
t
= 
r\ \cos\theta 
\\    
z=r\ \sin\theta
\end{cases}   
\ \ \ 
,    
\ \ \ 
r\overset{def.}{=}\sqrt{t^2+ z^2\ }\ \ \ , \ \ \ \theta\overset{def.}{=}\tan^{-1}\frac{\ z\ }{t}.   
\label{eq:newvar4}   
\ee   
The Euclidean action reads:   
\bea   
S_E         
&=&   
4\pi\ \phi^2r \omega^{\prime}\bigg|_{r_i}^{r_f}\  -2\pi\ \int\ dr \left[  2\left(\phi^2\right)^{\prime}\ r \omega^{\prime}-r U_o e^{2\omega}\right], 
\label{eq:Lagr1}    
\eea    
where the $2\pi$ factor comes from the integration over $\theta$ and $^{\prime}\overset{def.}{=}\partial_r$, with the equations of motion (e.o.m.)  for $\phi^2$ and $\rho$ now being:
\begin{equation}    
2 \left(r\ \omega^{\prime}\right)^{\prime} = -r\ e^{2\omega} \ \partial_{\phi^2}U_o, 
\ \ \ \ \ \ \ \ \ \ \ \    \ \ \ \ \ \ \ \ \ \ \ \    
( r(\phi^2)^{\prime})^ {\prime}    
=   
- r\ e^{2\omega}    \ U_o.
\label{eq:er}
\end{equation}     
Notice that the latter is exactly the e.o.m. for the scalar field obtained by CDL in the thin-wall approximation. Indeed, the only term that is missing w.r.t. the full equations is the $\frac{\rho^{\prime}\phi^{\prime}}{\rho}$-term that can be tuned to zero in the thin-wall approximation. Assuming $\theta$-independence of $\phi$ and $\omega$, \eqref{eq:3rdcinz} reads
\bea 
\partial_{r}\ \left(\frac{e^{-2\omega}}{r}\ \partial_{r}\ \phi^2 \right)  =0.
\label{eq:3rdcinr}    
\eea 
To obtain an explicit expression  for \eqref{eq:BTOT1}, we need to specify what we mean by the minima of the potential and the wall separating them within the Almheiri-Polchinski setup. We now turn to describing both, one at a time.

\subsubsection*{\texorpdfstring{${\left({\phi}^2\right)^{\prime}= 0}$}{} : defining the vacua\ }

For $\phi^2$=const., \eqref{eq:er} implies $U_o$=0 and \eqref{eq:3rdcinr} is trivially satisfied. Redefining $u\overset{def.}{=}\ln r$, \eqref{eq:er} becomes    

\begin{equation}    
2\ \ddot\omega = -r^2\ e^{2\omega} \ \partial_{\phi^2}U_o=- \partial_{\phi^2}U_oe^{2(\omega+u)} ,
\label{eq:ep1}
\end{equation}
where $^{\cdot}\overset{def.}{=}\partial_u$. Redefining $\omega+u\overset{def.}{=}f$, equation \eqref{eq:ep1} becomes     

\begin{equation}    
2\ \ddot f  =- \partial_{\phi^2}U_oe^{2f}.
\label{eq:ep2}
\end{equation}
The first equation in \eqref{eq:er} shows that $\partial_{\phi^2}U_o \overset{def.}{=}2\Lambda=$ constant. It defines the cosmological constant of the 2D spacetime. We can therefore determine the solutions corresponding to dS and AdS by respectively taking positive or negative values of such constant:


\medskip   

\underline{\ 1) $\Lambda>0$ \ } \ equation \eqref{eq:ep2} is solved by
\be  
 \rho 
= 
e^{\omega}    
= 
e^{f-u}    
= 
\frac{2aKr^{a-1}}{\sqrt{ 2\Lambda\  }\ (K^2\ r^{2a}+1)\ } . 
\label{eq:case1}    
\ee



\begin{figure}[ht!]    
\begin{center}    
\includegraphics[scale=0.5]{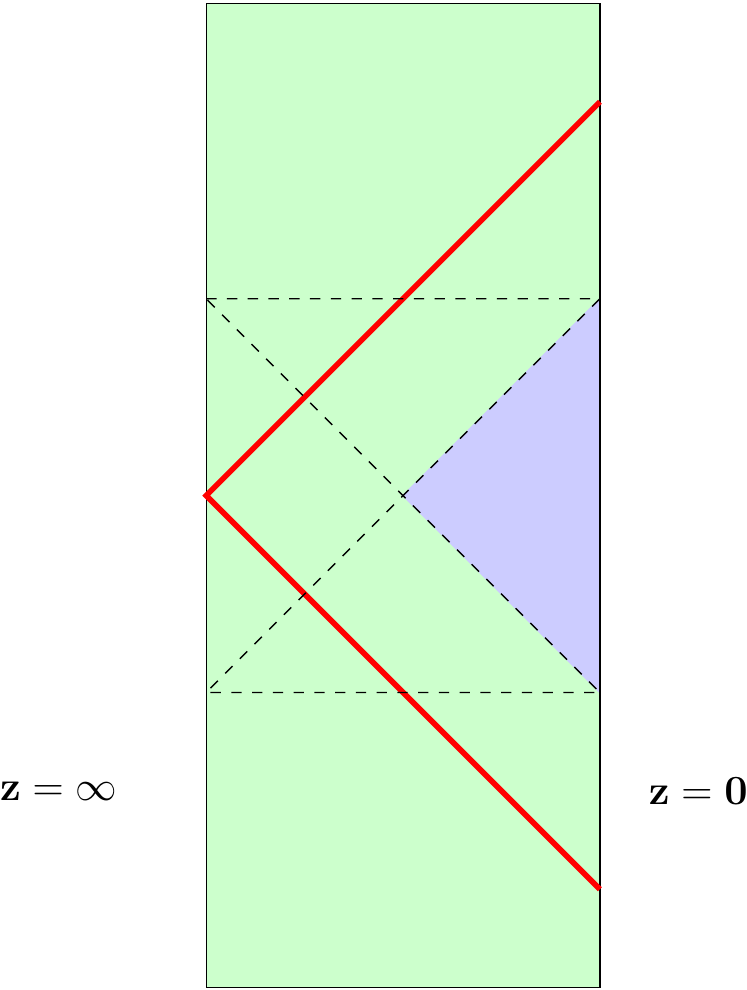} 
\ \ \     
\includegraphics[scale=0.5]{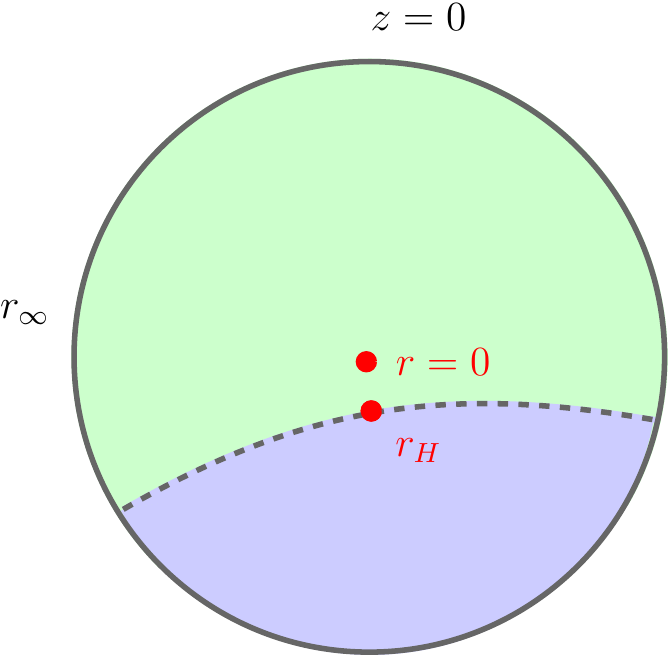}   
\ \ \ 
\includegraphics[scale=0.5]{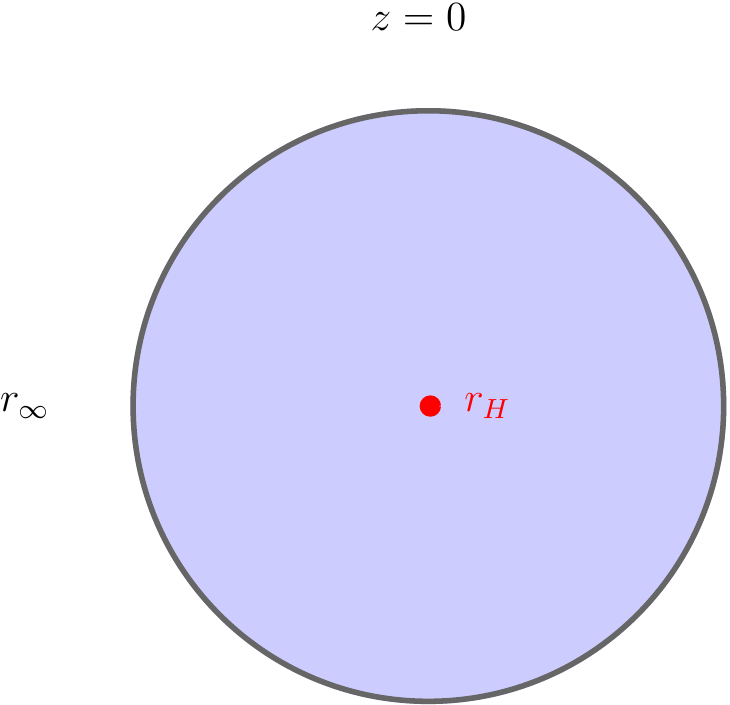}
\caption{\footnotesize The picture on the LHS shows the different regions of AdS$_2$ covered by the global (the whole rectangular strip), Poincarè (on the RHS of the red lines) and BTZ black hole (the shaded blue region). For the purpose of interest to us, we will only be using the latter, given that the first two are associated to the infinite regions of spacetime. The figures in the middle and on the right, show that the BTZ metric effectively covers only a portion of the Poincarè disk.   }   
\label{fig:3pict3} 
\end{center}   
\end{figure}

\underline{\ 2) $\Lambda<0$ \ }  \ equation \eqref{eq:ep2} is solved by  


\be  
\rho    
=    
e^{\omega}   
=    
e^{f-u}   
= \begin{cases}
\frac{a}{\sqrt{2\Lambda\ }\ (au+b)\ r} 
= 
\frac{a}{\sqrt{2\Lambda\ }\ (a\ \ln|r|+b)\ r }    \\       

\frac{a}{\ \sqrt{2\Lambda\ }\ \sin(\text{h})  (au+b) \ r} 
= 
\frac{2iaKr^{a-1}}{\ \sqrt{ 2\Lambda\  }\ (K^2\ r^{2a}-1)\ }    \\    
\frac{a}{\sqrt{2\Lambda\ }\ \sin  (au+b) \ r} 
= 
\frac{2aK^i\ r^{ia-1}}{\ \sqrt{ 2\Lambda\  }\ (K^{2i}\ r^{2ia}-1)\ } . 
\end{cases}    
\label{eq:cases2}    
\ee  
The three cases outlined in \ref{eq:cases2} correspond to the Poincar\'e patch of AdS$_2$, global AdS$_2$ and an AdS$_2$ black hole, respectively. Figure \ref{fig:3pict3} helps visualising the regions covered by the different coordinate systems. 
For either sign of the cosmological constant, the radial coordinate can be redefined, such that the line element takes the following form: 
\be 
ds^2    
=    
d\hat r^2+f(\hat r)\ d\theta^2 .    
\ee  
The corresponding redefinitions are   
\be 
r 
= 
\left(\ \frac{\ 1\ }{K}\ \tan\ \left(\frac{\sqrt{2\Lambda\ }\ \hat r}{2}\ \right)\ \right)^{1/a}    \ \ \  \text{for } \Lambda>0, \ \ \ \    
\label{eq:hatr1}
\ee   
and    
\be 
\frac{\ \ln\left|a\ \ln|r|+b\right|\ }{\ \sqrt{ 2\Lambda\  } \ }     
=         
\hat r   
\ \ \ ,\ \ \ r 
= 
\left(\ \frac{\ 1\ }{K}\ \tanh\ \left(\frac{\sqrt{2\Lambda\ } \hat r}{2}\ \right)\ \right)^{1/a}     \ \ \  \text{for } \Lambda<0, \ \ \ \   
\label{eq:hatr}     
\ee     
for the $1/r\ \ln r$ and $\frac{1}{\sinh}$ cases, respectively.

For $a=1, b=0$, \eqref{eq:case1} and \eqref{eq:cases2} reduce to: 

\be    
\boxed{\ \ 
\rho_{\text{dS}_2} 
=    
\frac{2}{\ \sqrt{2\Lambda\ }\ (r^2+1)\ } 
\ \ \ 
,    
\ \ \ 
\rho_{\text{AdS}_2} 
=    
\begin{cases}    
\frac{a}{\sqrt{2\Lambda\ }\  \ln|r|\ r \ }  \\    
\frac{2i}{\ \sqrt{2\Lambda\ }\ (r^2-1)\ } \\    
\frac{2i}{\ \sqrt{2\Lambda\ }\ (r^{2i}-1). \ }    
\end{cases}    \  }    
\label{eq:casesmetric}    
\ee    
Under a suitable coordinate transformation, these results are equivalent to the ones obtained by Almheiri and Polchinski.
The first and third solutions in \eqref{eq:cases2} are not relevant for our purposes, given that they correspond to the Poincarè patch, and global AdS. In figure \ref{fig:3pict3}, they correspond to the region inside the red triangle and the green shaded strip, respectively. Instead, we will restrict to the $\rho \sim \frac{1}{\sinh}$ case, corresponding to the BTZ black hole, \cite{Banados:1992wn}, (the region outside the black hole is shaded in blue), which is a quotient of the Poincarè patch.

\subsubsection*{\texorpdfstring{$(\phi^2)^{\prime}\neq 0$}{} : defining the wall }

We now turn to the case in which $(\phi^2)^{\prime}\neq0$. Integrating over \eqref{eq:3rdcinr} twice, we find, \cite{AP}
\be      
(\phi^2)^{\prime} 
= 
r\ c_1\ \ e^{2\omega}     
\ \ \ 
\Rightarrow   
\ \ \ 
\phi_+^2-\bar \phi^2=\frac{c_1}{2}\int_{0}^{\bar r} dr\ r\ e^{2\omega},      
\label{eq:follident122}    
\ee  
where $\bar\phi^2\overset{def.}{=}\frac{\phi_{+}^2+\phi_{-}^{2}}{2}$ is the mean value of the field between the two minima. This defines the turning point ($\bar\phi^2,\bar r$) in analogy with the Euclidean formalism of \cite{Coleman:1980aw}. Up to now, the constant $c_1$ is unspecified. However, its importance is central in our treatment, as will be explained later.

The radius of the nucleated bubble is parameterised by the $\hat r$-variable, defined in \eqref{eq:hatr1}  and \eqref{eq:hatr} for dS$_2$ and AdS$_2$, respectively. Setting $a=1, b=0$, it is a monotonic function and sgn\ $\hat r\equiv$ sgn$ r$. It therefore follows that we can trade the integration over $\hat r$ with that over $r$ without loss of generality.
From the e.o.m., we know that $U_o=0$ at any minimum, but different values of $\phi^2$ come with a corresponding value of $c_1$ triggering the flow described by \eqref{eq:follident122}. 
Equation \eqref{eq:follident122}  describes the flow from one value of the dilaton to that at the turning point   
\begin{figure}[ht!] 
\begin{minipage}[c]{0.25\textwidth}        
\includegraphics[scale=2]{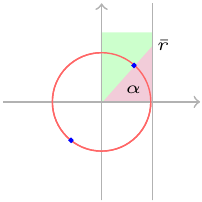}   
\end{minipage}\ \ \ \ \ \ \ \ \ \ \ \ \ \     
\begin{minipage}[c]{0.5\textwidth}
       \caption{\footnotesize The brane in 2D is mapped to a particle/antiparticle pair (blue dots). 
       The angle $\alpha\overset{def.}{=}\sqrt{\frac{\Lambda}{2}\ }\ \hat r$, denotes the location of the brane separating the background (green) from the interior of the bubble (purple). The corresponding value of $r$ is obtained by the intersection shown in the figure, namely from $\tan$(h)$\alpha$. Note that $\bar r=1$ corresponds to $\alpha=\frac{\pi}{4}$.}     \label{fig:2p}   
\end{minipage}   
\end{figure}

\bea   
\phi_+^2-\phi_-^2   
&=&     
c_1^+\int_0^{\bar r}\ dr\ r\ e^{2\omega}    
=      
\frac{c_1^+}{\Lambda_+}\ \int_0^{\bar r}\ dr\ \frac{2r}{ \ (r^2-1)^2\ }   =      
-\frac{c_1^+}{\Lambda_+}\ \left[\ \frac{1}{\ \bar r^2-1\ } +1\ \right].  
\label{eq:RGflow}      
\eea     
Equation \eqref{eq:RGflow} can also be interpreted as describing the running of the cosmological constant along the RG-flow originating from $\phi_+^2$ and parameterised by $c_1^+$.
Upon integrating over the reverse flow, defined by $c_1^-\equiv -c_1^+$, we get

\bea   
\phi_-^2-\phi_+^2   
&=&        
-\frac{c_1^-}{\Lambda_-}\ \left[\ \frac{1}{\ \bar r^2-1\ } +1\ \right].  
\label{eq:RGflow4}     
\eea    
 Consistency between \eqref{eq:RGflow}and \eqref{eq:RGflow4} leads to a constraint for the turning point $\bar r$, which in turn places constraints on the parameters of the theory of either vacua. A similar analysis can be carried out for dS and mixed AdS/dS transitions, and the respective constraints arising from this procedure can be summarised as follows

\begin{equation}  
\boxed{\ \ \ \begin{aligned}       
&\bar r^2 \bigg|_{  \text{AdS}_2\rightarrow\text{AdS}_2}
=    
1+\frac{\Lambda_-}{\Lambda_+} \ \ \ \ \ \ \  ,\ \ \ \ \  \ \    \bar r^2  \bigg|_{ \text{dS}_2\rightarrow\text{dS}_2} 
\frac{\Lambda_+}{\Lambda_-} -1  \\    
\\    
&\bar r^2   \bigg|_{\text{AdS}_2\leftrightarrow\text{dS}_2}    
=   
\frac{1}{2}\ \left(-\frac{\Lambda_++\Lambda_-}{\Lambda_+-\Lambda_-}\pm\sqrt{ \left(\frac{\Lambda_++\Lambda_-}{\Lambda_+-\Lambda_-}\right)^2+8\ \ }\right).  
\label{eq:constrads} 
\end{aligned}\ \ }         
\end{equation}  
For dS$_2\rightarrow$dS$_2$ processes, the constraint $\bar r>0$ requires $\Lambda_+>\Lambda_-$, implying only downtunnellling is allowed, wheras for transitions of the AdS$_2\leftrightarrow\ $dS$_2$ kind, only the the $+$ solution ensures $\bar r$ is physical.

\subsection*{Minkowski vacua     \ }

Before turning to the evaluation of the bounces, we wish to highlight that the system of equations \eqref{eq:er} and \eqref{eq:3rdcinr}, together with the following expression valid at local minima of the potential

\be   
U_o\bigg|_{min}    
=    
\phi^2\ \partial_{\phi^2}\ U_o+2c_1         
=   
0     
\label{eq:defuo}        
\ee     
implies that the Minkowski vacuum cannot provide a stable solution originating a nontrivial flow. Indeed, from \eqref{eq:defuo}, for $\Lambda=0$, it follows that $c_1=0$, implying the flow generated by such vacuum defines a flat direction in the space of metrics and does not end in a vacuum other than Minkowski itself. Because of this, at least for the moment, we will not be analysing any transitions involving such spacetimes, and we will only be looking at the flat limit of the bounces by taking $\underset{\Lambda_{\pm}\rightarrow0}{\lim}$.

\subsubsection*{The total bounce}

The total bounce is the result from the sum of 2 contributions,    

\bea 
B_{tot} 
= 
B_{wall}+B_{in}, 
\label{eq:tbe}   
\eea
given that $B_{out}=0$. The scalar field is constant in a local minimum of the potential, and therefore $U_o=0$ due to \eqref{eq:er}. From \eqref{eq:er}, the Euclidean actions for (A)dS$_2$, read

\bea    
S_E\bigg|_{\text{dS}_2}   
&=&       
-2\pi\ \phi_-^2\int_{0}^{\bar r}\ dr\ r\ e^{2\omega}\ \partial_{\phi^2}U_o  =      
4\pi\ \phi_-^2\ \left[\ \frac{1}{\ \bar r^2+1\ }-1\ \right],
\label{eq:dS2}   
\eea

\bea    
S_E\bigg|_{\text{AdS}_2}   
&=&           
-2\pi\ \phi_-^2\int_{0}^{\bar r}\ dr\ r\ e^{2\omega} \ \partial_{\phi^2}U_o  =     
4\pi\ \phi_-^2\ \left[\ \frac{1}{\ \bar r^2-1\ }+1\ \right],
\label{eq:AdS2}   
\eea      
from which the inner bounces associated to any possible transition between (A)dS$_2$ spacetimes, can be evaluated.
Inside the wall, the bounce reads   

\bea        
B_{wall}    
&=&     
2\pi\ \int_{0}^{\bar r} d\ r\ r\ e^{2\omega}\left[-\phi^2\ \partial_{\phi^2}\ U_o+U_o(\phi)+\phi_+^2\ \partial_{\phi^2}\ U_o(\phi_+)-U_o(\phi_+)\right] \nonumber\\   
&=&     
-2\pi\ \phi^2\ U_o\bigg|_{\phi_+^2}^{\phi_-^2}+\frac{4\pi}{c_1}\ \int_{\phi_-^2}^{\phi_+^2}d  \left(\phi^2\right)\ \left[U_o(\phi)+\phi_+^2\ \Lambda_+\right]\nonumber\\  
&=&        
\frac{4\pi}{c_1}\ S_1+4\pi\ \left(\phi_-^2-\phi_+^2\right),    
\label{eq:wallB}    
\eea           
where 

\be    
S_1    
\overset{def.}{=}    
 \int_{\phi_+^2}^{\phi_-^2}d  \left(\phi^2\right)\ U_o(\phi)    \ \ \ \text{with} \ \ U_o   
=   
-\frac{e^{-2\omega}}{r}\ \left(r(\phi^2)^{\prime}\right)^{\prime},   
\label{eq:tensionT}    
 \ee   
where the last expression follows from the e.o.m. for $\phi^2$. According to the specific configuration being studied, there are 4 possible cases:

\bea       
B_{wall}        
&=&    4\pi\Delta\phi^{^2}\cdot \begin{cases}  
 \left(2\frac{\phi_-^2}{\phi_+^2}+1\right)    \\
 \\
 \left(2\frac{\phi_-^2}{\phi_+^2}-1\right)    \\
 \\
\left(2\frac{\phi_-^2}{\phi_+^2}+1\right)     \\
\\
 \left(2\frac{\phi_-^2}{\phi_+^2}-1\right)    
    \end{cases}    
\Rightarrow \boxed{ B_{tot}= 4\pi \Delta\phi^{^2}\cdot \begin{cases}  
     \left(\frac{\phi_-^2}{\phi_+^2}+2\right)   \ \ \ [\text{AdS}_2\rightarrow \text{AdS}_{2}]\\
\\
 \left(3\frac{\phi_-^2}{\phi_+^2}-2\right)  \ \   [\text{dS}_2\rightarrow \text{dS}_{2} ] \\
\\
\left(3\frac{\phi_-^2}{\phi_+^2}+2\right)   \ \   [\text{AdS}_2\rightarrow \text{dS}_{2}]\\
\\
\left(\frac{\phi_-^2}{\phi_+^2}-2\right)    \ \ \ [\text{dS}_2\rightarrow \text{AdS}_{2}]
    \end{cases} }
\label{eq:wallnew1}    
\eea      
where $\Delta\phi^{^2}\overset{def.}{=}\phi_-^{^2}-\phi_+^{^2}$.  

Note that, upon taking $\underset{\Lambda_{\pm}\rightarrow 0}{\lim}$, all transition amplitudes vanish. In particular, there is no 2D counterpart for, either, dS$_4\rightarrow$\ Mink$_4$ nor Mink$_4\rightarrow$\ AdS$_4$ processes, which were originally addressed in \cite{Coleman:1980aw}.

\subsection*{CDL in the thin-wall approximation} \label{sec:thinwall}

In the above derivation of the total bounces,  a closed form expression could be found without needing to resort to the thin-wall approximation. It is therefore interesting to see what the result might be for the case in which we were to apply the standard procedure as in \cite{Coleman:1980aw}. 
The starting point is still an O(2)-symmeytric instanton solution 

\begin{equation} 
\begin{aligned}
ds^2=d\xi^2+\rho^2(\xi)d\theta^2.
\label{eq:O2}
\end{aligned}
\end{equation}
The radial coordinate used in our derivations, $r$,  is related to $\xi$ in \eqref{eq:O2} as follows\footnote{Notice that fo AdS the same arguments follow, the only difference being that trigonometric functions turn into hyperbolic functions due to the negative sign of $\Lambda$.}
\begin{equation} 
\begin{aligned}
\hat r
=\sqrt{\frac{2}{\Lambda}\ }\tan^{-1} r  
\ \ \ 
\Rightarrow    
\ \ \ 
d\hat r   
=     
\sqrt{\frac{2}{\Lambda}\ }\ \frac{dr}{1+r^2}
=     
\sqrt{\frac{2}{\Lambda}\ }\ \cos^2\left(\sqrt{\frac{\Lambda}{2}\ }\xi\right)\ dr   
=d\xi,
\end{aligned}
\end{equation} 
and to the function $\rho_{_{CDL}}$ in \eqref{eq:O2} as 
\begin{equation}
    \rho_{_{CDL}}    
    =\sqrt{\frac{2}{\Lambda}\ }\ \frac{r}{1+r^2}.
\end{equation}
In particular, we will be looking at the case of dS$_2\rightarrow$Mink$_2$ transitions.
In order to do so, we need to re-express the Euclidean action

\begin{equation}   
S_E  
=2\pi\int_{0}^{\bar r} dr\ \left[2\phi^2(r\omega^{\prime})^{\prime}+rU_oe^{2\omega}\right],   
\end{equation}   
in terms of the new variables $(\hat r, \rho)$ or $(\xi,\rho)$. 
For Minkowski, $\hat r\equiv r\equiv \xi\equiv \rho$. The Euclidean action therefore simply vanishes on the inside. For dS, we know that
$e^{2\omega}=\frac{2}{\Lambda(r^2+1)^2}$, 
as previously derived. On the inside, the dilaton is constant, and the Euclidean action  reduces to
\begin{equation} 
S_{in}  
= 
2\pi\phi^2r\omega^{\prime}   
\ \ \ \ \ \ \ \ \text{with} \ \ \ \ \ \ \ \  
\omega^{\prime}
= 
\frac{(e^{\omega})^{\prime}}{e^{\omega}}     =-2\rho_{_{CDL}}\sqrt{\frac{\Lambda}{2}\ }.  
\label{eq:rasrho}    
\end{equation}    
For ease of notation, we will simply write the radial coordinate as $\rho$. Inverting the last equality in \eqref{eq:rasrho}, we get     

\begin{equation} 
r 
= 
\frac{1-\sqrt{1-2\Lambda\rho^2\ }}{\sqrt{2\Lambda}\rho}   
\ \ \ \ \ \ \  
\Rightarrow\ \ \ \ \ \ \ S_{in}  
= 
-2\pi\phi^2\left(1-\sqrt{1-2\Lambda\bar\rho^2\ }\right).
\label{eq:Sin1}    
\end{equation} 
The wall's bounce, instead, reads  

\begin{equation}   
\begin{aligned}   
B_{wall}   
=&    
\pi\bar\rho\ \bar\phi^2\ S_1+4\pi(\phi_-^2-\phi_+^2),
\label{eq:newea}    
\end{aligned}   
\end{equation}
where we made use of the thin-wall approximation for extracting $\bar\rho$ from the integral and have also defined the tension of the wall in terms of the rescaled potential in the Einstein frame, $U_o^{^{new}}$ 
\begin{equation}    
S_1   =    
2\int d\xi \ U_o^{^{new}}  =    2\int d\xi \ \left[\phi^{-2}U_o-\partial_{\phi^2}U_o\right].   
\end{equation}   
For the case of a dS$_2\rightarrow$Mink$_2$ transition, the total bounce therefore reads   
\begin{equation}   
B_{tot}    
= 
-2\pi\phi_+^2\left(1+\sqrt{1-2\Lambda\bar\rho^2\ }\right)+\pi\bar\rho\ \bar\phi^2\ S_1 ,  
\label{eq:Btotdsmink} 
\end{equation}    
and it is extremised at
\begin{equation}      
\bar\rho=\frac{S_1\ \bar\phi^2}{4\Lambda\ \phi_+^2\sqrt{1+\frac{S_1^2\ \bar\phi^4}{8\Lambda\phi_+^4}\ }} .  
\label{eq:tp}
\end{equation}

\begin{equation}   
\Rightarrow\ \ \ \boxed{\ \ B_{tot}^{\ \text{dS}_2\rightarrow\text{Mink}_2}\bigg|_{extr}  
= 
2\pi\phi_+^2\ \sqrt{1+\frac{S_1^2\ \bar\phi^4}{8\Lambda\phi_+^4}\ }\left(1-\frac{3}{1+\frac{S_1^2\ \bar\phi^4}{8\Lambda\phi_+^4}}\right)  .}    
\end{equation}   
In a similar fashion, for the generic (A)dS$_2\rightarrow$(A)dS$_2$ case, the total bounce reads   

\begin{equation}   
\boxed{\ \ \ B_{tot}^{\ \text{(A)dS}_2\rightarrow\text{(A)dS}_2}    
= 
2\pi\phi_-^2\left(1+\sqrt{1-2\Lambda_-\bar\rho^2\ }\right)-2\pi\phi_+^2\left(1+\sqrt{1-2\Lambda_+\bar\rho^2\ }\right)+\pi\bar\rho\ \bar\phi^2\ S_1 \textcolor{white}{\Biggl [} \color{black} .}    
\label{eq:dstodstw}
\end{equation}

\subsection{Brown-Teitelboim in 2D from Almheiri-Polchinski}  \label{sec:BTFAP}

We now turn to the alternative path for determining the bounce describing 2D transitions, namely by removing the Ricci scalar and restoring the kinetic term in the AP action:

\begin{equation}   
S_{AP} 
=\int dx^2\ \left[\phi^2{\cal R}+U_o\right]e^{2\omega}. 
\label{eq:AP}   
\end{equation}   
The $\phi^{\prime\ 2}$-term can be restored by reabsorbing $\phi^2$ in the metric. In particular, under the rescalings $\rho\rightarrow\phi\rho$ and $U_o\rightarrow \phi^{-2}U_o\overset{def.}{=}16\Lambda_{_{BT}}$, \eqref{eq:AP} turns into

\begin{equation}   
\begin{aligned}
S_{AP}  
&=8\pi\int dz\ \left[-u^{\prime\ 2}+4\Lambda_{_{BT}}e^{2f}\right], 
\label{eq:AP1}   
\end{aligned}
\end{equation} 
with $f\overset{def.}{=}\ln\phi+\omega\overset{def.}{=}u+\omega$. 
Before delving into further calculations, some important remarks are in order. The e.o.m. undoubtably play a key role within the CDL formalism, as can be appreciated by direct inspection of their original work in 4D, \cite{Coleman:1980aw}. When performing the calculation for transition amplitudes within the AP setup (once having removed the kinetic term for the scalar field), the crucial relation that allowed us to circumvent the need to resort to the thin-wall approximation (upon which \cite{Coleman:1980aw} rely) was the change of variables resulting from requiring vanishing diagonal components for the energy momentum tensor, \eqref{eq:thc1}. As previously argued, in polar coordinates, they reduce to   

\begin{equation} 
\partial_r\left(\frac{e^{-2\omega}}{r}\ \partial_r\phi^2\right)   
=0 
\ \ \ 
\Rightarrow 
\ \ \ 
\partial_z\left( u\ \partial_z u\right)   =   (uu^{\prime})^{\prime}
=0,
\label{eq:encons}
\end{equation}
where $z\overset{def.}{=}\ln r$. At the same time, the term on the RHS in \eqref{eq:encons} also emerges upon integrating by parts the  kinetic term in  \eqref{eq:AP1}. Correspondingly, we are led to consider the fiollowing quantity, $M_{_{ADM}}\overset{def.}{=}uu^{\prime}\bigg|_{bdy}$, which can be identified with the ADM mass of the spacetime. When removing the kinetic term in ${\cal L}_{_{AP}}$, the dilaton is constant in each vacuum solution, therefore $M_{_{ADM}}$ is trivially vanishing to start with, as follows from the e.o.m.\footnote{As we shall see in greater detail in section \ref{sec:4.4}, this is consistent with the fact that such types of transitions take place in absence of gravity.} On the other hand, when performing the Weyl transformation of the metric leading to \eqref{eq:AP1}, this is no longer true. Indeed, when using \eqref{eq:AP1} to describe transitions, $\Delta M_{ADM}\overset{def.}{=}uu^{\prime}\bigg|_{bck}^{inst}\neq 0$. We thereby need to add such boundary term by hand to ensure energy conservation throughout the nucleation process as a whole.

From the considerations we have just made, \eqref{eq:AP1} should therefore be upgraded in the following way

\begin{equation}   
\begin{aligned}
S_{BT}^{\ grav} 
&=S_{AP}+16\pi uu^{\prime}\bigg|_{bdy}\\  
&=8\pi\int dz\ \left[-u^{\prime\ 2}+4\Lambda_{_{BT}}e^{2f}\right]  +16\pi uu^{\prime}\bigg|_{bdy},
\label{eq:BT1}   
\end{aligned}
\end{equation}
which is exactly equivalent to the BT action describing brane nuclaetion in 2D in presence of gravity, in which case the total bounce in conformal gauge (with  metric function parameterised by $e^{\varphi}$) reads, \cite{Brown:1988kg},  

\be  
B_{_{TOT}} 
= 
2\pi\ m \bar\rho+\frac{\pi}{2k}\ \int_{\bar r_{o}}^{{\infty}}\ dr\ r\ \left[-\left(\varphi_{,r}\right)^2\bigg|^{inst}_{bck}+4\Lambda_{I}e^{\varphi_{inst}}-4\Lambda_{o}e^{\varphi_{bck}}\right]+\frac{\pi}{k}r\varphi_{,r}\varphi\bigg|^{inst}_{bck},
\label{eq:origBT1}    
\ee  
where the \emph{effective} cosmological constants are defined as follows

\begin{equation} 
\Lambda_{o,I} 
\ 
\overset{def.}{=}      
\ 
\lambda+\frac{1}{2} k\ E_{o,I}^{2}.
\label{eq:Leff}
\end{equation} 
In BT's notation, $k$ is proportional to $G_{_{N}}$ and is kept fixed throughout their treatment. The dynamics is thereby carried by the change in the electric field, as follows from \eqref{eq:Leff}. On the other hand, on the case of AP, as discussed earlier on in the section, we can think of $k$ as being the running coupling ensuring the interpolation in between the two vacua parameterised by different values of the dilatonic field. Indeed, in this sense we argued earlier on that transitions in AP are equivalent to RG-flows.

The e.o.m. for $\phi$ is:   

\begin{equation}   
-u^{\prime\prime}=4\Lambda_{_{BT}}\ e^{2\omega+2u}.  
\label{eq:eompbt}
\end{equation}  
It admits  a solution of the kind

\begin{equation}     
u(z)=    
-\ln\left|\ \cos\text{(h)}\left(-2\sqrt{\Lambda_{_{BT}}}e^{2\omega}z+\text{const.}\right)\ \right| . 
\end{equation}  
Fixing $\omega\equiv 0$ and const.$\equiv 0$, the profile for the scalar field reads: 

\begin{equation}   
\phi(z)=\frac{1}{\ \cos\text{(h)}\left(-2\sqrt{\Lambda_{_{BT}}\ } \ z\right)\ },
\label{eq:eompbt1}
\end{equation}  
which is either a trigonometric or hyperbolic function, according to the sign of $\Lambda_{_{BT}}$.
Upon integrating the first term in \eqref{eq:BT1}    by parts, and using \eqref{eq:eompbt}, we get

\begin{equation} 
\begin{aligned}   
S       
&=8\pi\int_{\bar z}^{z_{_{h}}}\ dz \ u^{\prime\prime}\ (2u-1), 
\label{eq:CDLAP1}    
\end{aligned}
\end{equation}   
where $\bar z$ and $z_{_{h}}$ denote the location of the turning point and the horizon, respectively. Specifically, for AdS, the latter is simply $z_{_{h}}\equiv 0$, whereas for dS it is finite. 
For the instanton to exist, $\bar z>z_{_{_h}}$ is needed, and \eqref{eq:CDLAP1} therefore reads:

\begin{equation}   
\begin{aligned}
S_{tot}   
&=-16\pi\ \left[\sqrt{\Lambda_{_{BT}}}\tan\text{(h)}\left(-2\sqrt{\Lambda_{_{BT}}}  z\right)\ln\left|\ \cos\text{(h)}^2\left(-2\sqrt{\Lambda_{_{BT}}}\  z\right)\right|+\right.\\
&\ \ \ \ \ \ \ \ \ \left.+4\Lambda_{_{BT}}\   z+3\sqrt{\Lambda_{_{BT}}}\tan\text{(h)}\left(-2\sqrt{\Lambda_{_{BT}}}\  z\right)\right]\bigg|_{\bar z}^{z_{_{h}}}.
\label{eq:stotbtfromap}    
\end{aligned}
\end{equation}
Equation \eqref{eq:stotbtfromap}  holds for, both, the instanton and background solutions, the only difference being the value of $\Lambda_{_{BT}}$. From now on it is convenient to analyse the AdS and dS cases separately. 
\underline{1)\ \ AdS$_2\rightarrow$AdS$_2$}

\medskip

In this case $z_{_{h}}\equiv0$. Redefining 

\begin{equation}   
\sqrt{\Lambda_{_{\pm}}\ }\  \sqrt{1\pm\phi_\pm^2\ } \overset{def.}{=}\phi_\pm^{new}   
\ \ \ 
\Rightarrow    
\ \ \ 
\phi_{\pm}^2= \pm\left(1-\frac{\phi_\pm^{\ new\ 2}   }{\Lambda_{\pm}}\right),  
\label{eq:sametp}    
 \end{equation}    
 ensures that now $\phi_{tp}$ is the same on either side at $\bar z$, hence

\begin{equation} 
\begin{aligned}   
B_{in}^{\ \text{AdS}_2\rightarrow\text{AdS}_2}   
&= 16\pi \left[\phi_{tp}^{new} \ln\left|\ \frac{1-\frac{\phi_{tp,-}^{new\ 2}}{\Lambda_{_{-}}}}{1-\frac{\phi_{tp,+}^{new\ 2}}{\Lambda_{_{+}}}}\  \right|+3(\Lambda_{_{-}}-\Lambda_{_{+}})\ \bar z\right]. 
\label{eq:AP111}  
\end{aligned}   
\end{equation} 
Which follows from $\phi_{tp,-}^{new}\ \equiv\ \phi_{tp,+}^{new}$, by means of \eqref{eq:sametp}, whereas the wall's bounce reads:

\begin{equation} 
\begin{aligned}   
B_{wall}       
&=8\pi\int dz\ \left[- u^{\prime\ 2}+4\Lambda_{_{BT}}e^{u2}\right] \overset{def.}{=}32\pi\ \sigma   =  32\pi\ m\ \bar z,    
\label{eq:AP1111}  
\end{aligned}   
\end{equation} 
where $\sigma$ denotes the tension of the brane and $m\overset{def.}{=}\Lambda_+-\Lambda_-$. The total bounce therefore reads:

\begin{equation} 
\begin{aligned}   
\boxed{\ \ \ B_{tot}^{\ \text{AdS}_2\rightarrow\text{AdS}_2}       
=16\pi \left[\phi_{tp}^{new} \ln\left|\ \frac{1-\frac{\phi_{tp,-}^{new\ 2}}{\Lambda_{_{-}}}}{1-\frac{\phi_{tp,+}^{new\ 2}}{\Lambda_{_{+}}}}\  \right|-\frac{1}{3}m\  \bar z\right]. \color{white}\bigg].\color{black}\ \ }
\label{eq:AP111adsads}  
\end{aligned}   
\end{equation} 
\underline{2)\ \ dS$_2\rightarrow$dS$_2$}   
Similar considerations can be performed for the case of dS. The main difference is that now the horizons contribute with nontrivial terms to the actions. In particular, being $z_{_{h}}\neq0$, we get 

\begin{equation} 
\boxed{\ \ \ 
\begin{aligned}   
B_{tot}^{\ \text{dS}_2\rightarrow\text{dS}_2}              
&= 16\pi \left[\phi_{tp}^{new} \ln\left|\ \frac{1-\frac{\phi_{tp,-}^{new\ 2}}{\Lambda_{_{-}}}}{1-\frac{\phi_{tp,+}^{new\ 2}}{\Lambda_{_{+}}}}\  \right|+3\left(\Lambda_{_{-}}-\Lambda_{_{+}}+\frac{m}{6}\right)\ \bar z\right.+\\
&\left.+\phi_{h,-}^{new}\left(2-3\sqrt{1-\frac{\phi_{h,-}^{new\ 2}}{\Lambda_-}\ }\right)-\phi_{h,+}^{new}\left(2-3\sqrt{1-\frac{\phi_{h,+}^{new\ 2}}{\Lambda_-}\ }\right)\right]. 
\label{eq:AP111adsads1}  
\end{aligned}  \ \ }    
\end{equation} 
Once more, the terms linear in $\bar z$ provide an energy conservation relation and are therefore vanishing. The terms featuring in the last line, instead, provide the contributions from the horizons. The results \eqref{eq:AP111adsads}, \eqref{eq:AP111adsads1} agree with the expression of the total bounce for type-1 instantons in \cite{Brown:1988kg}

\begin{equation} 
\boxed{\ \ \ B 
= 
2 \pi m\bar\rho-\frac{4\pi}{k}\ \ln\frac{r_i}{r_o}  + \frac{2\pi\bar\rho}{k}\left(\Lambda_i r_i-\Lambda_o r_o\right)  \textcolor{white}{\Biggl [} \color{black} .\ \ }
\label{eq:BT}
\end{equation} 
prior to extremising w.r.t. $\bar\rho$. In \eqref{eq:BT}, we made use of the same notation as can be found in the literature, with

\be  
\bar r_{_{I,o}}  
\overset{def.}{=}  
\frac{2}{\Lambda_{_{I,o}}\bar\rho}\ \left[1-\sqrt{1-\Lambda_{_{I,o}}\bar\rho\ }\ \right].   
\ee   
Importantly, the last 2 terms in \eqref{eq:BT} arise only for the dS$_2\rightarrow $ dS$_2$ case, and are absent for AdS$_2\rightarrow$AdS$_2$, consistently with the results obtained in \eqref{eq:AP111adsads1} and \eqref{eq:AP111adsads}, respectively. Compatibility with our results follows from the following identifications

\begin{equation}   
\bar z\longleftrightarrow\ \bar\rho\ \ \ ,\ \ \ \phi_{_{tp}}^{^{new}}\overset{def.}{=}\frac{1}{k}=\frac{E_{o,I}^{2}}{2\left(\Lambda_{o,I}-\lambda\right)}\ \ \ ,\ \ \ \frac{\phi_-^{new\ 2}}{\Lambda_{_{-}}}\overset{def.}{=}1+\frac{\bar r_{i,o}^2\Lambda_{\pm}}{4} .   
\end{equation}

\subsection*{Key points and remarks}

\begin{itemize}

\item  Unlike the ordinary Schwinger process, in presence of gravity, the Euclidean action needs to be dressed with a boundary term corresponding to the ADM mass
\begin{equation}   
\begin{aligned}
S_{BT}^{\ grav} 
&=S_{AP}+16\pi uu^{\prime}\bigg|_{bdy} .
\label{eq:B1T}   
\end{aligned}
\end{equation}

\item In BT, the dynamical process of brane nucleation is encoded in the change in the electric field, while keeping $k$ (i.e. $G_{_N}$) fixed, as follows from \eqref{eq:Leff}. On the other hand, on the case of AP, $k$ is related to the running coupling, $\phi$, ensuring the interpolation in between the two vacua parameterised by different values of the dilatonic field.

\item Up-tunneling falls into the forbidden region of parameter space of the type-1 instanton, since, this configuration fails to satisfy the energy conservation relation associated to the brane nucleation process.

\item 
Interestingly, for either sign of the cosmological constant, the flat limit cannot be taken. This follows from the fact that such processes correspond to topology change, and would thereby fall within a different instanton type. The reason for this can be traced back to the extremality of the spacetimes involved, due to the definition of effective cosmological constants, 

\begin{equation} 
\Lambda_{o,I} 
\ 
\overset{def.}{=}      
\ 
\lambda+\frac{1}{2} k\ E_{o,I}^{2}.
\label{eq:LeffBT}
\end{equation}

\item As already highlighted in \cite{Brown:1988kg}, there are two different ways of calculating the bounce action. Either substituting the singular solutions to the field equations in the Euclidean action, or, alternatively, integrating over the 2 bulk spacetimes separately, in which case the $\delta$-function in the fields would be replaced by the Gibbons-Hawking boundary terms, with the membrane providing the natural boundary for both spacetimes. The 2$^{nd}$ procedure agrees with the first upon  extremising the bounce w.r.t. the membrane size, $\bar\rho$. According to \cite{Brown:1988kg}, this is preferable, since it ensures that the extremisation of the bounce on the classical instanton solution is not an artefact of the formalism\footnote{In section \ref{sec:2} we will argue that the starting point in the Hamiltonian formalism of FMP is precisely the first of the two methods outlined in \cite{Brown:1988kg}.}.

\end{itemize}   


  \section{Lorentzian transitions in 2D}   \label{sec:2}
  
  
As proved in an earlier work, \cite{DeAlwis:2019rxg}, vacuum transitions in 4D can be described in three equivalent ways, namely by means of the CDL, BT and FMP formalisms. The importance of their agreement in the well known case of dS$_4$ spacetime, is what led us to explore to what extent the obstruction to other types of transitions may be overcome. In particular, the role plawed by black holes in \cite{DeAlwis:2019rxg}, inspired by the original work of FMP, which had no counterpart in the BT and CDL methods, turned out to be crucial for enabling to define transitions involving Minkowski spacetimes. 

Given these motivations, we now turn to extending the Hamiltonian method of Fischler, Morgan and Polchinski (FMP) \cite{Fischler:1990pk} to its 2D counterpart, namely Jackiw-Teitelboim (JT)-gravity \cite{Jackiw:1984je,Teitelboim:1983ux}, whose action, up to boundary terms, reads

\begin{equation} 
S_{_{JT}} 
\ 
= 
\ 
\frac{\phi_{o}}{2\kappa^{2}}\ \int\ d^{2}x\ \sqrt{-g}\ {\cal R}+\frac{1}{2\kappa^{2}}\ \int\ d^{2}x\ \sqrt{-g}\ \phi( {\cal R}-2\Lambda),
\label{eq:JTact} 
\end{equation} 
with $\phi_{o}$ being a constant. The coupling of the dilaton $\phi$ to the Ricci scalar ensures gravity is nontrivial. We can think of (\ref{eq:JTact}) as being the 2D version of the Einstein-Hilbert action with a cosmological constant describing gravity in arbitrary dimension. $\phi$ plays the role of a Lagrange multiplier and ensures curvature is fixed. From a Field Theory point of view, the expression  (\ref{eq:JTact}) is that of a renormalised theory, with the dilaton accounting for the renormalisation of Newton's constant.


The setup consists of pure gravity with spherical symmetry and a wall (the bubble's surface) at $z=\hat z$, separating two spacetimes with two different cosmological constants $\Lambda_{+}, \Lambda_{-}$, similarly to Brown and Teitelboim \cite{Brown:1988kg}.
Given the above considerations, our proposed initial Lagrangian density, thereby takes the general form:

\bea
  {\cal L} 
&
   = 
&
   \sqrt{-g}\phi\left[{\cal R}-2\Lambda_{+}\Theta(z-\hat z)-2\Lambda_{-}\Theta(\hat z-z)\right]-\delta(z-\hat z)\sigma\sqrt{(N^{t})^2+L^{2}(N^{z}+ {\dot{\hat{z}}})^{2}} \ +    \nonumber \\
  & &
+  \sqrt{-g}\left[ \phi_{o}\chi_{_{{\cal M}}}-{\cal B}_{+}\Theta(z-\hat z)-{\cal B}_{-}\Theta(\hat z-z)\right],
   \label{eq:LagrTh1}  
\eea 
i.e. the action for JT-gravity coupled to matter, with the latter being provided by the brane term, the two cosmological constants $\Lambda_{+},\Lambda_{-}$  and two additional constants\footnote{These act as deformation parameters in the 2D theory, contributing to the Hamiltonian of the system in the form of constant energy terms. As we shall see in section \ref{sec:3}, they play the role of black hole masses, and will be subject to constraints specified in due course.}, ${\cal B}_{\pm}$. Following FMP we start with the metric:

  \begin{equation} 
  g_{\mu\nu}   
  \ 
  = 
  \ 
  \left( 
  \begin{matrix} 
  -(N^{t})^{2}+(N^{z})^{2}\ \ \  & \ \ \  LN^{z}\\
  LN^{z} \ \ \  &  \ \ \  L^{2}\\ 
  \end{matrix} 
  \right) .
  \\  
  \\    
  \label{eq:mans} 
  \end{equation}

   
\subsection{Actions with vanishing constant energies}


First, we will be considering the ${\cal B}_\pm=0$ case. The ${\cal B}\neq 0$ case will be dealt with in section \ref{sec:3}.
Substituting (\ref{eq:mans}) in (\ref{eq:LagrTh1}) with ${\cal B}_{\pm}=0$ and integrating by parts, we get\footnote{Notice that in the calculations we are omitting the topological term but we will add it back to the final expression for the action.}

\bea
  {\cal L} 
  &
   = &
 {\cal L}_{_{BTs}}+2\phi\left(\frac{N^{t\prime}}{L}\right)^{\prime}-\frac{2\dot\phi\dot L}{N^{t}}-\frac{2\phi^{\prime}N^{z}(N^{z\prime}-\dot L)}{LN^{t}}+\nonumber \\
  & & -2\Lambda_{+}\Theta(z-\hat z)-2\Lambda_{-}\Theta(\hat z-z)-\delta(z-\hat z)\sigma\sqrt{(N^{t})^2+L^{2}(N^{z}+ {\dot{\hat{z}}})^{2}} \nonumber\\
  &= &
   \ 
      {\cal L}_{_{BTs}}+ \pi_{L}\dot L+\pi_{\phi}\dot\phi-N^{t}{\cal H}_{t}-N^{z}{\cal H}_{z} ,
   \eea   
   where the total derivative terms read 
   
   \begin{equation}
    {\cal L}_{_{BTs}}    
    \ 
    = 
    \ 
    -2\left(\frac{\phi N^{t\prime}}{L}\right)^{\prime} + \left(\frac{2\phi\dot L}{N^{t}}\right)^{.}+\left(\frac{2 N^{z} (N^{z \prime}-\dot L)}{L N^{t}}\right)^{\prime} .   
    \end{equation} 
 The conjugate momenta are: 
 
   \begin{equation}
   \begin{aligned}    
&\pi_{L} 
\ 
= 
\ 
2\left(\frac{\phi^{\prime}N^{z}}{LN^{t}}-\frac{\dot\phi}{ N^{t}}\right), 
\ \ 
\ \ 
\pi_{\phi} 
\ 
= 
\ 
\frac{2(N^{z\prime}-\dot L)}{N^{t}},
\ \  
\ \  
\pi_{\hat z} 
\ 
= 
\ 
- \frac{\sigma L^{2}(N^{z}+{\dot{\hat{z}}})\delta(z-\hat z)}{\sqrt{(N^{t})^2+L^{2}(N^{z}+ {\dot{\hat{z}}})^{2}}},
\ \ \   
\end{aligned}
\end{equation} 
The Hamiltonian and total momentum are:

\begin{equation}
\begin{aligned}
   -{\cal H}_{t} 
   &
   =  
   \frac{\delta {\cal L}}{\delta N^{t}} 
= 
   -2\left(\frac{\phi^{\prime}}{L}\right)^{\prime}-\frac{2\dot\phi}{N^{t}}(N^{z\prime}-\dot L)+\frac{2\phi^{\prime}N^{z}(N^{z^{\prime}}-\dot L)}{N^{t2}L}+  \\ 
   & -2\phi L \left[\Lambda_{+}\Theta(z-\hat z)+\Lambda_{-}\Theta(\hat z-z)\right]-\frac{\sigma\ N^{t}\delta(z-\hat z)}{\sqrt{(N^{t})^2+L^{2}(N^{z}+ {\dot{\hat{z}}})^{2}}},\\    
   -{\cal H}_{z} 
   &
   =  
\pi^{\prime}_{L}-\phi^{\prime}\frac{\pi_{\phi}}{L}-\delta(z-\hat z)\pi_{\hat z}.
\label{eq:Hzc}
\end{aligned}   
\end{equation}
Away from the wall, the momentum constraint reads 

\begin{equation} 
\pi^{\prime}_{L} 
\ 
= 
\ 
\phi^{\prime}\frac{\pi_{\phi}}{L} ,
\ \ \ 
\end{equation} 
in terms of which ${\cal H}_t$ can be re-expressed\footnote{Imposing the Hamiltonian constraint is equivalent to selecting a specific energy for the configuration described by the Lagrangian of the theory. Because of this, the Lorentzian approach of FMP is naturally providing a microcanonical description of vacuum decays. The improvement of their method w.r.t. those outlined in section \ref{sec:4} is comparable to the one in \cite{Marolf:2022jra}, where the authors provide a microcanonical generalisation of the canonical treatment analysed in \cite{Marolf:2022ntb}.}, thereby leading to the following expression 
 
 \begin{equation} 
\boxed{\ \ \ 
 \left(\pi_{L}^{2}\right)\bigg|_{\hat z\pm\epsilon} 
 \ 
 = 
 \ 
4 \left[\left(\frac{\phi^{\prime}}{L}\right)^{2}\bigg|_{\hat z\pm\epsilon} + {\cal C}_{\pm} +\Lambda_{\pm}\phi^{2}\right],  \textcolor{white}{\Biggl [} \color{black} \ \ } 
 \label{eq:scale} 
 \end{equation} 
with ${\cal C }_{\pm}$ denoting an integration constants. A key feature emerging from this analysis is that (\ref{eq:scale}) is a dimensionful relation. This follows from the fact that $[\phi]=0$ in 2D, hence $[{\cal C}_{\pm}]=2$.  
 The junction conditions can be extracted upon integrating (\ref{eq:Hzc}) across the brane. In the rest frame of the latter, we get

\bea
\begin{cases}
 -\int_{\hat z-\epsilon}^{\hat z+\epsilon} dz\ {\cal H}_{z} 
 \ 
= 
 \ 
\pi_{L}\bigg|_{\hat z-\epsilon}^{\hat z+\epsilon} - \hat\pi_{\hat z}
 \ 
 = 
 \ 
 0 \nonumber \\ 
 -\int_{\hat z-\epsilon}^{\hat z+\epsilon} dz\ {\cal H}_{t} 
 \ 
=
 \ 
 -2\frac{\phi^{\prime}}{L}\bigg|_{\hat z-\epsilon}^{\hat z+\epsilon} -\frac{\sigma N^{t}}{\sqrt{(N^{t})^2+L^{2}(N^{z}+ {\dot{\hat{z}}})^{2}}} 
 \ 
 = 
 \ 
 0 
 \end{cases}\ \ \ \ \overset{N^{t}=1\ ,\ N^{z}=0}{\longrightarrow}\ \ \  \begin{cases}\frac{\phi^{\prime}}{L}\bigg|_{\hat z_{_{-}}}^{\hat z_{_{+}}}    
 \ 
 = 
 \ 
 -\frac{\sigma}{2} 
\\ 
\pi_{L}\bigg|_{\hat z_{_{-}}}^{\hat z_{_{+}}} 
 \ 
 = 
 \ 
 0,  
 \end{cases}
 \label{eq:jc}
 \eea
where the extrinsic curvatures on either side read

\begin{equation}  
\boxed{\ \ \ \frac{\phi^{\prime}}{L}\bigg|_{\hat z_{_{\pm}}} 
\ 
= 
\ 
\frac{(\Lambda_{+}-\Lambda_{-})\phi^{2}}{\sigma}\mp\frac{\sigma}{4} +\frac{{\cal C}_{+}-{\cal C}_{-}}{\sigma}. \textcolor{white}{\Biggl [} \color{black} \ \ } 
\ \ \ 
\label{eq:genexpr}    
\end{equation} 
Equating the new expression for $\pi_{L}$ with the one obtained from the variation of ${\cal L}$, we get

\begin{equation} 
\pi_{L}\bigg|_{\hat z_{_{+}}} 
\ 
= 
\ 
-2\dot\phi.
\end{equation} 
The constants ${\cal C_{+}}, {\cal C}_{-}$ can be absorbed by a rescaling of $\phi$ and $\pi_L$ . We will partially use this freedom to choose  ${\cal C_{+}}={\cal C}_{-}$ . Squaring both sides and
redefining, where the $b$-subindex denotes the value of the dilaton at the boundary. $\Phi_{b}=\phi_{b}/\sqrt{|{\cal C}|}$ and solving this quadratic  equation w.r.t. $\dot\Phi_{b}^{2}$  leads to the energy conservation relation 

\begin{equation} 
\boxed{\ \ \ \dot\Phi_{b}^{2}+V_{eff}
 \ 
 = 
 \ 
-1, \textcolor{white}{\Biggl [} \color{black} \ \ } 
 \end{equation} 
with effective quadratic  potential defined as

 \begin{equation} 
 \boxed{\ \ \ V_{eff} 
 \ 
 = 
 \ 
-\ \frac{1}{4}\left[\left( \frac{\Lambda_{+}-\Lambda_{-}-\frac{\kappa^{2}}{4}}{\kappa\sqrt{|{\cal C}|}}\right)^{2}+\Lambda_{+}\right]\Phi_{b}^{2}-\frac{3}{4},\textcolor{white}{\Biggl [} \color{black} \ \ }
\end{equation} 
where we have used that $\sigma=\kappa\phi_{b}$. The value of the dilaton at the turning point is found by setting $\dot\Phi_{_{b}}=0$, i.e. when the classical motion is reversed, and solving for the unique turning point 

\begin{equation} 
V_{eff} 
\ 
= 
\ 
-1 
\ \ \ 
\Rightarrow 
\ \ \ 
\Phi_{o}
\ 
= 
\ 
\left[\left(\frac{\Lambda_{+}-\Lambda_{-}-\frac{\kappa^{2}}{4}}{\kappa\sqrt{|{\cal C}|}}\right)^{2}+\Lambda_{+}\right]^{-\frac{1}{2}}. 
\label{eq:phio}    
\end{equation} 
Notice the similarity with the 4D case from FMP where the dilaton is playing the role of the compactification radius.    


\subsubsection*{Metric and dilaton profiles} \label{subsec:422a} 


We now determine the solutions to the e.o.m. arising from the variational principle. These are needed for evaluating the bulk and boundary actions, addressed next.     
In conformal gauge, the constraint ${\cal R}=2\Lambda$ can be solved in order to recover an expression for $L$ in terms of the $z$-coordinate. In the Lagrangian density we started from, $\phi({\cal R}-2\Lambda_{\pm})$ plays the role of a perturbation around the background action which, for the 2D case is purely topological. For the given gauge choice, the metric and dilatonic profile can be expressed as

\begin{equation} 
 ds^{2} 
\ 
= 
\ 
\frac{1}{z^{2}}\left(-dt^{2}+dz^{2}\right) 
\ \ \ 
, 
\ \ \ 
\text{with} 
\ \ \ L 
\ 
= 
\ 
\frac{1}{z}     
\ \ \ 
, 
\ \ \ 
\phi 
\ 
= 
\ 
\text{const.} 
\label{eq:Pcads}    
\end{equation} 
The e.o.m. arising from ${\cal L}_{_{JT}}$, i.e. 

\begin{equation} 
\phi\left(-\frac{L^{\prime\prime}}{L}+\frac{2L^{\prime2}}{L^{2}}\right)    
\ 
= 
\ 
2\phi\Lambda_{\pm},     
\ \ \ \ \ \ \ \ 
\ \ \ \ \ \ \ \ 
\phi^{\prime\prime} 
\ 
= 
\ 
2\phi \Lambda_{\pm} L^{2},
\label{eq:1eom}    
\end{equation} 
are solved by  

\be      
\boxed{\ \ \ L 
\ 
= 
\ 
\frac{c}{\sinh(bz)} 
\ \ \ 
, 
\ \ \ 
\phi 
 \ 
 = 
 \ 
 2 a\coth(bz),\textcolor{white}{\Biggl [} \color{black}  \ \ } 
 \ \ \ 
 \text{with } 
 \ \ \ 
 \Lambda_{\pm}    
 \ 
\overset{\text{def.}}{=}     
 \ 
\frac{b^{2}}{c^{2}}.
\label{eq:sol1}     
\ee   
 For the case of  pure AdS$_{2}$ in Poincarè coordinates, (\ref{eq:Pcads}), the metric diverges at the location of the conformal boundary, placed at $z=0$. The latter can be mapped to the horizon at $\rho=\gamma$ via the following coordinate transformation      

\begin{equation} 
\boxed{\ \ \ z 
\ 
\overset{\text{def.}}{=}     
\ 
\frac{1}{b}\coth^{-1}\left(\frac{\rho}{\gamma}\right) \textcolor{white}{\Biggl [} \color{black} \ \ } 
\ \ \ \Rightarrow 
\ \ \ 
dz 
\ 
= 
\ 
\frac{\gamma}{b}\frac{d\rho}{\gamma^{2}-\rho^{2}},     
\label{eq:coordtr1} 
\end{equation} 
whereas the dilatonic profile becomes linearly-dependent on the redefined spatial coordinate 

\be 
\boxed{\ \ \ \phi 
\ 
= 
\ 
2a\frac{\rho}{\gamma}.\textcolor{white}{\Biggl [} \color{black}  \ \ }    
\end{equation} 
Notice that, going from $\rho$ to $z$ in (\ref{eq:coordtr1}), is analogous to the analytic extension of the metric beyond the horizon. In particular, such transformation takes place by introducing an additional scale, $\mu$.
Upon substituting (\ref{eq:coordtr1}) in the original line element (\ref{eq:Pcads}), we get

\begin{equation} 
ds^{2} 
\ 
= 
\ 
\frac{\rho^{2}-\gamma^{2}}{\gamma^{2}}\left(-dt^{2}+\frac{\gamma^{2}}{b^{2}}\frac{d\rho^{2}}{(\rho^{2}-\gamma^{2})^{2}}\right).    
\label{eq:newle} 
\end{equation} 
The main feature of this result is the explicit emergence of a horizon at $\rho=\gamma$ if $\gamma^{2}>0$. For the $\gamma^{2}<0$ case, instead, there is no horizon. Rewriting (\ref{eq:newle}) in terms of $\phi$, with $\gamma=1/b$, we get    

\begin{equation} 
\boxed{\ \ ds^{2} 
\ 
= 
\ 
\left(\frac{\phi^{2}}{4a^{2}}-1\right)\left(-dt^{2}+\frac{1}{4a^{2}b^{2}}\frac{d\phi^{2}}{\left(\frac{\phi^{2}}{4a^{2}}-1\right)^{2}}\right). \ \ } 
\label{eq:newlee} 
\end{equation} 
The vanishing of the conjugate momentum w.r.t. $L$ imply:   

\be
\ \ \ \ \boxed{\ \ \ \phi^{2} 
\ 
= 
\ 
-\frac{{\cal C}}{\Lambda_{\pm}} \ \overset{def.}{=}\ \mu_{\pm}^2, \color{white}\bigg]\color{black}\ \ }    
\label{eq:pil3} 
\ee   
where the system is understood to be evaluated at the turning point in the $N^{z}=0$ gauge.
Equation \eqref{eq:pil3} enables to identify the parameter in (\ref{eq:newlee}) $2a=\sqrt{-{\cal C}/\Lambda_{\pm}}=\mu_{\pm}$ as the cosmological horizon in a dS$_{2}$ metric. Furthermore, from (\ref{eq:sol1}), we get: 

\be 
\phi^{\prime} 
\ 
=  
 \ 
 2ab\left(1-\coth^{2}(bz)\right) 
\ 
=  
 \ 
 2ab-\frac{b}{2a}\phi^{2} 
 \ 
 = 
 \ 
 \frac{b}{2a}\left(4a^{2}-\phi^{2}\right),
 \label{eq:2v}    
 \ee 
 from which it can be appreciated that the change in sign of the extrinsic curvature ($\phi^{\prime}$) takes place when crossing the cosmological horizon at $\phi=2a$. 


\section*{Bulk and boundary actions}   


In the Hamltonian formalism, the transition rate is defined in terms of the extremised total action including, both, bulk and wall terms 
\be
\Gamma(A\rightarrow A\oplus W\oplus B)= \exp\left[-S_{tot}+S_{bckgr}\right],
\ee
where $A,B$ denote the spacetimes involved separated by the brane $W$, and  

\begin{equation} 
S_{tot}^{\ \text{A}/\text{B}}  
\ 
= 
\ 
S_{top}+S_{bulk}+S_{brane} \ \ \ ,\ \ \  S_{bckgr} 
\ 
 \overset{def.}{=} 
\ 
 \frac{\phi_{bckgr}}{16\pi G} \chi_{_{\cal M}}.
\label{eq:totAB}
\end{equation}
The topological term $S_{top}$ cancels with the background contribution
when evaluating the transition rate. 
The extremised action is obtained by integrating over the Hamilton-Jacobi equations. By imposing the secondary constraints derived above, the variation of the bulk action therefore reads 

\begin{equation} 
\delta S_{bulk} 
\ 
= 
\ 
\int dz\left[\delta L(z)\pi_{L}+\delta\phi(z)\pi_{\phi}\right] 
\ 
= 
\ 
2\int dz\ \delta L(z)\pi_{L} 
\label{eq:var} 
\end{equation} 

\begin{equation} 
\begin{aligned}  
\Rightarrow 
\ \ \ 
S_{bulk}^{(+)} 
\ 
&=    
\ 
\frac{2\eta}{G}\int dz\ \int dL \sqrt{\left(\frac{\phi^{\prime}}{L}\right)^{2}+{\cal C}+\Lambda_{+}\phi^{2}} \\    
&= 
\frac{2\eta}{ G}\int_{0}^{\hat z} dz\ \left[ L\  \sqrt{\left(\frac{\phi^{\prime}}{L}\right)^{2}+{\cal C}+\Lambda_{+}\phi^{2}}  -\phi^{\prime}\ \cosh^{-1}\left(\frac{\phi^{\prime}}{L\sqrt{-{\cal C}-\Lambda_{+}\phi^{2}}}\right)  \right].
\label{eq:bulk1} 
\end{aligned}   
\end{equation}
Similarly, for the other vacuum, 
\begin{equation} 
S_{bulk}^{(-)} 
\ 
= 
\ 
\frac{2\eta}{ G}\int_{\hat z}^{\infty} dz\ \left[ L\  \sqrt{\left(\frac{\phi^{\prime}}{L}\right)^{2}+{\cal C}+\Lambda_{-}\phi^{2}}  -\phi^{\prime}\ \cosh^{-1}\left(\frac{\phi^{\prime}}{L\sqrt{-{\cal C}-\Lambda_{-}\phi^{2}}}\right)  \right].
\label{eq:bulk2} 
\end{equation}

The bulk action is extremised when the integrand featuring in the first term of (\ref{eq:bulk1}) and (\ref{eq:bulk2}) vanishes, therefore leaving with the corresponding contribution from the value of the dilaton at the turning point and at the horizons characterising spacetimes being involved. For such values, the argument of $\cosh^{-1}$ is imaginary, therefore, we can rewrite it as $\cos^{-1}$, which provides an overall $\pi$ factor. The extremes of integration for, both, (\ref{eq:bulk1}) and (\ref{eq:bulk2}) are dictated by the relative position of the innermost horizon in a given spacetime w.r.t. the turning point $\phi_{o}$ , whose expression was explicitly derived in the previous subsection, cf. eq. (\ref{eq:phio}). From equations (\ref{eq:2v}) we therefore get the following contributions

\begin{equation} 
S_{bulk}^{(-)} 
\ 
= 
\ 
\frac{2\pi\eta}{ G}\int_{z_{tp}}^{z_{h,-}} dz\ \phi^{\prime} 
\ 
= 
\ 
\frac{2\pi\eta}{ G}\int_{\mu_{-}}^{\phi_{o}\ \epsilon_{-}} d\phi\ 
\ 
= 
\ 
\frac{2\pi\eta}{ G}\left[\phi_{o}\ \Theta\left(-\frac{\phi^{\prime}}{L}\bigg|_{-}\right)-{\mu_{-}}\right],   
\label{eq:bulk3} 
\end{equation} 
with $\Theta\left(-\frac{\phi^{\prime}}{L}\bigg|_{-}\right)$ being the $\Theta$-function that is non-vanishing only if the extrinsic curvature on the inner side of the brane is negative, thereby indicating that the turning point is placed beyond the horizon . The same argument follows for the bulk integral over the outer vacuum.For the case involving pure AdS, instead, due to the absence of horizons, only the $\phi_{o}$-terms will be contributing as relevant extremes of integration. Furthermore, precisely because of this, the extrinsic curvature on either side will also preserve the sign over the entire spacetime, thereby leading to the mutual cancellation of the $\Theta$-terms. Combining the claims just made, the corresponding result for mixed AdS/dS configurations follows.  
For the 3 cases of interest, the total bulk action therefore reads 

\begin{equation} 
\boxed{\ \ \ \begin{aligned}
&S_{bulk}^{\  \text{dS}\rightarrow\text{dS}} 
\ 
= 
\ 
S_{bulk}^{(+)}+S_{bulk}^{(-)} 
\ 
= 
\ 
\frac{2\pi\eta}{G}\left[\phi_{o}\ \Theta\left(-\frac{\phi^{\prime}}{L}\bigg|_{-}\right)-{\mu_{-}}-\phi_{o}\ \Theta\left(-\frac{\phi^{\prime}}{L}\bigg|_{+}\right)+{\mu_{+}}\  \right]   \\
\\    
&S_{bulk}^{ \ \text{AdS}\rightarrow\text{AdS}} 
\ 
= 
\ 
S_{bulk}^{(+)}+S_{bulk}^{(-)} 
\ 
= 
\ 
0\\
\\    
& S_{bulk}^{\  \text{dS}\rightarrow\text{AdS}} 
\ 
= 
\ 
S_{bulk}^{(+)}+S_{bulk}^{(-)} 
\ 
= 
\ 
\frac{2\pi\eta}{G}\left[\phi_{o}\ \ \Theta\left(-\frac{\phi^{\prime}}{L}\bigg|_{-}\right)-\phi_{o}\ \Theta\left(-\frac{\phi^{\prime}}{L}\bigg|_{+}\right)+{\mu_{+}}\  \right].
\label{eq:actdsds}   
\end{aligned}\ \ \ } 
\end{equation}

The brane action is obtained by considering the variation of  (\ref{eq:var}) at the location of the wall $z=\hat z$. From the junction conditions, $\hat \pi_{L}\bigg|_{+}=\hat \pi_{L}\bigg|_{-}$, it follows that only the second term in, both,  (\ref{eq:bulk1}) and  (\ref{eq:bulk2}) contributes nontrivially  

\bea   
S_{brane} 
&
= 
 &
\frac{2}{G}\int_{0}^{{\phi}_{o}}\ d\phi_{b}\left[\cosh^{-1}\left(\frac{\phi_{b}^{\prime}(\hat z-\epsilon)}{L\sqrt{-{\cal C}-\Lambda_{-}\phi_{b}^{2}}}\right)-\cosh^{-1}\left(\frac{\phi_{b}^{\prime}(\hat z+\epsilon)}{L\sqrt{-{\cal C}-\Lambda_{+}\phi_{b}^{2}}}\right)   \right]\nonumber \\
&
= 
&
\frac{2}{G}\int_{0}^{{\phi}_{o}}\ d\phi_{b}\left[\cosh^{-1}\left(\frac{A_{1}\phi_{b}}{\sqrt{1-\frac{\Lambda_{-}}{|{\cal C}|}\phi_{b}^{2}}}\right)-\cosh^{-1}\left(\frac{A_{2}\phi_{b}}{\sqrt{1-\frac{\Lambda_{+}}{|{\cal C}|}\phi_{b}^{2}}}\right)   \right] ,
\label{eq:brane2in1}
\eea
where

\be
A_{1} 
\ 
\overset{\text{def.}}{=}
\ 
\frac{\Lambda_{+}-\Lambda_{-}+\frac{\kappa^{2}}{4}}{\kappa\sqrt{-{\cal C}}} 
\ \ \ 
, 
\ \ \ 
A_{2} 
\ 
\overset{\text{def.}}{=} 
\ 
\frac{\Lambda_{+}-\Lambda_{-}-\frac{\kappa^{2}}{4}}{\kappa\sqrt{-{\cal C}}}.
\end{equation} 
The brane action associated to the cases listed in \eqref{eq:actdsds} reads:

\bea
S_{brane}^{\  \text{dS}/\text{dS}} 
&
= 
\frac{2\pi\eta}{G}\left[\phi_{o}\ \Theta\left(-\frac{\phi^{\prime}}{L}\bigg|_{-}\right)-\phi_{o}\ \Theta\left(-\frac{\phi^{\prime}}{L}\bigg|_{+}\right)+\right.\\
& \left.+\frac{{\mu_{-}}}{2\pi} \ln\bigg|\frac{{\mu_{-}}A_{1}+1}{{\mu_{-}}A_{1}-1}\bigg| -\frac{{\mu_{+}}}{2\pi} \ln\bigg|\frac{{\mu_{+}}A_{2}+1}{{\mu_{+}}A_{2}-1}\bigg| \right],  \nonumber
\label{eq:vr} 
\eea

\begin{equation}    
S_{brane}^{\  \text{AdS}/\text{AdS}}
\ 
= 
\     
\frac{\eta}{G}\bigg[{\mu_{-}} \ln\bigg|\frac{{\mu_{-}}A_{1}+1}{{\mu_{-}}A_{1}-1}\bigg| -{\mu_{+}}\ln\bigg|\frac{{\mu_{+}}A_{2}+1}{{\mu_{+}}A_{2}-1}\bigg| \bigg] .    
\label{eq:vr1} 
\end{equation}   
 The $\Theta$-terms in (\ref{eq:vr1}) cancel each other out for the same argument outlined when dealing with the bulk action. For the case of mixed transitions of the dS$\rightarrow$AdS-kind, the brane action would be structurally equivalent to that of the dS$\rightarrow$ dS case, once having suitably accounted for the difference in sign of the cosmological constants involved.


\subsection{Transitions among dS, AdS and Minkowski vacua}


In summary we can write the transitions so far as:
\begin{equation}  
\boxed{\ \ \ \begin{aligned}    
\\
&\ln\Gamma_{_{\ \text{AdS}\rightarrow\text{AdS}}  }    
\ 
= 
\ 
\frac{\eta}{G}\left[{\mu_{-}} \ln\left|\frac{{\mu_{-}}A_{1}+1}{{\mu_{-}}A_{1}-1}\right| -{\mu_{+}} \ln\left|\frac{{\mu_{+}}A_{2}+1}{{\mu_{+}}A_{2}-1}\right| \right]  \\ 
\\
&  \ln\Gamma_{_{\ \text{dS}\rightarrow\text{dS}} }     
\ 
= 
\ 
\frac{2\pi\eta}{G}\left[{\mu_{+}}-{\mu_{-}}+\frac{{\mu_{-}}}{2\pi} \ln\left|\frac{{\mu_{-}}A_{1}+1}{{\mu_{-}}A_{1}-1}\right| -\frac{{\mu_{+}}}{2\pi} \ln\left|\frac{{\mu_{+}}A_{2}+1}{{\mu_{+}}A_{2}-1}\right| \right]\\
\\
& \ln\Gamma_{_{\ \text{dS}\rightarrow\text{AdS}}  }    
\ 
= 
\ 
\frac{2\pi\eta}{G}\left[{\mu_{+}}+\frac{{\mu_{-}}}{2\pi} \ln\left|\frac{{\mu_{-}}A_{1}+1}{{\mu_{-}}A_{1}-1}\right| -\frac{{\mu_{+}}}{2\pi} \ln\left|\frac{{\mu_{+}}A_{2}+1}{{\mu_{+}}A_{2}-1}\right| \right].
\\
\label{eq:joined1} 
\end{aligned} \ \ }    
\end{equation} 
Their equivalence with the corresponding expressions in BT becomes manifest under the following parametric redefinitions

\begin{equation} 
\begin{aligned}
\bar z  
&=-\frac{1}{4\sqrt{\Lambda_+\ }}\ \ln\left|\ 1-\frac{\phi_{tp,+}^{new\ 2}}{\Lambda_{_{+}}}\ \right|\ \ \ ,\ \ \  
\phi_{tp}^{new\ 2}    
\overset{def.}{=}    
\Lambda_{\pm}A_{_{1,2}}\mu_{\pm}  
= 
A_{_{1,2}}\sqrt{\frac{-{\cal C}}{\Lambda_\pm}\ }\ \Lambda_{_{BT}}.\   
\end{aligned}
\end{equation}

\subsubsection*{Comparing dS, AdS and Minkowski transitions}

Some important observations can be drawn when comparing (\ref{eq:joined1}). 
\begin{itemize}

\item{\it {\bf Horizons' contribution} }

The first main difference between the AdS and dS transitions is the horizon contribution to the amplitude, which is crucial when considering the flat limit. A detailed explanation and interpretation of these findings will be outlined in section \ref{sec:5}.

    \item{\it {\bf Bounds on wall tension}} 
    
    For the AdS$_2\rightarrow$AdS$_2$ case, the classical turning point $\phi_{o}$ is real iff the tension of the brane is constrained within the following range

\begin{equation} 
\boxed{\ \ \    
\kappa      
\ 
<    
\ 
 2\left|\sqrt{-\Lambda_{+}}-\sqrt{-\Lambda_{-}}\right| .   \textcolor{white}{\Biggl [} \color{black}         
\ \ }    
\label{eq:range}    
\end{equation}

\item{\it {\bf The flat limit from AdS}} 

Interestingly, upon taking, either, $\Lambda_{\pm}=0$ in AdS$_2\rightarrow$AdS$_2$ processes, the amplitude is still finite

\begin{equation} 
\boxed{\ \ \ \begin{aligned} 
&S_{tot}^{\ \text{Mink.}\rightarrow \text{AdS}} 
\ 
= 
\ 
\frac{\eta}{G}\ {\mu_{-}}\ \ln\bigg|\frac{{\mu_{-}}A_{1}+1}{{\mu_{-}}A_{1}-1}\bigg|   \\   
\\    
& S_{tot}^{\ \text{AdS}\rightarrow \text{Mink}} 
\ 
= 
\ 
-\frac{\eta}{G}\ {\mu_{+}}\ \ln\bigg|\frac{{\mu_{+}}A_{2}+1}{{\mu_{+}}A_{2}-1}\bigg|
\label{eq:mads} 
\end{aligned}
\ \ } 
\ \ \ \ \text{with}\ \ \ A_{1} 
\ 
\overset{\text{def.}}{=}
\ 
\frac{-\Lambda_{-}+\frac{\kappa^{2}}{4}}{\kappa\sqrt{-{\cal C}}} \ \ ,\ \ A_{2} 
\ 
\overset{\text{def.}}{=}
\ 
\frac{\Lambda_{+}+\frac{\kappa^{2}}{4}}{\kappa\sqrt{-{\cal C}}}.
\end{equation} 
Therefore up and down-tunneling are allowed in this case. Finiteness follow from the fact that the brane action vanishes on the side where the cosmological constant is turned off.



\item{\it {\bf Detailed balance}}

Under the exchange $\mu_{+}\leftrightarrow\mu_{-}$, with ${\cal C}_{+}={\cal C}_{-}$,  we get $A_1 leftrightarrow A_2$ therefore, from  (\ref{eq:vr}), we obtain:

\be
\boxed{\ \ \ \frac{\Gamma_{1\rightarrow 2}}{\Gamma_{2\rightarrow 1}}= e^{\frac{8\eta\pi}{G}\left(\phi_{+}^{h}-\phi_{-}^{h}\right)}=e^{\eta(S_1-S_2),}  \textcolor{white}{\Biggl [} \color{black} \ \ }    
\label{eq:detbal}
\ee   
proving that the principle of detailed balance is satisfied, once we identify   the value of the dilaton at the horizon with the entropy. It is interesting to notice that detailed balance is still working even in this setup, thereby consistently proving that the value of the dilaton at the horizon is playing the role of the black hole entropy. 
However, the particular nature of the cosmological horizon leads to unitarity issues upon taking the flat spacetime limit from a pure dS setup.


\item{\it {\bf The flat limit from dS}}

The main difference w.r.t. the AdS case is that, this time, the definition of the turning point, $\phi_o$, is not subject to any parametric constraint, since $\Lambda_{\pm}>0$. However, the presence of an additional term proportional to the cosmological horizon in $S_{_{TOT}}$, inevitably leads to a divergent action upon taking the flat limit on eiter side of the brane

\begin{equation} 
\boxed{\ \ \ 
\begin{aligned}
&S_{tot}^{\ \text{Mink.}\rightarrow \text{dS}} 
\ 
= 
\ 
\frac{2\eta\pi}{G}\left[\ -{\mu_{-}}+\frac{{\mu_{-}}}{2\pi}\ \ln\bigg|\frac{{\mu_{-}}A_{1}+1}{{\mu_{-}}A_{1}-1}\bigg|+{\mu_{+}}\right] \rightarrow \infty \\
\\
&S_{tot}^{\ \text{dS}\rightarrow \text{Mink.}} 
\ 
= 
\ 
\frac{2\eta\pi}{G}\left[\ {\mu_{+}}-\frac{{\mu_{+}}}{2\pi}\ \ln\bigg|\frac{{\mu_{+}}A_{2}+1}{{\mu_{+}}A_{2}-1}\bigg|-{\mu_{-}}\right] \rightarrow -\infty, 
\end{aligned}\ \ }    
\end{equation} 
for up and down-tunneling, respectively. In particular, this proves that there is no 2D counterpart of the CDL transition for dS$\rightarrow$ Mink. It is an important feature of the 2D theory that the divergence does not cancel, thereby signalling the need for a deeper understanding. Indeed, when taking the $\underset{\mu_{+}\rightarrow \infty}{\lim}$, we are really moving very far away w.r.t. the perturbative regime where the near-horizon approximation of JT-gravity is valid. As the horizon moves deeper inside the IR bulk, so too does the turning point. However, $\phi_{o}$ will at most asymptote to a finite value, ultimately reemerging from the horizon in the form of a naked singularity. At this point, the causal structure of the spacetime prevents the nucleation of a new bulk phase.  





\end{itemize}


\section{Lorentzian transitions with black holes}\label{sec:3}


\subsection{Actions with non-vanishing constant energies} \label{sec:aed0}

We will now turn to the more general case in which a constant term (w.r.t. $\phi$) is being added to the Lagrangian density, which now reads

\bea
  {\cal L} 
& 
= 
& 
   \sqrt{-g}\phi\left[{\cal R}-2\Lambda_{+}\Theta(z-\hat z)-2\Lambda_{-}\Theta(\hat z-z)\right]-\delta(z-\hat z)\sigma\sqrt{(N^{t})^2+L^{2}(N^{z}+ {\dot{\hat{z}}})^{2}}\nonumber\\
&+& 
\sqrt{-g}\left[ \phi_{o}\chi_{_{{\cal M}}}-{\cal B}_{+}\Theta(z-\hat z)-{\cal B}_{-}\Theta(\hat z-z)\right].
\eea    

The additional $\cB$-terms play the same role as the black hole mass in the 4D analysis carried out in \cite{Fischler:1990pk}. Indeed, such term contributes to the 2D metric function by means of a linear term in $\phi$, therefore featuring the appropriate scaling for a 2D black hole. However, unlike the 4D case, where $M$ emerges as an integration constant, the lower dimensionality of the current analysis requires it to be defined as a parameter of the theory. 

The procedure for determining the total action  is very similar to the one carried out in section \ref{sec:2}, hence we will just highlight the main differences. The Hamiltonian constraint picks up additional ${\cal B}_{\pm}$ terms, whereas the momentum constraint remain unaltered. This follows from the fact that the determinant of the metric is $N^{z}$-independent. In particular, the latter ensures that the relation in between the conjugate momenta $\pi_{L}$ and $\pi_{\phi}$ is preserved. Upon integrating the Hamiltonian constraint across the wall, the conjugate momentum w.r.t. the metric reads

 \begin{equation} 
 \boxed{\ \ \left(\pi_{L}^{2}\right)\bigg|_{\hat z_{_{\pm}}} 
 \ 
 = 
 \ 
 4\ \left[\left(\frac{\phi^{\prime}}{L}\right)^{2}\bigg|_{\hat z_{_{\pm}}}  + {\cal C}_{\pm} + {\cal B}_{\pm}\phi +\Lambda_{\pm}\phi^{2}\right], \ \ } 
 \label{eq:dual}    
 \end{equation} 
and the extrinsic curvatures 

\begin{equation} 
\frac{\phi^{\prime}}{L}\bigg|_{\hat z_{_{\pm}}} 
\ 
= 
\ 
\frac{(\Lambda_{+}-\Lambda_{-})\phi^{2}}{\sigma}\mp\frac{\sigma}{4} +\frac{{\cal C}_{+}-{\cal C}_{-}}{\sigma} +\ \frac{({\cal B}_{+}-{\cal B}_{-})\phi}{\sigma} 
\ \ \ 
,
\label{eq:genexpr1}    
\end{equation} 
from which the junction conditions follow
similarly to the ${\cal B}_{\pm}=0$ case. Upon performing the same procedure outlined in section \ref{sec:2}, and still requiring ${\cal C}_{+}={\cal C}_{-}$, the effective potential reads 

 \begin{equation} 
 V_{eff} 
 \ 
 = 
 \ 
-\ \frac{1}{4}\left[\left(\left( \frac{\Lambda_{+}-\Lambda_{-}-\frac{\sigma^{2}}{4}}{\sigma} \right)\Phi_{b}^{2}+\frac{({\cal B}_{+}-{\cal B}_{-})\Phi_{b}}{\sqrt{|{\cal C}|}\sigma} \right)^{2}-1+\frac{{\cal B}_{+}}{\sqrt{|{\cal C}|}}\Phi_{b}+\Lambda_{+}\Phi_{b}^{2}\right]-1.
\end{equation} 
We will restrict to the case ${\cal B}_{+}={\cal B}_{-}={\cal B}$. The general case  ${\cal B}_{+}\neq{\cal B}_{-}$ can be considered without changing the conclusions. 
 For ${\cal B}_{+}={\cal B}_{-}$, the turning points are

\begin{equation}     
\boxed{\ \ \  \Phi_{1,2}   
\ 
\overset{\text{def.}}{=}     
\ 
-\frac{{\cal B}\Phi_{o}^{2}}{2\sqrt{|{\cal C}|}}\ \left[\ 1\pm\sqrt{1+\frac{4{\cal C}}{{\cal B}^{2}\Phi_{o}^{2}}}\ \right],  \textcolor{white}{\Biggl [} \color{black}\ \ }    
\label{eq:phi12}    
\end{equation} 
which, in the ${\cal B}\rightarrow0$ limit, consistently reduces to $|\phi_{1}|=|\phi_{2}|=\phi_{o}$, thereby correctly recovering the result obtained in section \ref{sec:2}. From (\ref{eq:phi12}), the existence of 2 physical turning points is ensured upon requiring
 the following parametric constraints 
 
\be 
\boxed{\ \ \ -\frac{1}{4}<\frac{{\cal C}}{{\cal B}^2\Phi_{o}^2}<0, \ \ \ \ \ \ \text{with}\ \ \ {\cal B}, {\cal C}<0, \textcolor{white}{\Biggl [} \color{black}\ \ }   
\label{eq:constrm}
\ee 
which can also be rewritten as follows   

\be 
\boxed{\ \ \ -2\sqrt{|\Lambda{\cal C}|\ }<{\cal B}<0.  \textcolor{white}{\Biggl [} \color{black}\ \ }   
\label{eq:constrm1}
\ee 
In particular, \eqref{eq:constrm1} defines a bound for ${\cal B}$, which, as will be argued in section \ref{sec:fl}, plays an essential role in our analysis. Indeed, as shown next, this parameter is related to the black hole mass.  


\subsubsection*{Metric and dilaton profiles}


Let us generalise the results of section  \ref{subsec:422a}  to the case $\cB\neq 0$.  Here, the dilatonic e.o.m. pick up an additional constant term, and, consequently, so too does the field, 

\begin{equation} 
 \phi^{\prime\prime} 
\ 
= 
\ 
2\phi L^{2}\Lambda_{\pm} \pm  {\cal B}L^{2} 
\ \ \ 
\Rightarrow 
\ \ \ 
\phi 
\ 
= 
\    
2a\coth(bz)\pm\frac{{\cal B}}{2\Lambda_{\pm}}. 
\label{eq:2eom}        
\end{equation}    
Under the coordinate transformation (\ref{eq:coordtr1}), the line element turns into 

\begin{equation} 
\boxed{\ \ ds^{2} 
\ 
= 
\ 
\frac{\left(\phi\pm\frac{{\cal B}}{2\Lambda_{\pm}}\right)^{2}-4a^{2}}{4a^{2}}\left(-dt^{2}+\frac{4a^{2}}{b^{2}}\frac{ d\phi^{2}}{\left(\left(\phi\pm\frac{{\cal B}}{2\Lambda_{\pm}}\right)^{2}-4a^{2}\right)^{2}}\right), \ \ } 
\label{eq:newlee1} 
\end{equation} 
which is exactly of the S(A)dS-kind. The parameters featuring in (\ref{eq:newlee1}) can be related to the ones we have and will be using in the expressions for the total actions via the following identification 

\be 
\frac{{\cal B}^{2}}{4\Lambda_{\pm}^{2}}-4a^{2}  
\ 
= 
\ 
\frac{{\cal  C}}{\Lambda_{\pm}} 
\ \ \ 
\Rightarrow 
\ \ \ 
{\cal B} 
\ 
= 
\ 
\pm 2\Lambda_{\pm}\sqrt{4a^{2}+\frac{\cal C}{\Lambda_{\pm}}}    
\ \ \ 
. \label{eq:conv}    
\ee      
Note that, in this case,  the metric has 2 horizons, associated to the roots of the metric function in (\ref{eq:newlee1}) that read

\be 
\phi_{h,1,2}^{\pm}   
\ 
 = 
 \ 
-\frac{|{\cal B}|}{2\Lambda_{\pm}}\left[\text{sign}_{\pm}({\cal B})\pm\frac{4a\sqrt{\Lambda_{\pm}}}{{\cal B}}\right]    
\ 
 \overset{(\ref{eq:conv})}{=}   
 \ 
-\frac{|{\cal B}|}{2\Lambda_{\pm}}\left[\text{sign}_{\pm}({\cal B})\pm\sqrt{1-\frac{4{\cal C}\Lambda_{\pm}}{{\cal B}^{2}}}\right].
\ \ \ 
\ee    
This shows the importance of introducing $\cal B$. Which sign is being identified with the innermost or outermost horizon of the given spacetimes entirely depends on the signs of the parameters of the theory $({\cal B}, \Lambda_{\pm})$. In particular, ${\cal B}$ will have opposite sign on the 2 sides to ensure the presence of 2 horizons for the $\Lambda_{\pm}>0$ case. Setting the outer horizon as a reference, the innermost horizon for ${\cal B}<0$\footnote{This condition follows from requiring both turning points to be physical, see eq. (\ref{eq:phi12}).} is given by the choice of the $+$ root. As a consequence of this, the outermost horizon on the inner vacuum is given by the$-$ sign, with ${\cal B}>0$. The change in sign for ${\cal B}$ in the 2 vacua ensures that we can sistematically use the same coordinate $\phi$ on both sides of the brane. Indeed, this is not something surprising, given that the value of the dilaton changes sign in different causal patches of a maximally-extended metric, and therefore it is in agreement with the description outlined above given that the nucleation process requires crossing the horizon of a given spacetime in static coordinates.

In the 4D case, FMP showed that the black hole mass arises as an integration variable, while in the 2D case we show that it is part of the theory, since it features at the level of the Lagrangian density. However, one should not be misled by this, in the sense that, unlike the Euclidean formalisms referred to in section 2, the main feature of the Hamiltonian method is precisely that of implementing constraints at the level of the Lagrangian, such that the total action describing the transition is effectively accounting for a joined system of two vacua separated by a wall. The black hole mass will therefore enter in the joined system's Lagrangian inside the Hamiltonian constraint, and the matching conditions will therefore act as a selection rule assigning to each spacetime a notion of state labelled by their respective value of the black hole mass, ${\cal B}_{\pm}$.


\subsection{Transitions with conical defects} 


The evaluation of the extremised total action can be carried out in complete analogy with the procedure applied in \ref{sec:2}. The bulk and brane actions read 

\bea   
S_{bulk}^{^{(\pm)}} 
\ 
= 
\ 
\frac{2\eta}{ G}\int dz\ \left[ L\  \sqrt{\left(\frac{\phi^{\prime}}{L}\right)^{2}+{\cal C}+{\cal B}\phi+\Lambda_{\pm}\phi^{2}}  -\phi^{\prime}\ \cosh^{-1}\left(\frac{\phi^{\prime}}{L\sqrt{-{\cal C}-{\cal B}\phi-\Lambda_{\pm}\phi^{2}}}\right)  \right],
\nonumber
\eea    

\bea   
S_{brane} 
&
= 
& 
\frac{2\eta}{G}\int_{\phi_{1}}^{\phi_{2}}\ d\phi_{b}\left[\cosh^{-1}\left(\frac{\phi_{b}^{\prime}(\hat z-\epsilon)}{L\sqrt{-{\cal C}-{\cal B}\phi_{b}-\Lambda_{-}\phi_{b}^{2}}}\right)-\cosh^{-1}\left(\frac{\phi_{b}^{\prime}(\hat z+\epsilon)}{L\sqrt{-{\cal C}-{\cal B}\phi_{b}-\Lambda_{+}\phi_{b}^{2}}}\right)   \right]. 
\nonumber
\eea
Once more, extremisation of the bulk actions occurs at the horizons according to the range of the mass parameter. This time, we have two horizons on either side, defined by 
\begin{equation} 
\phi_{h,1,2}^{+} 
\ 
\overset{\text{def.}}{=} 
\ 
-\frac{|{\cal B}|}{2\Lambda_{+}}\left[\text{sgn}({\cal B})\mp\sqrt{1-\frac{4\Lambda_{+}{\cal C}}{{\cal B}^{2}}}\right]    
\ \ \ 
, 
\ \ \ 
\phi_{h,1,2}^{-} 
\ 
\overset{\text{def.}}{=} 
\ 
-\frac{|{\cal B}|}{2\Lambda_{+}}\left[\text{sgn}({\cal B})\mp\sqrt{1-\frac{4\Lambda_{-}{\cal C}}{{\cal B}^{2}}}\right],
\label{eq:hor2}
\end{equation} 
with ${\cal C}_{\pm}<0$. Given that $\Lambda_{\pm}\ge 0$, for both horizons to be physical, it must be that $\text{sgn}({\cal B})=-1$. The bulk action is extremised at the outermost and innermost horizons on the 2 sides of the brane, respectively, (\ref{eq:hor2}), and at the turning point $\phi_{2}$, provided the argument of $\cosh^{-1}$ changes sign in $S_{bulk}$. 

The total bulk action reads
\bea               
S_{bulk} 
=     
\frac{2\eta}{G}\ \left[\ -\phi_{h,2}^{-}+\phi_{h,1}^{+}+\phi_{2}\ \left[\Theta\left(-\frac {\phi^{\prime}}{L}\bigg|_{-}\right)-\Theta\left(-\frac {\phi^{\prime}}{L}\bigg|_{+}\right)\right]\right],   
\ \ \ 
\eea 
and therefore we can write the transitions as:
 
\begin{equation}
\boxed{\ \ \      
\begin{aligned}      
S_{_{tot}}^{\text{(A)dS}\rightarrow\text{(A)dS}} 
& =
\frac{2\eta}{G} \left[ -\frac{{\cal B}}{2\Lambda_{+}} \left[1- \sqrt{1-\frac{4{\cal C}\ \Lambda_{+}}{{\cal B}^{2}}}-\ln\left|\frac{y_{4}}{y_{3}}\sqrt{\frac{1-y_{3}^{2}}{1-y_{4}^{2}}}\right| \right] +\right. 
\nonumber\\    
&   
+ 
\frac{{\cal B}}{2\Lambda_{-}}\ \left[1- \sqrt{1-\frac{4{\cal C}\ \Lambda_{-}}{{\cal B}^{2}}}-\ln\left|\frac{y_{2}}{y_{1}}\sqrt{\frac{1-y_{1}^{2}}{1-y_{2}^{2}}}\right|\right] +               
\nonumber\\  
& 
-  
\phi_{2}\ln\left|\frac{A_{2}\ y_{2}}{A_{4}y_{4}}\sqrt{\frac{1-y_{4}^{2}}{1-y_{2}^{2}}}\right|+\phi_{1}\ln\left|\frac{A_{2}\ y_{1}}{A_{4}y_{3}}\sqrt{\frac{1-y_{3}^{2}}{1-y_{1}^{2}}}\right| + 
\label{eq:totadsads} \\     
&    
-     
\left.\frac{1}{2}\left[\sqrt{\frac{\frac{{\cal B}^{2}}{4\Lambda_{-}}-{\cal C}}{\Lambda_{-}}}\ \ln\ \left|\frac{1-y_{2}}{1+y_{2}}\frac{1+y_{1}}{1-y_{1}}\right| - 
\sqrt{\frac{\frac{{\cal B}^{2}}{4\Lambda_{+}}-{\cal C}}{\Lambda_{+}}}\ \ln\ \left|\frac{1-y_{4}}{1+y_{4}}\frac{1+y_{3}}{1-y_{3}}\right|\right] \right],        
\end{aligned}       \  \ }
\end{equation}

where $y_{1,2} 
\equiv y_{3,4} 
\ 
= 
\ 
y(\phi_{1,2})$, 

\begin{equation} 
y 
\ 
\overset{\text{def.}}{= }    
\ 
\sqrt{\frac{\Lambda}{\frac{{\cal B}^{2}}{4\Lambda}-{\cal C}}}\left(\frac{{\cal B}}{2\Lambda}+\phi_{b}\right) 
\ \ \ 
,   \ \ \  \phi_{1,2}    
\ 
= 
\ 
-\frac{{\cal B}\phi_{o}^{2}}{2\sqrt{|{\cal C}|}}  \left[1\pm\sqrt{1+\frac{4|{\cal C}|}{{\cal B}^{2}\phi_{o}^{2}}}\right] 
\end{equation} 
and, for pure notational convenience, the following parametric redefinitions were performed

\begin{equation} 
\begin{aligned}
&A_{1} 
\ 
= 
\ 
\frac{\Lambda_{+}-\Lambda_{-}+\frac{\kappa^{2}}{4}}{\kappa\sqrt{{\cal C}-\frac{{\cal B}_{-}^{2}}{4\Lambda_{-}}}}, 
\ \ \  
\ \ \ 
A_{3} 
\ 
= 
\ 
\frac{\Lambda_{+}-\Lambda_{-}-\frac{\kappa^{2}}{4}}{\kappa\sqrt{{\cal C}-\frac{{\cal B}_{+}^{2}}{4\Lambda_{+}}}}  \\
\\
&A_{2}
\ 
\overset{\text{def.}}{= } 
\ 
\frac{A_{1}}{\sqrt{\Lambda_{-}}}\sqrt{\frac{{\cal B}^{2}}{4\Lambda_{-}}-{\cal C}},  
\ \ \  
\ \ \ A_{4}
\ 
\overset{\text{def.}}{= } 
\ 
\frac{A_{3}}{\sqrt{\Lambda_{+}}}\sqrt{\frac{{\cal B}^{2}}{4\Lambda_{+}}-{\cal C}}.  
\ \ \ 
\ \ \ 
\label{eq:a1a3}    
\end{aligned}
\end{equation} 
Setting ${\cal B}=0$, (\ref{eq:totadsads}) reduces to the results obtained in section \ref{sec:2}. Upon exchanging $\Lambda_{-}\longleftrightarrow \Lambda_{+}$, the turning points on the 2 sides are correspondently mapped to $y_{2}\rightarrow -y_{4}$ and $y_{1}\rightarrow -y_{3}$. From this follows that the arguments of all the $\ln$-terms are inverted. The ratio of direct and reverse processes, lead to an expression for detailed balance which reads

\be         
\boxed{\ \ \ln \frac{\Gamma_{\Lambda_{+}\rightarrow\Lambda_{-}}}{\Gamma_{\Lambda_{-}\rightarrow\Lambda_{+}}}
\ 
=    
\ 
\frac{4\eta}{G}\ \left[\ \phi_{h}^{+} -\phi_{h}^{-}\ \right],    \ \ }       
\label{eq:next101} 
\ee
 with $\phi_{h}^{\pm}$ denoting the value of the dilaton at the BH horizons on either side. Notice that this holds for, either, (A)dS$\rightarrow$(A)dS.

 
\subsection{Flat limit}   \label{sec:fl}


In the last part of this section, we will show that, upon taking the flat limit on either side of the brane, the 2D transition amplitude is still well defined. Let us consider the $\Lambda_-\equiv 0$ case. In the flat limit, the value of the dilaton at the outermost horizon on the inner vacuum's side becomes constant

\be
\underset{\mu_{-}\rightarrow \infty}{\lim}\ \frac{{\cal B}}{2\Lambda_{-}}\ \left[1-\sqrt{1-\frac{4{\cal C}\Lambda_{-}}{{\cal B}^{2}}}\right]   
\ \equiv\ 
\boxed{\ \ \ \frac{{\cal C}}{{\cal B}}     
\overset{\text{def.}}{=}    
\phi_{f}.    \textcolor{white}{\Biggl [} \color{black}   \ \ } 
\label{eq:constphi}   
\ee   
The constraint on the black hole mass ensuring the existence of 2 physical turning points, \eqref{eq:constrm}, (derived in section \ref{sec:aed0}) can be re-expressed in terms of \eqref{eq:constphi} as follows 

\be  
\boxed{\ \ \ 0>{\cal B}>-\frac{\phi_f}{\ 4\Phi_{o}^{2\ }}.\textcolor{white}{\Biggl [} \color{black}   \ \ } 
\label{eq:ncm}   
\ee
In the $\underset{{\cal B}\rightarrow 0}{\lim}$, (\ref{eq:constphi}) diverges and therefore we correctly recover the behaviour of the divergent dilaton at the horizon encountered in section \ref{sec:2}. The rate  (including background subtraction) becomes:

\begin{equation}
\boxed{\ \ \    
\begin{aligned}          
B_{tot}^{\text{\ \ Mink.}\rightarrow\text{dS}} 
& 
=      
\frac{2\eta}{G}\ \left[\ -\phi_{h,2}^{-}+\phi_{f}- \sqrt{\frac{\frac{{\cal B}^{2}}{4\Lambda_{-}}-{\cal C}}{\Lambda_{-}}}\ \left[\ y_{2} \ln\left|\frac{A_{2}y_{2}}{\sqrt{1-y_{2}^{2}}}\right| - y_{1} \ln\left|\frac{A_{2}y_{1}}{\sqrt{1-y_{1}^{2}}}\right| \right]+ \right.
\nonumber\\     
&      
-        
\left. 
\sqrt{\frac{\frac{{\cal B}^{2}}{4\Lambda_{-}}-{\cal C}}{\Lambda_{-}}}\ \left[ 
\tanh^{-1} y_{2} -
\tanh^{-1} y_{1}\right] \right],       
\label{eq:eq13} 
\end{aligned}\ \ }    
\end{equation}       
and is symmetric under the exchange of inner and outer vacua. Detailed balance is still satisfied:

\be         
\boxed{\ \ \ \ln \frac{\Gamma_{\text{Mink.}\rightarrow\text{(A)dS}}}{\Gamma_{\text{(A)dS}\rightarrow \text{Mink.}}}
\ 
=    
\ 
 \frac{4\eta}{G}\ \left[\ \phi_{f}-\phi_{h}\ \right].  \textcolor{white}{\Biggl [} \color{black}    \ \ }       
\label{eq:next1011} 
\ee

\subsection*{Summary of results and comparison with the 4D case} 

The above calculations show that:

\begin{itemize} 

\item The constant $\cB$-term added to ${\cal L}$ is playing the role of the black hole mass. This introduces a second turning point from the 2D dS point of view, as long as \eqref{eq:constrm} is satisfied.

\item The flat spacetime limit for (A)dS$\rightarrow$ (A)dS transitions, in the presence of a black hole, is finite. The dilaton at the horizon reaches a constant value, consistent with unitarity.

\item In particular, the Mink$\rightarrow$AdSBH transition we obtain is in agreement with arguments provided in \cite{Maldacena:2010un}, supporting compatibility of the process with the holographic principle. 

\end{itemize} 
The key differences w.r.t. the 4D case analysed in \cite{DeAlwis:2019rxg} are:

\begin{itemize} 

\item The divergence of dS$_{2}\rightarrow$ Mink$_{2}$ actions in absence of black holes 

\item The black hole mass is a parameter of the theory, explicitly featuring in the Lagrangian density, and is not an integration constant, albeit subject to the constraint \eqref{eq:constrm}.

\end{itemize}

\section{Holographic interpretation}\label{sec:4.4}

The importance played by black holes in ensuring the finiteness of transition amplitudes upon taking the flat limit, motivates seeking for a deeper physical  interpretation of our findings. In doing so we rely upon holography, with the main reason for this being that holographic techniques in 2D have played a key role for addressing the information loss paradox within the context of black hole physics, \cite{JM1}.
The importance of understanding decay processes involving Minkowski and AdS spacetimes from a holographic perspective was already addressed in \cite{Maldacena:2010un}, and one of our  findings is proving agreement between the arguments outlined in  \cite{Maldacena:2010un} and ours in the lower-dimensional setup.

In the present section, we argue that 2D vacuum transitions can be embedded within a similar holographic setup as the one describing black hole evaporation in braneworld models. The main reason being the role played by JT-gravity in both settings. In pursuing such task, we argue that:

\begin{itemize}

\item An entropy interpretation may  be assigned to the total action or bounce and not only to the ratio of decay rates as in detailed balance.

\item Unitarity issues emerging in the flat limit can be overcome in the Hamiltonian formalism in presence of non-extremal spacetimes in a way which is analogous to the island proposal. This follows from the fact that the total action can be rewritten as a difference of generalised entropies, each one defined as \cite{JM1, JM2}\footnote{This expression turns out to be applicable to dS$_2$ spacetimes as well, and our results are therefore in agreement with those of \cite{iic}.}

\be    
S_{\text{gen}} (R) 
=
\underset{\partial I}{\text{{min\ extr}}}\ \left[\frac{\text{Area}(\partial I)}{4G_{N}}\ +\ S_{_{EE}}(I\cup R)\right].  
\label{eq:genentr1}   
\ee

\item Transitions described by means of the BT and FMP methods are actually \emph{local}.

\item For the case of vacuum transitions, the entanglement in \ref{eq:genentr1} is \emph{internal}.

\item From the boundary perspective, a vacuum transition in the bulk may correspond to a phase transition in the boundary such as a deconfinement to confinement transition.

\end{itemize}

\subsection{Islands and horizons in holographic black hole evaporation}

\small{We will start by briefly reviewing the recent work on black holes and the information paradox that will be useful for addressing the vacuum transitions in a similar way. Our first motivation in searching for a holographic embedding of vacuum transitions, is the importance of the role played by horizons, with the Holographic Principle being its prototypical example. Before delving into a detailed analysis of our findings, we first provide some further arguments in support of the similarity between the information loss paradox, and the diveregence of the entropy associated to the cosmological horizon in the static dS patch. To the best of our knowledge, the arguments we propose provide an original motivation towards applying new holographic techniques, as outlined from section \ref{sec:KRvt} onwards.

\subsubsection*{Holographic black hole evaporation and gravitating baths }    \label{sec:KR}

In the first formulation of the information paradox, namely within the context of the evaporating (1-sided) black hole, spacetime is asymptotically flat. In its corresponding holographic embedding, this assumption is mapped to having a \emph{non-gravitating bath}. In the KR/HM-construction realising it,\cite{JM2,JM1,ATC1} (as depicted on the RHS of figure \ref{fig:RS}), the bath (denoted in blue) lives on the conformal boundary at $z=0$, and, in this sense, is non-gravitating.

\begin{figure}[ht!]        
\begin{center} 
\includegraphics[scale=0.95]{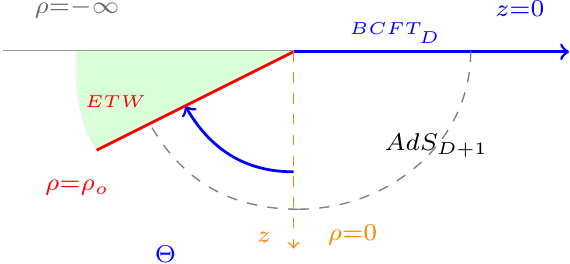} 
\ \ \ \ \ \ \ \ \ \ \ 
\includegraphics[scale=0.95]{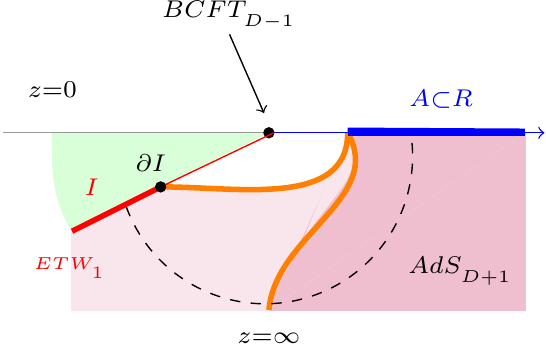} 
\caption{\footnotesize{The dashed vertical line (on the LHS) denotes the \emph{Cardy branes} within the KR/HM-construction, \cite{JM1,JM2,ATC1}, separating the causal wedge of the \emph{bulk} CFT$_{D}$ from that of the black hole QMs living on the defect BCFT$_{D-1}$. The purple shaded regions on the RHS denote the entanglement wedges for the radiation region $A\subset R$ as time evolves. The orange curves on the RHS correspond to the RT-surfaces asociated to the choice of the boundary subregion $A$. }}    
\label{fig:RS}    
\end{center} 
\end{figure}

The KR/HM-construction is able to simultaneously realise three different, albeit physically equivalent, descriptions of the same system, \cite{JM1}, namely:

\begin{enumerate} 
\item{From the \emph{bulk}, the system is described by an AdS$_{_{D+1}}$ with an ETW brane.} 

\item{From the \emph{brane}, it is given by a CFT$_{D}$ with a UV-cutoff + gravity on the ETW brane (asymptotically AdS$_{_{D}}$) which in turn is coupled with transparent boundary conditions to a BCFT$_{_{D}}$. }

\item{From the \emph{boundary}, it is a BCFT$_{_{D}}$ with nontrivial BCs. } 

\end{enumerate}

The key quantity to keep track of is the ratio between the central charges of the bulk and defect CFTs defining the BCFT$_D$, \cite{BB3456}, 

\be 
\boxed{\ \ \ F 
\ 
\overset{\text{def.}}{=} 
\ 
\frac{c_{bdy}}{c_{bulk}} 
\ 
= 
\ 
\frac{6\ln g}{c_{bulk}}, \textcolor{white}{\Biggl [} \color{black}\ \ } 
\label{eq:F}   
\ee 
from which the tension of the brane, $T$, and brane angle, $\Theta$, can be determined as follows, \cite{BB3456}, 

\be 
e^{F} 
\ 
= 
\ 
\frac{1+T}{\sqrt{1-T^{2}}} 
\ 
= 
\ 
\frac{1+\sin\Theta}{\cos\Theta} 
\ \ \ 
, 
\ \ \ 
1-e^{-2F}
\ 
= 
\ 
\frac{2\sin\Theta}{1+\sin\Theta},  
\label{eq:coordf}    
\ee 
with BH evaporation being described by the mutual exchange of d.o.f. between the bath and the black hole. Constraints on the CFT parameters are holographically dual to the allowed ranges for $T$ and $\Theta$ by virtue of \eqref{eq:coordf}.  From this follows that $t_{Page}$ is mapped to a critical value of $F$, beyond which an island is expected to arise. The second orange line in figure \ref{fig:lr}, which joins $R$ with $I$, becomes the dominant channel\footnote{The study of phase transitions in holography initiated from the first work of \cite{Witten:1998qj}.} for $t>t_{Page}$. The endpoint of $I$ is the quantum extremal surface (QES) contributing with the area law term to $S_{gen}$

\be    
\boxed{\ \ \ S_{\text{gen}} (R) 
=
\underset{\partial I}{\text{{min\ extr}}}\ \left[\frac{\text{Area}(\partial I)}{4G_{N}}\ +\ S_{_{EE}}(I\cup R)\right].  \textcolor{white}{\Biggl [} \color{black} \ \ } 
\label{eq:genentr} 
\ee

\begin{figure}[ht!]    
\begin{center} 
  \includegraphics[scale=1]{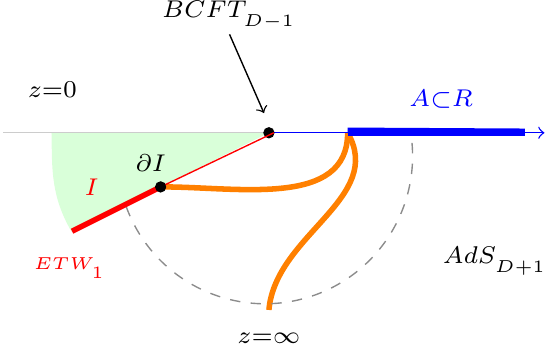}
\ \ \ \ \ \ \ \ \ \ \ 
\includegraphics[scale=1]{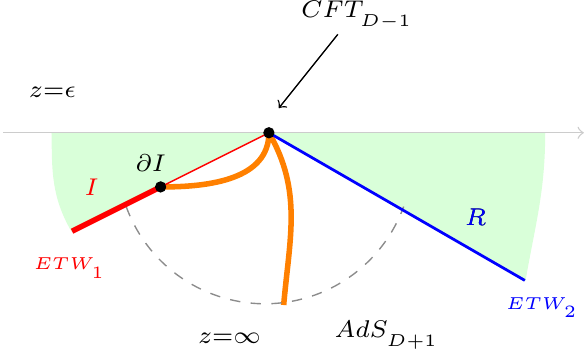}      
\caption{\footnotesize{Non-gravitating (left) and a gravitating bath (right) KR/HM-constructions.} }       
\label{fig:lr}   
\end{center}     
\end{figure}

The main caveat featuring in the non-gravitating bath setup, namely the presence of a massive graviton \cite{BBK}, $m_{_{grav}}^{2}\sim c_{bulk}/c_{bdy}$, can be overcome by introducing a gravitating bath, as shown on the RHS of figure \ref{fig:lr}. The 2 configurations can be smoothly interpolated under the action of an RG-flow driving the conformal boundary at a finite cutoff in the AdS bulk, thereby rendering it gravitating. In terms of the parameter $F$, \eqref{eq:F}, the RG-flow corresponds to the limit in which $c_{_{bdy}}>>c_{_{bulk}}$ such that, to sufficiently good approximation, the BCs of the BCFT$_D$ encode all the d.o.f. of the boundary theory. Correspondingly, the only part of the conformal boundary that is left is a codimension-2 theory, CFT$_{D-1}$, located at $z=\epsilon$, as shown on the RHS of figure \ref{fig:lr}, where the issue of the \emph{flat entanglement spectrum} can be circumvented if calculating the entanglement on a subsystem of the relic codimension-2 boundary theory, \cite{BBK}.

\section*{Cosmological horizon\ }

\begin{figure}[ht!]      
\begin{center} 
\includegraphics[scale=0.55]{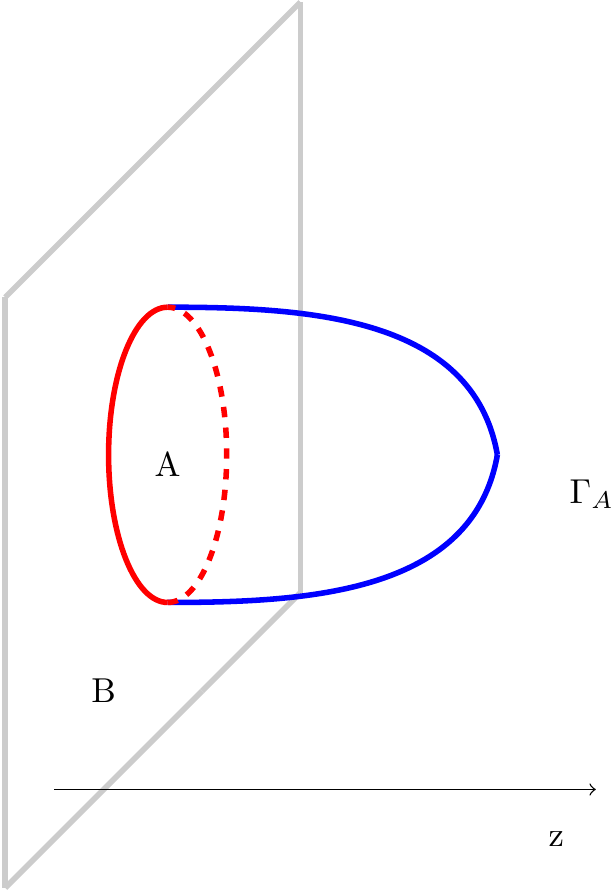}
\ \ \ \ \ \ \ \ \ \ \ \ \ \ \ \ \ 
\includegraphics[scale=1.2]{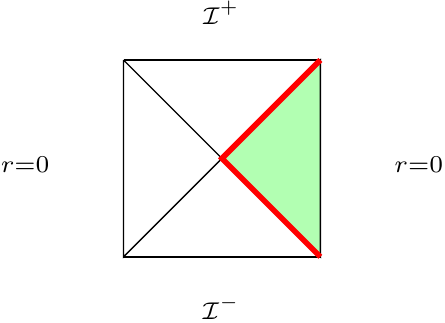} 
\caption{\footnotesize{The picture on the left is at a fixed time slice. The plane is where the CFT lives and the bulk is AdS, with $z$ realised holographically. The subregion of bulk spacetime lying between $A$ and $\Gamma_{A}$ defines the \emph{entanglement wedge} of $A$. Global dS$_{_{D}}$ for $D>2$ (on the RHS). An observer at $r=0$ can only access d.o.f. within the cosmological horizon $r=1/H$. }} 
\label{fig:dS}
\end{center} 
\end{figure} 

Equipped with this insight, we now turn to the case of the cosmological horizon, the main point of this digression being that of outlining the similarity between the unitarity issues in global dS and black hole evaporation in asymptotically flat spacetimes.

The cosmological horizon is placed at a finite distance\footnote{Each point along the red line in figure \ref{fig:dS} is a codimension-2 surface of radius $r_{h}$.}, $r_{h}=1/H$, w.r.t. an observer located at the origin of the static patch, thereby implying that there is only a finite amount of spacetime that she/he can be in causal contact with. Because of this, one might be led to conclude that the area of the cosmological horizon should somehow be related to the entropy of the static patch itself, as the Holographic Principle would suggest. However, it is also known that the dS Hilbert space is infinite-dimensional, \cite{B}. The reason for this follows from the fact that the dS isometry group is noncompact, and therefore has no finite dimensional unitary representations. This provides one of the main obstacles towards extending the holographic dictionary of \cite{JM} to the case of a positive cosmological constant.

The configuration just described seems to share similar unitarity issues as encountered for the case of the evaporating black hole. Assuming the horizon surrounding the static patch arises from partial tracing over the global spacetime, c.f. figure \ref{fig:dS},  when taking $\underset{\Lambda\rightarrow 0}{\lim}$, the cosmological horizon will correspondingly be pushed towards ${\cal I}^{+}$, at which the entropy should vanish. Intuitively, this is in agreement with the partial-tracing argument outlined above, since, as the cosmological horizon is pushed further away from the observer in the static patch, the d.o.f. which were previously ``hidden'' (i.e. traced-over), are expected to re-enter the horizon, thereby becoming accessible to the observer. However, this falls short from being true given the divergence of the entropy defined by the area-law, which, e.g., in 4D scales as $\sim{\cal O}\left(\frac{1}{H^2}\right)$. The area of the codimension-2 surface located at such distance can be interpreted by holographic arguments as the entropy of the static patch, which, in the von Neumann formulation, follows from having traced over the spacetime beyond the cosmological horizon, in analogy with the black hole picture. As such, it is not a QES. 

Given the monotonic behaviour of, both, the background action for dS$\rightarrow$dS transitions (as found in \cite{DeAlwis:2019rxg}) and the von Neumann entropy, we suggest that the divergence issue arising from the flat spacetime limit of de Sitter might be solved by introducing a more suitable definition of the de Sitter entropy, in a similar fashion as $S_{gen}$ enables to recover unitarity within the context of black hole evaporation. As anticipated at the beginning of this section, this identification turns out to be possible\footnote{Interesting new developments towards realising a dS$_4$/CFT$_3$ correspondence have recently been presented in \cite{Cotler:2023xku}, where the interpretation of the maximal entanglement between the north and south pole in global dS shares the same motivation as the argument presented in our work.}.


\subsection{Holographic embedding of vacuum transitions}    \label{sec:KRvt}

Here we present our proposal for the holographic description of the vacuum transitions. As highlighted throughout the previous sections, the three methods used for deriving transition amplitudes exhibit unexpected behaviours. In the present section, we wish to provide a unifying holographic framework describing all processes, mostly motivated by the fact that:

\begin{enumerate}  
\item The action for the system of 2D spacetimes joined by a wall is analogous to the JT-gravity coupled to a CFT$_2$ setup adopted to formulate the island proposal within the context of the information paradox. 

\item The \emph{bulk-brane-boundary} description of the same phenomenon in the KR/HM setup accounts for complementary features of the same underlying process.

\end{enumerate}

It is indeed in the spirit of the BCFT formulation of the information paradox, that we propose a holographic embedding of the transitions described analytically in the first three sections of the present work, after having implemented suitable adaptations due to the specifics of the configurations being analysed.

\begin{figure}[ht!]
\begin{center}
\includegraphics[scale=0.9]{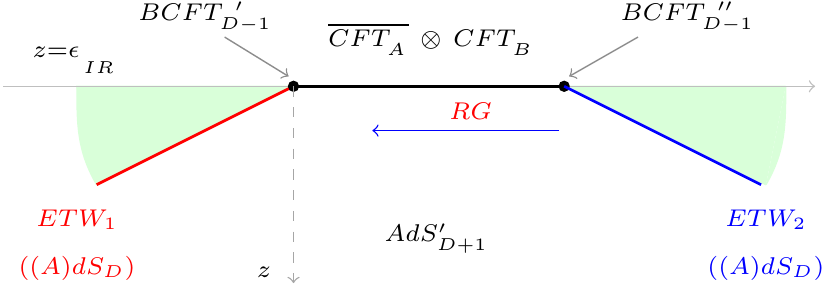}  
\ \ \ \ \ \ \ \ \ \ \ \ 
\includegraphics[scale=0.9]{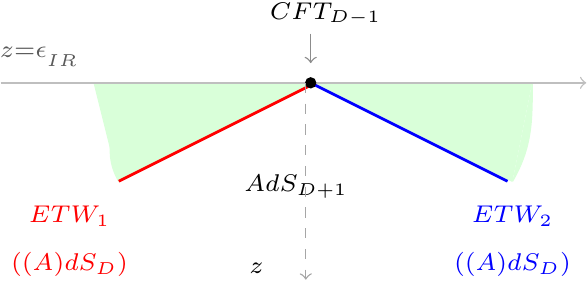}
\caption{\footnotesize{The LHS shows a generalisation of figure \ref{fig:lr} for vacuum transitions. A deformation drives the RG-flow on the boundary, assuming $c_{bulk}<<c_{bdy}$ at both endpoints. In the IR of the RG-flow (as shown on the RHS), the vacuum transition takes place in between subregions of different $AdS_{_{D}}$'s, each one lying on a different ETW. More detailes will be outlined in section \ref{sec:5}. }} 
\label{fig:foldtr221} 
\end{center} 
\end{figure}
       
Figure \ref{fig:foldtr221} shows our proposal for the holographic embedding of 2D vacuum transitions:
\begin{itemize} 

\item Given that the process involves two different spacetimes, each one being associated to a different JT-action, the holographic embedding of the transition can be described by two ETW branes (each one corresponding to one of the two vauca, $ETW_{_{1,2}}$) with the brane separating them (mediating the decay process) being geometrically realised by a composite CFT (denoted by $\overline{CFT}_{_{A}}\otimes CFT_{_{B}}$ in figure \ref{fig:foldtr221}), which can be achieved by performing the folding trick.

\item The radiation region lies on the conformal boundary (denoted by the tensor product of CFTs arising from the folding trick), which in turn is brought at a finite cutoff, $\epsilon_{_{IR}}$, with nontrivial BCs, the latter denoted by $BCFT_{_{D-1}}, BCFT_{_{D-1}}^{\prime}$. The theory living on the gravitating bath interpolates between the values of the $D$-dimensional cosmological constants. As such, the wall plays the role of the entangling surface in between the spacetimes involved, with the entanglement being \emph{internal}.

\item  Attached to the radiation region are the two ETW branes. Each one of them accommodates one of the two spacetimes, (A)dS$_{_{D}}$.

\item The composite CFT on the gravitating bath (depicted on the LHS of figure \ref{fig:foldtr221}) is ultimately driven by an irrelevant operator to the configuration on the RHS of figure \ref{fig:foldtr221}, which is reminiscent of the wedge holographic construction, \cite{BB-1}.  

\item As explained at multiple stages throughout our treatment, the total bounce and action for a given vacuum transition, result once having suitably accounted for background subtraction. As we shall see, it is particularly instructive to assign a corresponding holographic realisation of the background configuration as well. Our proposal is represented in figure \ref{fig:foldtr2211} on the RHS. We now turn to explaining it in more detail, albeit most arguments follow through from the ones outlined when explaining the RHS of figure \ref{fig:foldtr221}. The background action is a pure CFT. As such, it can be equally described independently of the choice of the UV cutoff. Hence, starting from a pure AdS$_{_{D+1}}$/CFT$_{_{D}}$ setup, with the conformal boundary defined on the whole real line at $z=0$, we can perform a suitable conformal transformation enabling to restrict the boundary domain to an interval with BCs dual to two ETW branes characterised by the same $D$-dimensional cosmological constant. Given that the theory is still conformal, the width of the segment can be arbitrary; for the purpose of interest to us, we need to bring the conformal boundary at the same IR cutoff as the two-spacetime joined system on the RHS of figure \ref{fig:foldtr221}, such that the background subtraction exactly overlaps with the instanton.

 \end{itemize}

 \begin{figure}[ht!]
\begin{center}
\includegraphics[scale=0.9]{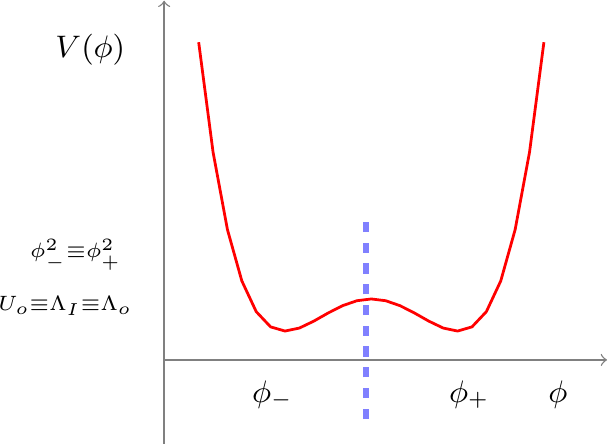}  
\ \ \ \ \ \ \ \ \ \ \ \  \ \ \ \ \ \ \ \ \ \ \ \  
\includegraphics[scale=0.9]{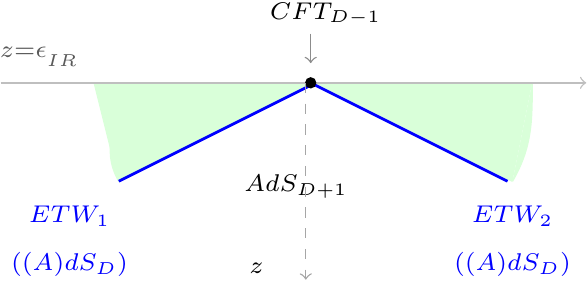}
\caption{\footnotesize{The background configuration, w.r.t. which the decay rate is normalised, is drawn on the right. From the boundary theory, CFT$_{D-1}$, this corresponds to the deconfined phase of the theory. }} 
\label{fig:foldtr2211} 
\end{center} 
\end{figure} 

\subsubsection*{Concrete proposal for the vacuum transition from the CFT side}

If there is a holographic dual formulation of vacuum transitions, a natural question to ask is: what is the corresponding physical effect that happens on the CFT side  that describes the vacuum transition in the bulk. Here we proposed that on the CFT side, the vacuum transition corresponds to a field theoretical phase transition such as a confinement/de-confinement phase transition. For this we need to identify the dual of the background spacetime and the dual of the final composite spacetime. For concreteness let us concentrate on a bulk AdS.
As we described before the vacuum transition  corresponds not to the decay of a full AdS but a portion of an AdS. For the dual we may represent this in terms of a double-well scalar potential that originally has two vacua related by $\phi \rightarrow -\phi$ separated by a domain wall. Modding out by the $\phi \rightarrow -\phi$ symmetry this reduces the bulk spacetime to a portion  with a boundary. This can provide a description of the dual of the original AdS. The description of the final spacetime can be seen in this way as a similar scalar potential but with the two minima being non-degenerate corresponding to two different spacetimes joined by the boundary. The question is how are the theories in the two boundaries related.

In figure \ref{fig:foldtr2211}  the background  can be thought of as being given by a degenerate double well potential, with the two vacua identified. In their holographic realisation in terms of ETW branes (as shown on the RHS of figure \ref{fig:foldtr2211}), only one of them will be undergoing the transition, leading to the configuration depicted on the RHS of figure \ref{fig:foldtr221}.
 
The interpolation between the deconfined and the confined phase of the theory living on the boundary can be realised by adding to the deconfined phase corresponding to figure \ref{fig:foldtr2211}, a  deformation, leading to degeneracy breaking between the two vacua, as shown on the RHS of figure \ref{fig:SSB}. Given that our analysis focuses on 2D, $T\bar T$-deformations will turn out being the relevant ones. 

The reason why we can effectively interpret vacuum transitions in terms of a deconfinement/confinement transition can be understood making use of the  analysis carried out in \cite{Klebanov:2007ws} and \cite{Komargodski:2020mxz}. In the former, it is shown that the difference between the two phases is dictated by the different dependence of the entanglement entropy on $N$, which is the key parameter in the holographic dictionary\footnote{Indeed, $N$ is related to the central charge and the AdS radius.}. In particular, following the RT prescription, this amounts to a diffference in the behaviour of the expectation value of the Wilson line connecting two arbitrary points on the conformal boundary: for the deconfined phase, $S_{_{EE}}\sim {\cal O}(N^{^2})$, wheras for the confined phase, $S_{_{EE}}\sim$ const. What this practically means is that, the former is actually associated to a conformal theory at $z=0$, for which the Wilson line diverges unless a UV-cutoff is introduced. Bulk reconstruction, in such setup, allows to recover the entire (infinite) AdS$_{_{D+1}}$ bulk volume upon removing the cutoff. On the other hand, the confined phase is characterised by a constant value of the entanglement entropy w.r.t. $N$, meaning that bulk reconstruction can effectively account for a fixed AdS$_{_{D+1}}$ bulk volume. In figure \ref{fig:foldtr2211}, this corresponds to having brought the conformal bundary at $\epsilon_{_{IR}}$, resulting only in a finite (A)dS$_{_{D}}$ volume being involved in the process.  

In a similar, and somewhat complementary fashion, the CFT description of the processes dealt with in our treatment find realisation in a recent work \cite{Komargodski:2020mxz}, where scalar potentials of the kind depicted in figure \ref{fig:SSB} have been identified with degenerate and non-degenerate vacua in studies of 2d QCD. In \cite{Komargodski:2020mxz} the degenerate case can be identified with a deconfinement phase, whereas the non-degenerate case with a confined phase\footnote{See also \cite{MVRHFC} for an interesting discussion of confinement and cosmology.}. In their treatment, the expectation value of the Wilson loop separating different vacua exhibits, either, an area- or perimeter-law like behaviour according to, whether, the system is in the confined or deconfined phase, repsectively. More efficiently, such classification can be further performed by analysing the ratio of partition functions of the gauged and ungauged theory in the infinite volume\footnote{Namely the large-$N$ limit.} limit. If such ratio is order unity, it means the system is deconfined. On the other hand, if the ratio vanishes it is confined.

 \begin{figure}[ht!]
\begin{center}
\includegraphics[scale=0.9]{figures/deconfined.pdf}  
\ \ \ \ \ \ \ \ \ \ \ \  \ \ \ \ \ \ \ \ \ \ \ \  
\includegraphics[scale=0.9]{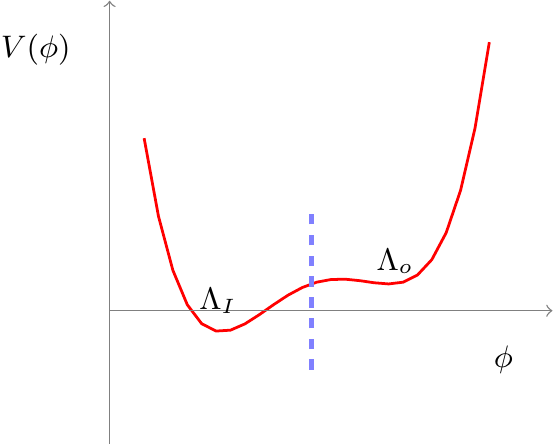}
\caption{\footnotesize{Under the action of an irrelevant operator, the boundary theory can be smoothly interpolated from the deconfined (left) to the confined phase (right). The latter is the field theory realisation of the RHS of figure \ref{fig:foldtr221}. In such description, the interpolating scalar, $\phi$, plays the role of a monopole field in a deconfinement/confinement transition, in a similar fashion as in \cite{Komargodski:2020mxz}.} } 
\label{fig:SSB} 
\end{center} 
\end{figure}

In drawing a comparison between the processes dealt with in the present work and those of  \cite{Komargodski:2020mxz}, it is important to notice that, in such reference, their analysis embraces, both, vacua and universes. While often used interchangeably, the main difference between vacua and universes is that the latter are separated by infinite barriers, whereas the former might allow finite tension domain walls interpolating between different superselection sectors of the same universe, and are therefore the ones relevant for our treatment. The systems analysed in \cite{Komargodski:2020mxz} exhibit multiple vacua, distributed between different universes, and study confinement and deconfinement of Wilson lines interpolating between them. Their counterpart for transitions of the kind associated to the RHS of figure \ref{fig:SSB},  are therefore those involving confined Wilson lines interpolating between vacua belonging to the same universe. 

Clearly more work needs to be done in order to fully describe the vacuum transitions from the CFT side.


\subsubsection{Local transitions,  entanglement and islands in wedge holography}  \label{sec:5}     


As a first concrete application of the tools outlined in the first part of this section, we now turn to the holographic interpretation of the FMP results obtained in sections \ref{sec:2} and \ref{sec:3}. In particular, we prove that:

\begin{itemize} 

\item Under suitable parametric redefinition, the results obtained by means of the FMP method are found to agree with the expression provided by \cite{MVR} for describing mutual approximation of boundary states belonging to different CFTs.
\item 
The corresponding expression for the transition rate in presence of gravity, and in absence of black holes, is given by the difference of entropies of $T\bar T$-deformed CFTs, hence proving the locality of the nucleation process. Given its interpretation as being an internal entanglement, such transitions provide an example of an AdS$_2$/CFT$_1\subset$ AdS$_3$/CFT$_2$. \footnote{The lower-dimensional holographic setup involved in these processes shares similar features as to the one involving spacetimes emerging from matrix QMs, as analysed, e.g. in \cite{Anous:2019rqb}, and their higher dimensional counterparts in \cite{VanRaamsdonk:2021duo}.}

\item Upon adding black holes, instead, the total action can be expressed as the difference of generalised entropies, with an island emerging beyond a critical value of the black hole mass. As such, this is an example of an  AdS$_2$/CFT$_1\not\subset$ AdS$_3$/CFT$_2$.     

\end{itemize}


\subsection*{ICFTs and wedge holography} \label{subsec:3}


In this subsection, we prove that quantum transitions involving different subregions of spacetime can be described by means of dual CFTs interacting via an interface, \cite{MVR}. The authors of the latter describe mutual approximation of boundary states, $\Psi_{_{A,B}}$, belonging to the Hilbert spaces of different CFTs, (as shown in figure \ref{fig:compcft}).

The correspondence adopted in \cite{MVR}, specifying to the case of AdS$_3$/CFT$_2$, relates the following set of parameters on either side of the interface

\be    
\boxed{\ \ \ L_{1}\ ,\ L_{2}\ ,\ \kappa 
\ \ \ 
\longleftrightarrow 
\ \ \ 
c_{_{A}}\ ,\ c_{_{B}},\ c_{bdy}\sim\ln g,\color{white} \textcolor{white}{\Biggl[} \color{black} \ \ } 
\ee
with $L_{1,2}$ being the AdS radii, $\kappa$ the tension of the wall separating the bulk spacetimes, $c_{_{A,B}}$ the central charges, and $\ln g$ denoting the entropy of the ICFT, $S_{ICFT}$, with $g$ defining the degeneracy of the ground state.

\begin{figure}[ht!]      
\begin{center} 
\includegraphics[scale=0.8]{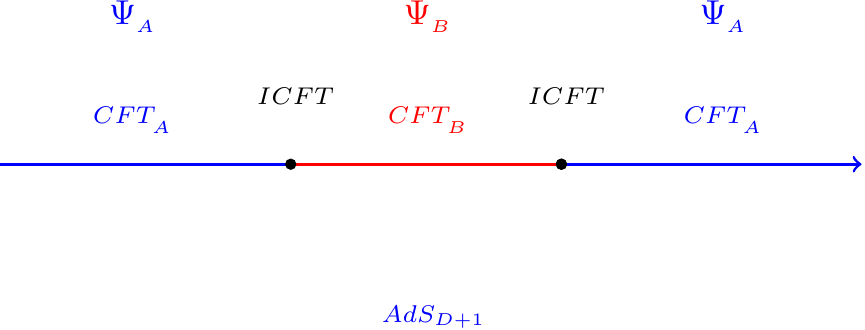}   
\caption{\footnotesize{Replacing a small portion of the background CFT$_{_{A}}$ with a different theory, CFT$_{_{B}}$ implies a change in the energy spectrum. Because of this, for a given boundary state, the corresponding $S_{_{EE}}$ leads to a different effective bulk-reconstruction of the AdS dual. }}   
\label{fig:compcft}  
\end{center}     
\end{figure}

Our first finding is that AdS$_2\rightarrow$AdS$_2$ transitions are equivalent to the case where the mutual approximation described in \cite{MVR} only involves the ground states. Indeed, under suitable parametric redefinition, the action for the transition coincides with the one obtained by \cite{MVR}, as explicitly shown below

\bea
S_{tot}^{\ \text{AdS}\rightarrow\text{AdS}}     
&
= 
&
\frac{\eta}{G}\left[\ {\mu_{-}}\ln\bigg|\frac{{\mu_{-}}A_{1}+1}{{\mu_{-}}A_{1}-1}\bigg|-\mu_{+}\ln\bigg|\frac{{\mu_{+}}A_{2}+1}{{\mu_{+}}A_{2}-1}\bigg|\ \right] 
\nonumber \\
& = &
\frac{2\eta}{G}\bigg[\left({\mu_{+}}-{\mu_{-}}\right)\tanh^{-1}\left(\frac{\Delta_{-} }{\kappa}\right) -\left({\mu_{+}}+{\mu_{-}}\right)\tanh^{-1}\left(\frac{\kappa}{\Delta_{+}  }\right)     
\bigg]  \nonumber \\ 
&
=   
&
2\eta\ S_{_{ICFT}}    
\ 
= 
\ 
2\eta\ \ln g (\kappa) \ =\ 
-2\eta\ F_{\partial},   
\label{eq:bricft}
\eea    
where we defined

\begin{equation} 
\Delta_{\pm}  
\ 
\overset{\text{def.}}{= } 
\ 
2\left(\sqrt{-\Lambda_{+}}\pm\sqrt{-\Lambda_{-}}\right),   
\end{equation} 
and $g (\kappa)$ denotes the ground state degeneracy and $F_{\partial}$ the \emph{boundary freee energy}, which is a monotonic function of $\kappa$. The entire transition is hence described by means of a codimension-2 theory as depicted in figure \ref{fig:lngk1}, with the bulk emerging via wedge holography.
\begin{figure}[ht!] 
\begin{center} 
\includegraphics[scale=0.8]{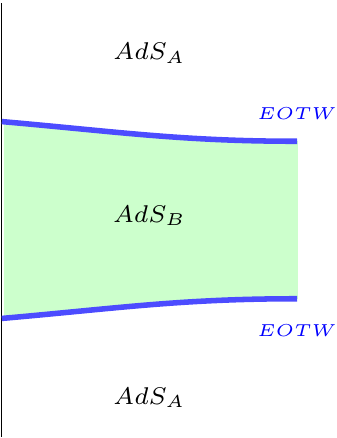}     
\ \ \ \ \ \ \ \ \ \ \ \ \ \ \ \ \ \ 
\includegraphics[scale=0.8]{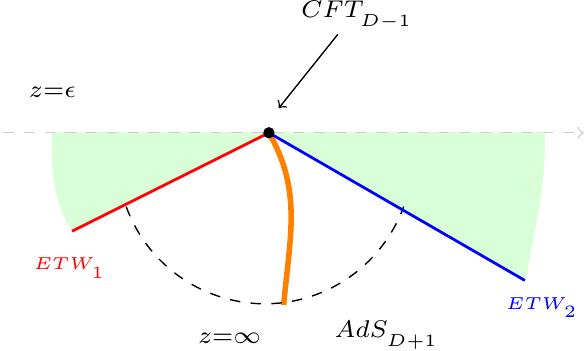} 
\caption{\footnotesize{As shown in \cite{MVR}, the trajectory of the interface separating pure AdS$_{3}$ phases has no turning point (LHS), and, correspondingly, the degeneracy of the ground state is a monotonic function of the tension $\kappa$, as found in section \ref{sec:4}. No island emerges in absence of conical defects (cf. picture on the RHS). The bulk spacetime is completely determined by means of wedge holography from the $CFT_{_{D-1}}$. $S_{_{EE}}$ is evaluated on the codimension-2 surface at the conformal boundary.  }}    
\label{fig:lngk1}
\end{center} 
\end{figure} 
Making use of the following parametric redefinitions

\begin{equation} 
\mu 
\ 
\overset{def.}{=}     
\ 
\sqrt{\frac{\lambda c}{12}\  }
\ \ \ 
, 
\ \ \ 
\phi_{o}   
\ 
\overset{def.}{=}    
\ 
r
\ \ \ 
, 
\ \ \ 
\lambda   
\ 
\overset{def.}{=}   
\ 
2\sqrt{-{\cal C}\Lambda,\ }      
\label{eq:2termssttb}    
\end{equation} 
the total action can be rewritten as

\be 
\boxed{\ \ S_{tot}^{\ \text{(A)dS}\rightarrow\text{(A)dS}}
\ 
= 
\ 
2\pi\eta\left[\ S_{_{T\bar T}}^{^{-}}- S_{_{T\bar T}}^{^{+}}   \ \right]_{_{\text{univ.}}}   
\ 
= 
\ 
-2\pi\eta\   F_{\partial} 
= 
\ 
2\pi\eta\ S_{_{ICFT}}, \textcolor{white}{\Biggl[} \color{black} \ \ }     
\label{eq:chofrel}     
\ee 
where    
\begin{equation} 
\begin{aligned}
S_{_{T\bar T}} 
\ 
&= 
\ 
\left(1-\frac{r}{2}\frac{\partial}{\partial r}\right) \ln {\cal Z}_{S^{2}}^{^{\ \ AdS, T\bar T}} 
\ 
= 
\ 
\pi \frac{c}{3}\  \sinh^{-1}\ \left(\sqrt{\frac{12}{c\lambda}}\ r\ \right)    \\
S_{_{T\bar T}} 
&=   
\left(1-\frac{r}{2}\ \frac{\partial  }{\partial r}\right)\ln {\cal Z}_{S^{2}}^{^{\ \ dS,  T\bar T}} 
\ = 
 \ 
 \frac{c}{6}+\frac{c}{3}\ \ln\left|\ \sqrt{\frac{\lambda c}{12 }}\frac{1}{r}\ \right|+\frac{c}{3}\ \tan^{-1}\ \left(\frac{1}{\sqrt{\frac{\lambda c}{12r^{2}}-1}  }\right)   +\text{const.}   \nonumber\\  
&=
 S_{\epsilon} + S_{_{\text{univ.}}},   
 \label{eq:sttds}    
  \end{aligned}
\end{equation}    
for the duals of AdS$_3$ and dS$_3$, respectively, \cite{BB70, BB48}. $S_{_{\text{univ.}}}$ and $S_{\epsilon}$ denote the \emph{universal} and \emph{cutoff-dependent} parts of \eqref{eq:sttds}, respectively, with

\be   
 S_{\epsilon}   
 \overset{def.}{=}   
 \frac{c}{3}\ \ln\left|\ \sqrt{\frac{\lambda c}{12 }}\frac{1}{r}\ \right|,
 \ \ \ \ \ \ 
 \text{and}\ \ \ \epsilon\overset{def.}{=}\sqrt{\frac{\lambda c}{12 }.\ }    
 \ee
According to the Casini-Huerta-Myers prescription, the universal part is the one contributing to the definition of the boundary free energy for a BCFT, and therefore this is the only term we need to retain for our purposes. Indeed, $S_{\epsilon}$ can be removed by simply setting $r=\epsilon$, namely choosing the cutoff to be equal to the localisation radius $r$.
From \eqref{eq:chofrel}, we deduce that:    

\begin{enumerate} 

\item This chain of equivalence relations, \eqref{eq:chofrel}, suggests an interesting direct correspondence between the mutual approximation of CFT states, \cite{MVR}, and $T\bar T$-deformed CFTs.
By virtue of the identification with $S_{gen}$, equation (\ref{eq:bricft}) implies the absence of a QES, and therefore of an island, consistently with the fact that the spacetimes involved in the transitions have no event horizons. 

\item The direct relation between the extremised action and $S_{_{T\bar T}}$ in 2D, ensures the locality of the process being described. However, as also suggested from \eqref{eq:2termssttb}, upon taking the flat limit on either side, the cutoff radius diverges ($r\rightarrow\infty$), and so too does the turning point, meaning the process is forbidden. We therefore conclude that, as long as the parameters of the theory are kept constant, no issue arises, and the transition remains local.

\item The cutoff nature of the cosmological horizon suggested by \eqref{eq:chofrel}, proves that its contribution to $S_{gen}$ should be understood as being part of $S_{_{T\bar T}}$. The motivation that led us to connect this issue with the one encountered in the black hole information paradox, precisely resided in the need to correctly define the dS entropy. Indeed, its divergence resembles that of the monotonic von Neumann entropy in the BH evaporation process, and, holographically, this corresponds to the standard divergence of $S_{_{EE}}$ upon removing the UV-cutoff. Just as in the black hole evaporation process the event horizon is not the QES extremising $S_{gen}$, so too can be claimed for the cosmological horizon, therefore proving the absence of islands in pure de Sitter spacetimes. 

\item Localisation techniques play a key role in evaluating the partition function for a given CFT. Given its key role in determining $S_{_{EE}}$, the extremisation procedure rooted in the formalism is compatible with the method defining $S_{_{TOT}}$ by means of the FMP method. In this section, we prove that these quantities are indeed proportional to each other.

\end{enumerate}

\subsubsection*{The emergence of the island}  \label{sec:5.5}

In presence of a non-extremal black hole, the total action for AdS$_2\rightarrow$AdSBH$_2$ reads: 

\bea               
S_{tot}^{\text{\ AdS$\rightarrow$AdSBH}} 
&      
= 
&         
\frac{2\pi\eta}{G}\ \left[\   
\frac{{\cal B}}{2\Lambda_{-}}\ \left[1-\sqrt{1-\frac{4{\cal C}\ \Lambda_{-}}{{\cal B}^{2}}}- \ln\left|\frac{y_{2}}{y_{1}}\sqrt{\frac{1-y_{1}^{2}}{1-y_{2}^{2}}}\right| \right]+\  \right.    
\nonumber\\ 
& 
- 
& 
\phi_{2}\ln\left|\frac{A_{2}\ y_{2}\sqrt{1-y_{4}^{2}}}{A_{4}\ y_{4}\sqrt{1-y_{2}^{2}}}\right|+\phi_{1}\ln\left|\frac{A_{2}\ y_{1}\sqrt{1-y_{3}^{2}}}{A_{4}\ y_{3}\sqrt{1-y_{1}^{2}}}\right| +\nonumber\\ 
& 
- 
&
\left.\frac{1}{2}\sqrt{\frac{\frac{{\cal B}^{2}}{4\Lambda_{-}}-{\cal C}}{\Lambda_{-}}}\ \ln\ \left|\frac{1-y_{2}}{1+y_{2}}\frac{1+y_{1}}{1-y_{1}}\right| +\mu_{+}\ \ln\ \left|\frac{1-y_{4}}{1+y_{4}}\frac{1+y_{3}}{1-y_{3}}\right| \  \right],      
\label{eq:next2} 
\eea 
where the notation is the same as in section \ref{sec:3}. \eqref{eq:next2} coincides with the result by \cite{MVR} for the case of an excited state of the new CFT approximating the ground state of the background theory under
the following identifications

\bea 
S_{brane} 
& 
= 
&  
(1-\mu)\Delta t_{2} -\Delta t_{1} \nonumber\\   
&   
= 
&      
- \frac{1}{2}\sqrt{\frac{\frac{{\cal B}^{2}}{4\Lambda_{-}}-{\cal C}}{\Lambda_{-}}}\ \ln\ \left|\frac{1-y_{2}}{1+y_{2}}\frac{1+y_{1}}{1-y_{1}}\right|    +\mu_{+}\ \ln\ \left|\frac{1-y_{4}}{1+y_{4}}\frac{1+y_{3}}{1-y_{3}}\right| ,    
\\ 
S_{hor}^{^{(-)}} 
&      
=     
&    
(1-\mu) \frac{\beta}{2}
\ 
= 
\    
- \frac{1}{2}\sqrt{\frac{\frac{{\cal B}^{2}}{4\Lambda_{-}}-{\cal C}}{\Lambda_{-}}} ,  
\label{eq:shordt}   
\eea  
with the latter, \eqref{eq:shordt},  being the 2$^{nd}$ term featuring in the total action \eqref{eq:next2}. The terms proportional to $\phi_{1,2}$ are $\sim{\cal O}({\cal B})$, and so too are the first and third terms featuring in the first line of the expression for the total action. The results in \cite{MVR} are in agreement with ours upon expanding the total action w.r.t. the deformation parameter, ${\cal B}$. As also mentioned by the authors of \cite{MVR}, their expression does not contain terms that vanish upon removing the cutoff, thereby explaining the missing terms in performing the comparison. We also made a consistency check with the results obtained in section \ref{sec:3}, and found that subleading terms in the l.o. ${\cal B}$-expansion vanish in the $\underset{{\cal B}\rightarrow 0}{\lim}$.

\begin{figure}[ht!]     
\begin{center}
\includegraphics[scale=0.8]{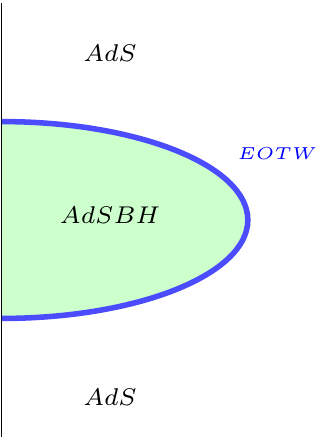} 
\ \ \ \ \ \ \ \ \ \ \ \ \ \ \ \ \ \ 
\includegraphics[scale=0.8]{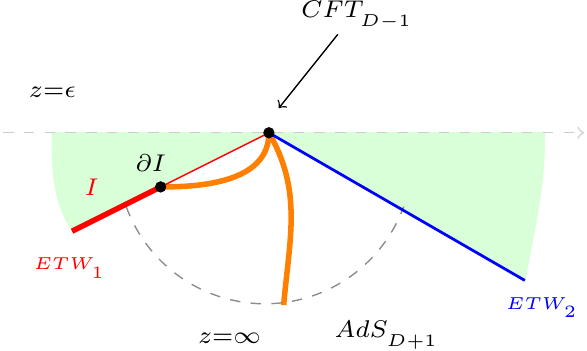} 
\caption{\small{By adding a black hole in the 3D bulk (LHS), instead, a turning point emerges, resulting in the ETW brane closing up and rejoning the conformal boundary. Correspondingly, an island emerges in the wedge holographic picture (RHS). This is the key difference w.r.t. figure \ref{fig:lngk1} on the LHS, where there is no bulk turning point, and therefore, the RT surface doesn't intersects the ETW brane.  }}    
\label{fig:VR1}    
\end{center} 
\end{figure}

The integration variables in the descriptions outlined in section \ref{sec:3} and \cite{MVR} are related as follows    

\be 
\boxed{\ \ y 
\ 
\overset{\text{def.}}{=}  
\ 
\sqrt{\frac{1-\mu}{\lambda}}\ r. \ \ }    
\label{eq:newident1}   
\ee 
In terms of which, the 3D BTZ geometry, defining an orbifold of the universal cover of the Poincarè metric, can be associated with the line element

\be 
ds^{2} 
\ 
= 
\ 
-\frac{r^{2}-r_{h}^{2}}{L^{2}}\ dt^{2}+\frac{L^{2}}{r^{2}-r_{h}^{2}}\ dr^{2}+\frac{r^{2}}{L^{2}}\ dx^{2}, 
\label{eq:BTZM3}      
\ee 
and becomes 2D conformally flat at the cutoff $r=r_{c}$, where the $T\bar T$-deformed CFT of inverse temperature $\beta\overset{def.}{=}2\pi L^{2}/r_{h}$ is located. Notice that $\underset{r_{h}\rightarrow 0}{\lim}\ \beta=\infty$, as expected for the T=0 theory. 
Recall that the variable $y$, featuring in \eqref{eq:newident1}, introduced in section \ref{sec:3}
parametrises the value of the dilaton, $\phi_b$, on the brane interpolating between the 2D vacua involved in the decay process. As previously argued, in the holographic interpretation of the process, the 1D brane is mapped to a bulk CFT$_2$ on which $S_{_{T\bar T}}$ is evaluated. 

In the same fashion, one might be led to believe that the role played by the cutoff $r_c$ is analogous to $\phi_b$, since all other parameters in the definition of $y$ are held fixed by the formalism. However, we will now be arguing that extra care is needed in building the correspondence between the 2 setups. For ${\cal B}=0$ , $y=\frac{\phi_b}{\mu}    
$, and, due to the absence of horizons, $
y_{tp} 
\ 
=1$ is unique. On the other hand, in presence of a black hole, the following identification needs to be made
\be 
\tanh x 
\ 
\overset{\text{def.}}{=}     
\ 
y 
\ \ \ 
\Rightarrow 
\ \ \ 
\sinh x 
\     
= 
\ 
\frac{r_{h}}{r_{c}}    
\ 
= 
\ 
\frac{\sqrt{1-y^{2}} }{y}    
\ \ \ \Rightarrow\ \ \ 
 r_{c} 
\overset{\text{def.}}{= } 
\ 
{r_{h}}\frac{y}{\sqrt{1-y^{2}} }  
\ \ \ 
\overset{y\rightarrow 1}{\longrightarrow } 
\ \ \    
\infty,    
\label{eq:divergcutoff}    
\ee 
from which the following information can be extracted:     

\begin{enumerate}   
\item 
If ${\cal B}=0$, then $y_{tp}=1$ is unique, and $r_c=\infty$. Correspondingly, \eqref{eq:BTZM3} reduces to a pure AdS$_3$ metric, and the bulk configuration is the same as the one drawn on the LHS of figure \ref{fig:lngk1}. 

\item As proved in section \ref{sec:fl}, $y_{tp}=1$ is also the limiting value of the turning points upon taking the flat limit on either side of the 2D transition. 
The last passage in \eqref{eq:divergcutoff} thereby shows that the effect of removing the cutoff in \cite{MVR} is equivalent to taking $\underset{{\cal B}\rightarrow 0}{\lim}$ given the identification of the turning point $y_{tp}$ with $r_{c}$. 

\item These observations imply the need to add a black hole in the 3D bulk in order to to keep $r_{c}$ finite, thereby ensuring the ETW brane separating the two phases can actually close up, as shown on the RHS of figure \ref{fig:VR1}. 

\end{enumerate}

In order to provide further justification for the claims made so far, we will now show that: 

\begin{itemize} 

\item Transition in between pure (A)dS spacetimes described by \eqref{eq:chofrel} can also be obtained in presence of a suitable value of ${\cal B}\neq 0$, thereby justifying the identification of the black hole mass with the $T\bar T$-deformation ensuring the locality of the transition taking place\footnote{Our finding is hence in agreement with \cite{Morvan:2022ybp}, where the authors show that dS black holes behave as localized, constrained, states, and therefore feature a lower entropy w.r.t. pure dS.}.

\item For arbitrary values of ${\cal B}$, the brane action undergoes a phase transition when the black hole mass decreases below a certain critical value. 

\end{itemize}

\section*{How the number of turning points depends on ${\cal B}$}

For particular values of ${\cal B}$, the turning point reduces to

\be   
   \boxed{\ \ \  \phi_{2} 
 \ 
 = 
 \ 
    \begin{cases}
\ 0\ \ \ \ \ \text{for  ${\cal B}=0$ \ \ \ \ \ \ \ \ \ \ \ \ \ $\equiv$ \ \ pure dS, 1 horizon, $y_{2}=0$} \\
\\
\ \mu_{\pm}\ \ \   \text{for ${\cal B}=-2\sqrt{|{\cal C}|\Lambda}$ \ \ $\equiv$ \ \ extremal dSBH, 2 degenerate horizons, $y_{2}=1$} \\    
\\
\ \in\ \ ]0,\mu_{\pm}[\   \text{for  $-2\sqrt{|{\cal C}|\Lambda}<{\cal B}<0$  $\equiv$  non-extremal dSBH, 2 horizons, $y_{2}\neq\{0,1\}$.} \\
    \end{cases} \textcolor{white}{\Biggl[} \color{black}}    
    \nonumber
    \label{eq:cases1}   
\ee    
This shows that, in terms of the coordinate $y$, there is no turning point in absence of a conformal symmetry breaking term in ${\cal L}$, thereby forbidding the transition to take place. On the other hand, for ${\cal B}_{\text{dS}}=-2\sqrt{|{\cal C}|\Lambda}$, the turning point is unique, and therefore coincides with the results obtained in section \ref{sec:2}. For any other value ranging in between, i.e. ${\cal B}\ \in\ ]-2\sqrt{|{\cal C}|\Lambda},0[$, there are 2 physical turning points in terms of $\phi_b$, consistently with the constraint derived in section \ref{sec:3}, \eqref{eq:constrm1}.

Having said this, we therefore conclude that:    

\begin{itemize} 

\item The role of ${\cal B}$ is effectively equivalent to that of an irrelevant operator on a CFT$_2$.   

\item The results obtained in section \ref{sec:2} should be understood of as arising from the IR limit of the RG-flow of an irrelevant operator, whose imprint in the transition amplitude is given by the finiteness of the turning point at which nucleation takes place. 

\item The cosmological horizons and the cutoff radius at which localisation takes place are proportional to each other, implying the flat limit is equivalent to removing the cutoff altogether; this explains the divergences encountered in section \ref{sec:2}. 

\item According to the value of ${\cal B}$, the nucleation process features a different number of physical turning points. Only in presence of two distinct physical values of the turning points, the flat limit can be taken (as explained in section \ref{sec:fl}). 

\item In light of further considerations made in the remainder of our work, it is important to notice that, due to the specifics of the formalism, the parameter ${\cal B}$ is not simply part of the theory, but, rather specifies the state involved in the transition. This follows from the Hamiltonian constraint and the definition of the transition amplitude. 

\end{itemize}

\section*{Relation to $S_{gen}$ and phase transition of $S_{brane}$}

The systems described in section \ref{sec:3} experience a phase transition as the black hole mass decreases, thereby providing further evidence of the fact that, in some cases, 2D vacuum tranistion can only take place in presence of islands.
The actions obtained in sections \ref{sec:2} and \ref{sec:3} can be recast to the difference of two generalised entropies, \eqref{eq:genentr}, 

\be
\boxed{\ \ \ \ S_{tot}^{\text{\ \ (A)dS}\rightarrow\text{(A)dS}} 
\ 
= 
\ 
2\pi\eta\ \left[S_{gen}-S_{gen}\right]
\ 
= 
\ 
2\pi\eta\ \left[\frac{\phi_{h,+}}{G}-\frac{\phi_{h,-}}{G}+S_{_{EE,-}}^{^{(A)dS, T\bar T}}-S_{_{EE,+}}^{^{(A)dS, T\bar T}}\right].  \textcolor{white}{\Biggl[} \color{black}  \ \ }    
\label{eq:phirho14}      
\ee 
In agreement with arguments outlined in \cite{EW1}, the island only emerges beyond a certain threshold value of the black hole mass, which, as previously shown, is accounted for by the parameter ${\cal B}$. Each part of the brane action is an entanglement entropy of two disjoined intervals on a  CFT$_2$ with deformation parameter ${\cal B}$. Expanding to leading order in ${\cal B}$, both terms can be brought to the following form

\bea   
S_{_{EE}}  
&=&
S_{_{EE}}^{^{(2)}}\bigg|_{{\cal B}=0}+\delta S_{_{EE}}^{\ s}\nonumber\\   
&\sim& 
 -\frac{8c_-}{3}\ln\left|\frac{1}{\epsilon}\ \sinh\left(\frac{2\pi r_-}{\beta_-}\right)\right|-\left(1-\frac{\phi_{o}^2}{\mu_{-}^2}\right)\frac{{\cal B}c_{-}^2}{9\Lambda_-}\coth^2\left(\frac{2\pi r_-}{\beta_-}\right)+\frac{\frac{c_-}{3}}{\sinh^2\left(\frac{2\pi r_-}{\beta_-}\right)}.   
 \label{eq:sl}    
\eea
In the second line of \eqref{eq:sl}, we redefined the parameters feturing in our original expressions in order to make the comparison with the standard CFT calculation more explicit. Notice that this interpretation is fully legitimate, since the parameter ${\cal B}$ specifies the states on either vacuum being involved in the transition process.

\begin{figure}[ht!] 
\begin{center} 
\includegraphics[scale=0.3]{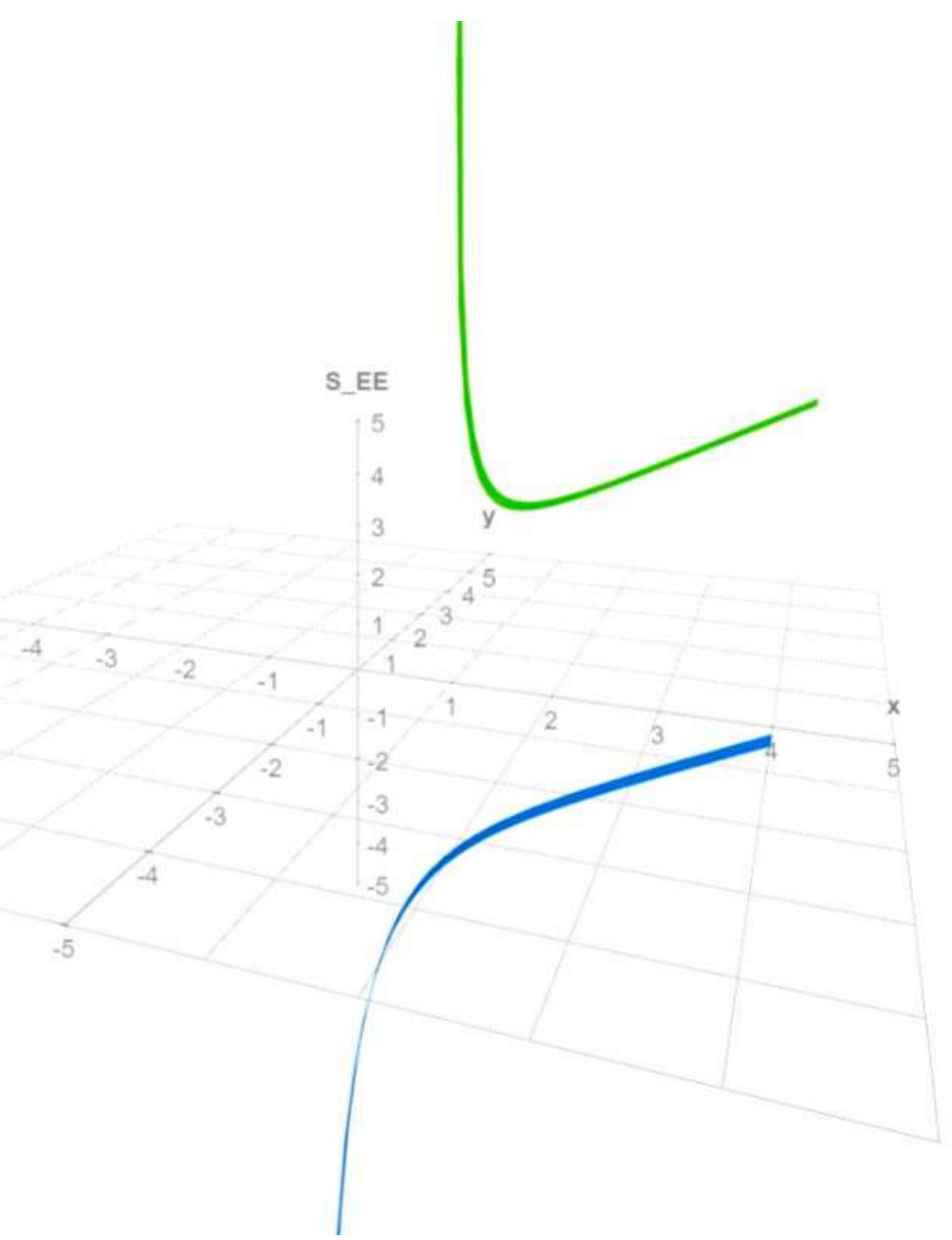}      
\ \ \ \ \ \ \ \ \ \ \ \ \ \    
\includegraphics[scale=0.28]{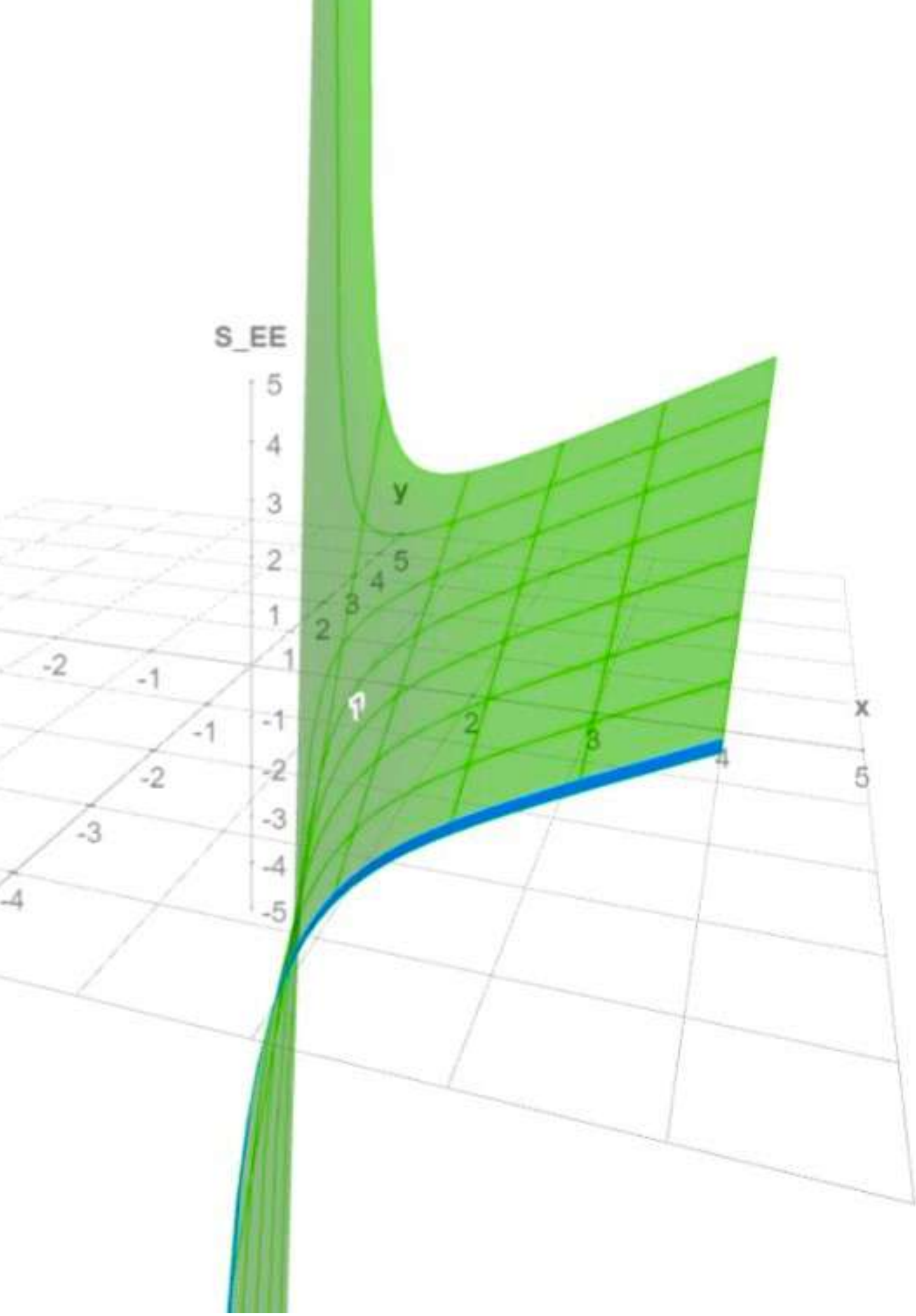} 
\caption{\footnotesize{A 3D plot of $S_{_{EE}}({\cal B}, x)$. $x$ denotes the portion of the CFT$_2$ on which $S_{_{EE}}$ is evaluated. As ${\cal B}$ changes, so too does the slope of \eqref{eq:sl}, as shown on the LHS. The figure on the RHS shows the smooth interpolation in between the monotonic behaviour of $S_{_{EE}}$ as one would expect if evaluated on the disconnected channel over 2 intervals, to a profile admitting a minimum. The phase transition occures at ${\cal B}<0$, corresponding to a upper bound to the black hole mass below which an island needs to be taken into account.  }}    
\label{fig:lngk}
\end{center} 
\end{figure}

In section \ref{sec:3}, we showed that ${\cal B}<0$ and, from section \ref{sec:2}, we know that $\phi_{o}^2>\mu_{-}^2$, therefore $y>0$ is the range of interest for the decay process described by \eqref{eq:phirho14}. As shown on the RHS of figure \ref{fig:lngk}, the entanglement entropy, $S_{_{EE}}$, experiences a phase transition, for $y>0$. In particular, the latter takes place at $y=1$. It is possible to argue that this is indeed consistent with our finding in section \ref{sec:3} that the black hole mass features in the Lagrangian density as a $T\bar T$-deformation, which needs to be constrained between a certain range, to ensure the existence of two turning points, \eqref{eq:constrm}. 

Indeed, figure \ref{fig:lngk} further supports our argument that, the phase tranistion undergone by the brane action corresponds to a Page transition beyond a suitable value of ${\cal B}$, with the latter playing a similar role as $t_{Page}$ within the context of the information loss paradox. Consequently, this justifies the emergence of the island in \eqref{eq:phirho14} upon interpreting the total action as being the difference of generalised entropies.

\subsection*{An emergent holographic embedding of 2D vacuum transitions with gravity}

In summary, from the holographic interpretation of the FMP results we proved that:

\begin{itemize}

\item The flexibility of the FMP method relies upon the fact that such kinds of \emph{exactly integrable deformations} ($T\bar T$s) can be accounted for at the level of the Lagrangain density, as explicitly outlined in section \ref{sec:3}. Because of this, this setup can potentially accommodate an island, as long as the black hole mass lies within a certain range. 

\item In absence of black holes, the total action for the transition is proportional to the defect entropy, $S_{_{ICFT}}$, which is codimension-2 w.r.t. the auxiliary AdS$_3$ bulk. This can also be re-expressed as a difference of $S_{_{EE}}^{^{T\bar T}}$s, thereby ensuring the locality of the transition as well as the fact that $S_{_{TOT}}$  is given in terms of \emph{internal} entanglement entropies evaluated on a gravitating bath.

\item In presence of black holes, $S_{_{brane}}$ behaves as a quantity experiencing a phase transition beyond a certain value of ${\cal B}$, which can be thought of as playing the role of $t_{_{Page}}$ within the island formulation. Furthermore, given the fact that the $S_{_{EE}}$s defining $S_{_{brane}}$ are basically describing an \emph{internal} entanglement, due to the role of the dilaton in the given setup, our findings are compatible with the claims of \cite{BB44}.

\end{itemize}

Having said this, we are now able to further motivate our proposal, i.e. figure \ref{fig:foldtr221}, as being an appropriate adaptation of figure \ref{fig:lr} for describing vacuum transitions in presence of gravity.  

\begin{itemize}   

\item    The total action for AdS$_2\rightarrow$AdS$_2$ transitions calculated by means of the FMP method, is a codimension-2 quantity w.r.t. its AdS$_3$ embedding, and therefore is to be asssociated to the IR-limit of an RG-flow, in agreement with wedge-holography. The same picture can be drawn for transitions involving dS spacetimes, given their realisation via $T\bar T$-deformations in the IR of an AdS.

\item   The portions of spacetime involved in the transition lie on different ETW branes, which, in turn, are holographically dual of the defects BCFT$_{D-1}$s. The interval in between the two defect theories now accommodates a composite CFT$_{D}$ with nontrivial BCs provided by the defect theories themselves. 

\item  Under the assumption that $c_{bdy}>>c_{bulk}$, the RG-flow will drive the pair of defect CFT$_{D-1}$s in the UV to a single CFT$_{D-1}$ in the IR. An island emerges if the resulting CFT$_{D-1}$ alone is unable to realise the bulk AdS$_3$ by wedge holograhy.

\end{itemize}

\subsubsection*{Implications for up-tunnelling}  \label{sec:5.4}

Given the chain of relations \eqref{eq:chofrel}, which in turn relies upon the parametric identification \eqref{eq:2termssttb}, we can now appreciate the fact that the obstructions to up-tunneling outlined in section \ref{sec:2} share a common origin. In particular, we notice that:     

\begin{itemize} 

\item The vanishing brane action in the flat limit, is compatible with the fact that $S_{_{EE}}^{^{T\bar T}}\rightarrow 0$ as $\epsilon\equiv\frac{r}{\sqrt{c\lambda}}\rightarrow0$, thereby further justifying \eqref{eq:chofrel}.

\item The divergence of the bulk action in the flat limit of dS can be recast to the need for redefining the universal part of $S_{_{EE}}^{^{T\bar T}}$ for dS including the $\frac{c}{6}$ term, which, for the decay process analysed in section \ref{sec:2} corresponds to $\mu$, namely the value of the dilaton at the horizon. Given the cutoff nature of the cosmological horizon for pure dS spacetimes (due to \eqref{eq:chofrel}), it therefore follows that, upon taking the flat limit, the transition is no longer local.

\item AdS$_2\rightarrow$AdS$_2$ up-tunneling, AdS$_2\rightarrow$Mink$_2$ and Mink$_2\rightarrow$dS$_2$ can only be achieved as long as we are able to ensure the locality of the decay processes, or, equivalently, the finiteness of the turning point associated to them. For this to happen, the localisation radius, $r$, and the cosmological constant of a given vacuum need to disentangle. This can be achieved by introducing an additional scale, ultimately ensuring the finiteness of the transition under arbitrary change of the cosmological constant.

\item This new scale, namely the black hole mass, is responsible for an additional contribution to the total action which cannot be reabsorbed within the definition of $S_{_{EE}}^{^{T\bar T}}$, thereby signalling the emergence of a QES in \eqref{eq:phirho14}, and proving the need for an island to be present for up-tunneling to take place.

\end{itemize}

\subsubsection{A BCFT\texorpdfstring{$_2$}{} perspective: transitions between ETW branes}\label{sec:ETW}

In section \ref{sec:3} we proved agreement between the BT and FMP results in absence of black holes. This subsection is meant to provide further supportive evidence of their matching from a holographic point of view, showing that transitions described via the BT formalisms provide an example of an AdS$_2$/CFT$_1 \subset\ $AdS$_3$/CFT$_2$. In doing so, we will be focussing on AdS$_2\rightarrow$ AdS$_2$ processes, showing that the total bounce can be mapped to the entanglement entropy on a BCFT$_2$.

As already explained in section \ref{sec:4}, in the BT setup, the starting point is analogous to that of a type-II Einstein-Maxwell-dilaton gravity theory in 2D, which was proved to arise from dimensional reduction of a BTZ black hole in \cite{AA}. 

In performing this comparison, the key quantity of interest is the boundary free energy, $F_{\partial}$, \cite{CHM}, which, for a BCFT$_2$ reads

\be    
-F_{\partial} 
= 
S_{CFT_{_{2}}}+S_{bdy} 
= 
\frac{T}{\ \sqrt{1-T^2\ }\ } - \ln\left|\frac{\ 1-T\ }{\ 1+ T\ }\right| ,  
\label{eq:freeE}    
\ee    
where $T$ is the tension of the ETW brane and is related to its opening angle w.r.t. the Cardy brane as $\Theta\overset{def.}{=}\tan^{-1} T$, as depicted on the LHS of figure \ref{fig:plot2f}.

\begin{figure}[ht!]        
\begin{center}   
\includegraphics[scale=0.8]{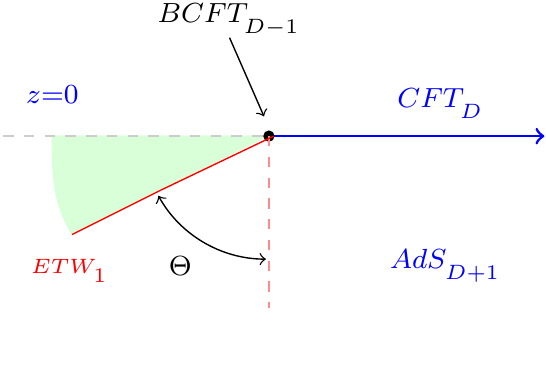}  
\ \ \ \ \ \ \ \ 
\includegraphics[scale=0.9]{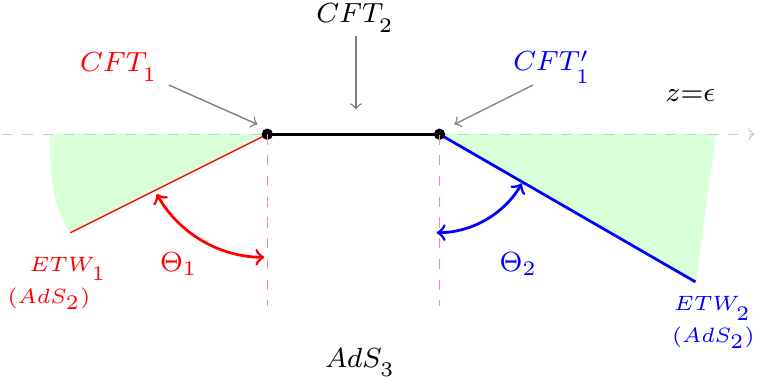}   
\caption{\footnotesize  The CFT$_1$s are dual to the AdS$_2$ spacetimes involved in the transition living on the two ETW branes and the wedges of opening angles $\Theta_{1,2}$. Their corresponding tensions, result in the value of $\Lambda_{\pm}$ on the two sides of the wall.}    
\label{fig:plot2ff}    
\end{center}    
\end{figure}

For type-1 instantons, \eqref{eq:sbdy} is nontrivial for both vacua, i.e. $T,\Theta\neq0$, and the total bounce can be re-expressed as follows:

\be 
\boxed{\ \ \ B_{type-1} 
= 
2\pi\ S_{BCFT_2}=2\pi\ \left[S_{CFT_2}+S_{bdy}^{-}-S_{bdy}^{+}\right] \textcolor{white}{\Biggl [},\ \ }    
\label{eq:BThol}    
\ee    
where

\bea    
r_o^{-1}\overset{def.}{=}\bar\rho\ \ \ ,\ \ \  r_H\overset{def.}{=}\sqrt{\Lambda_{o,I}\ }\ \ \ ,\ \ \ 
T    
\overset{def.}{=}    
\sqrt{1-\frac{\ r_H^2\ }{r_o^2}\ }    
\ \ ,\ \ \            
S_{bdy}     =
- 2\ln\left|\frac{1-\sqrt{1-\Lambda\ \bar\rho^2\ }  }{\ \sqrt{\Lambda\ }\ \bar\rho\ }\right|.\nonumber  
\label{eq:sbdy}      
\eea       
Equation \eqref{eq:BThol} indicates that either vacuum can be associated to a defect CFT$_1$ living at the endpoints of a bulk CFT$_2$. The overall system therefore defines a BCFT$_2$ as depicted on the RHS of figure \ref{fig:plot2ff}, with $S_{bdy}$ being evaluated on the lower-dimensional defect CFTs.  In particular, there are two different values of $T$, one for each vacuum. The mismatch in between the two gives rise to the holographic central charge, which counts the d.o.f. of the ground state of the codimension-2 theory living on the boundary.

\subsubsection{An internal CFT\texorpdfstring{$_2$}{}  }\label{sec:2.31}

We now turn to the holographic interpretation of the results obtained in section \ref{sec:4.1} by means of the CDL method, further supporting the emergent analytic complementarity between the CDL and BT formalisms in 2D. In doing so, two main features will become manifest: 

\begin{enumerate} 

\item The entanglement along the dilatonic direction is effectively \emph{internal}. 

\item The CDL method (in 2D) describes vacuum decays in \emph{absence} of gravity.

\end{enumerate}

A CFT$_2$ is uniquely defined by its central charge and the conformal dimensions of the operators living in the theory, $(c,\Delta_i)$. Combined together, they define the energy spectrum of the states of a given CFT. 

From the fundamental parameters of the CFT$_2$, the entropy of an excited state is defined in terms of the \emph{Cardy formula},

\be   
S_{_{Cardy}  }    
=      
2\pi \sqrt{ \frac{c}{6}\ \left(\Delta-\frac{c}{24}\right),\ } 
\label{eq:CardyS}    
\ee    
where $\Delta$ and $\Delta_o=\frac{c}{24}$ denote the black hole and ground state energies, respectively. 


For the case of a CFT$_2$ where the degeneracy of the ground state is 1, the following definitions hold

\be     
S_{CFT_2} 
= 
\frac{c}{3}\ \sqrt{\mu-1\ }, 
\ \ \  
\ \ \ 
E_o    
= 
\frac{c}{12}\ \mu, 
   \ \ \ \ \ \ 
c       
= 
6\frac{\ \partial S_{CFT}^2\ }{\partial E_o},
\label{eq:entr1}    
\ee  
with $\mu\overset{def.}{=}\frac{\Delta}{c}$ being related to the black hole mass in the excited state. 
Focussing on the AdS$_2\rightarrow$ AdS$_2$ process, with associated extremised bounce that can be rewritten as follows

\bea    
B_{tot}^{\ \text{AdS}_2\rightarrow \text{AdS}_{2}}    
&=&       4\pi\ S_\pm^2 =
4\pi\ c_{hol}^{\ {^{CFT_2}}}\ M_{_{BH}},  
\label{eq:fc1}    
\eea

\be    
c_{hol}^{\ ^{CFT_2}} 
\overset{def.}{=}         
\phi_-^2+2\phi_+^2 ,
\ \ \ \ \ \ 
M_{_{BH}}    
\overset{def.}{=}  
\frac{\phi_-^2}{\phi_+^2}-1,
\ \ \ \ \ \     
S_{\pm}
\overset{def.}{=}     
\sqrt{\phi_-^2+2\phi_+^2 \ }\ \sqrt{\frac{\phi_-^2}{\phi_+^2}-1\ },     
\label{eq:ep11}    
\ee    
with $M_{_{BH}}  $ being a conserved charge along the flow and $c_{hol}^{\ ^{CFT_2}}$ its conjugate chemical potential. The last passage in \eqref{eq:fc1} follows from the relation between the entropy product $S_+S_-$ and the central charge.  Our result \eqref{eq:ep11} goes beyond the original formulation of $c$ by Cardy because it is associated to the defect lying in between the two vacua.

A key feature of \eqref{eq:fc1} is the fact that it defines an entropy product, and (apparently) not a difference in between entropies, as one might have expected to start with. The main reason for this is that the background configuration in CDL is providing a reference state, the \emph{ground state} of the CFT$_2$, w.r.t. which the newly-nucleated spacetime behaves as an excited state, namely an extremal black hole. Comparison with \eqref{eq:entr1} enables us to deduce that

\be    
\boxed{\ \ \ \frac{\phi_-^2}{\phi_+^2}\overset{def.}{=}\mu,\textcolor{white}{\Biggl[}\ \ \ }    
\label{eq:boxed1}    
\ee    
indicating that the newly nucleated vacuum is to be identified with an extremal black hole spacetime embedded in the backround vacuum. 
Extremality follows from the fact that the mass of the BH is related to the cosmological constant itself. Indeed, from \eqref{eq:ep11}, $M_{_{BH}}$ is a function of $\Lambda_{\pm}$.

Relying upon arguments inspired by black hole microstate cosmology, \cite{MVR1}, a further remark is in order. We have shown that transitions of the kind AdS$_2\rightarrow$Mink$_2$ are forbidden. This is perfectly in agreement with the identification we have just made, namely \eqref{eq:boxed1}, since, from the CFT point of view, we can think of Mink$_2$ as being obtained from $\underset{c\rightarrow \infty}{\lim}$ of a CFT$_2$. The nucleation process would therefore correspond to the ground state of a CFT$_2$ insertion attempting to mimic an excited state of the background CFT$_2$. However, by definition, there is no horizon in the new vacuum that could attempt to screen the infinite number of d.o.f. associated to a divergent central charge, and therefore the relative entropy \eqref{eq:entr1} cannot be defined. In conclusion, due to the extremality nature of the spacetime suggested by \eqref{eq:boxed1}, such transition is forbidden, since it corresponds to an overcounting of the microstates defining the entropy.

\section*{Final remarks}

To finish this section  we briefly collect some relevant points and mention the potential relation of our results  with other recent developments.

\begin{itemize} 

\item

In summary, in absence of black holes, the three methods lead to the following results in 2D

\be                             
\begin{cases}            
B_{_{CDL/AP}}^{\ \ wo.kt}   
=    
2\pi\ S_{_{CFT_2}}^2  
=    
2\pi\ c_{hol}^{\ {^{CFT_2}}}\ M_{_{BH}}\\    \\    
S_{_{FMP}}^{^{\ {\cal B}=0}}=B_{_{BT} }   
= 
B_{_{CDL/AP}}^{\ \ w.kt}+M\bigg|^-_+   
=   
2\pi\ S_{_{BCFT_2}}
=2\pi\ \left[S_{CFT_2}+S_{bdy}^{-}-S_{bdy}^{+}\right]
\end{cases}     ,
\label{eq:BT1BT1}    
\ee              
where $B_{_{CDL/AP}}^{\ \ wo.kt}$ and $B_{_{CDL/AP}}^{\ \ w.kt}$ denote the bounces calculated in absence and in presence of the kinetic term for $\phi$, respectively, whereas 

\be   
S_{bdy}    
=    \ln\  <0|{\cal B}>\overset{def.}{=}    
\ln g= 
\ln\left|\frac{1+T}{1-T}\right|  ,
\ee  
in turn highlighting the consistency and mutual complementarity of their holographic interpretations.

\item The identifications featuring in \eqref{eq:BT1BT1}, signal the emergence of an \emph{entropic hgierarchy}, resulting in an interesting correspondence between our findings and those of \cite{EW2,EW3} within the realm of von Neumann algebras (VNAs). In particular, the fact that the CDL calculation describes transitions in absence of gravity and the total bounce can be recast in the form of a relative entropy, is compatible with the fact that a type-III VNA only admits a relative entropy definition, rather than a von Neumann entropy. Turning on gravity, namely going beyond large-$N$, we have the BT and FMP results (in absence of black holes), therefore corresponding to type-II$_{\infty}$ VNAs. Upon adding a black hole, instead, which at the level of the Lagrangian density corresponds to turning on an irrelevant deformation, the UV theory is different w.r.t. that of the case with ${\cal B}=0$, leading us to conclude that the projection operator ensuring the reduction from type- II$_{\infty}$ to type-II$_1$ is mapped to the choice of the deformation in the FMP setup, hence to the choice of the black hole mass.

\item Vacuum transitions are defined in terms of amplitudes, and we have gathered supportive evidence throughout this work that such processes actually resemble a statement of mutual approximation of CFT states, \cite{MVR}. We emphasised that the emergence of the island can only be achieved in presence of non-extremal black holes, and that, because of this, upon taking the flat limit, the corresponding action remains finite and experiences a phase transition beyond a critical value of the black hole mass. Rephrasing this in the language \cite{Susskind,EW1}, the flat limit corresponds to the large-$N$ limit of the dual gravitational theory at hand. Inspired by the arguments of \cite{EW1}, regarding the importance of scale separation for reconciling connected amplitudes from disconnected boundaries (CADBs) with holography, we therefore claim that, below a critical value of the black hole mass, the flat limit leads to a divergent action due to the lack of scale separation between the cosmological constant and the black hole mass.

\end{itemize}


\section{Conclusions and Outlook}

The present work was devoted to understanding the subject of vacuum transitions in the simple setting of 2-dimensional gravity. Due to the peculiarities of the latter, the general analysis is not a straightforward extension of the 4D cases. Our main results are the calculation of transition amplitudes obtained following three known approaches, namely the two Euclidean methods of CDL and BT,  and the Lorentzian one of FMP. We presented explicit expressions for the transition rates in each method, and for the different signs of the vacuum energies, and took the first steps towards  understanding the results. Interestingly, in comparing the Euclidean and Lorentzian prescriptions, we find many similarities, as well as differences. 

Specifying to the case of (A)dS$_2\rightarrow$(A)dS$_2$ transitions, we derived the corresponding expressions for the total bounce or action in the AP setup following the CDL method in thick, \eqref{eq:wallnew1}, and thin-wall approximation, \eqref{eq:dstodstw}, the BT, \eqref{eq:AP111adsads}, and FMP method without black holes, \eqref{eq:joined1}. As emphasised in section \ref{sec:2}, the BT and FMP results agree under suitable parametric redefinition, thereby proving expected agreement in between the 2 methods. Such redefinition, though, is highly nontrivial, and can be recast to the different way in which the cosmological constants are being defined in the 2 setups, which in turn is due to the difference in between the Lagrangian densities defining the 2 theories. 

On the other hand, the result obtained by means of the CDL setup in the thin-wall approximation, \eqref{eq:dstodstw}, is equivalent to part of the BT result. In particular, all terms can be identified (under suitable redefinition), apart from the ones arising from the boundary terms associated to the ADM mass. As also argued in section \ref{sec:4}, these provide a nontrivial contribution to the total bounce upon turning on gravity. However, as a consequence of the particular nature of the 2D theory, the e.o.m. for the AP setup (when removing the kinetic term for the dilaton from the Lagrangian density), imply $M_{_{ADM}}=0$. Therefore, 2D transitions described by means of the thin-wall approximated CDL method, should really be interpreted as describing nucleation processes in absence of gravity. 

Furthermore, the thick-wall analysis leads to an expression for $B_{_{TOT}}$ with unique features, in the sense that there seems to be no suitable reparametrisation enabling to recast it in any of the expressions obtained by means of other procedures. However, as argued in section \ref{sec:2}, under suitable parametric redefinitions, we can still provide a holographic interpretation of our result, which is complementarily in agreement with the others. 

We may summarise the holographic interpretation of our results as follows:

\begin{figure}[h!]    
\begin{center}    
\includegraphics[scale=0.9]{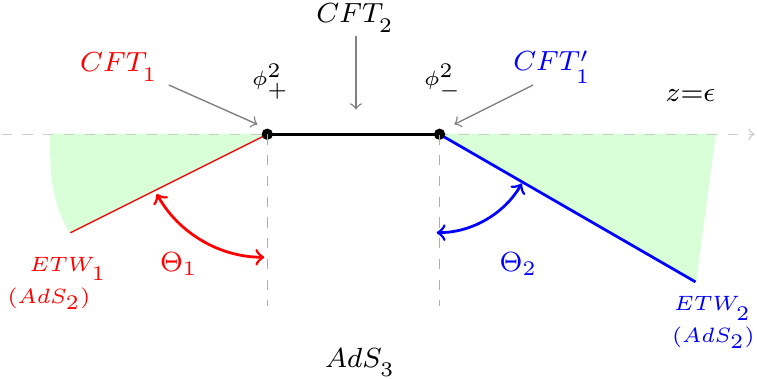}    
\caption{\small The 3 holographic descriptions are complementary to each other.}    
\label{fig:plot2f} 
\end{center}      
\end{figure}

\begin{itemize}

\item  In section \ref{sec:4.4} we provide a possible holographic interpretation of the total bounces and actions calculated throughout our work, showing mutual compatibility and complementarity. Figure \ref{fig:plot2f} shows how these processes can be understood  as taking place in a KR/HM setup with two ETW branes and one \emph{internal} gravitating bath, geometrising the brane mediating the nucleation process. In particular, we find that the corresponding expression for the transition rate in presence of gravity, and in absence of black holes, is given by the difference of entropies of $T\bar T$-deformed CFTs, hence illustrating the \emph{locality} of the nucleation process. 

\item  Upon adding black holes of suitable mass, \eqref{eq:constrm}, instead, the total action can be expressed as the difference of generalised entropies, with an island emerging beyond a critical value of the black hole mass. Furthermore, under suitable parametric redefinition, the results obtained by means of the FMP method are found to agree with the expressions derived in \cite{MVR} for describing mutual approximation of boundary states belonging to different CFTs.  

\item Inspired by field theoretic arguments, we propose that vacuum transitions in the ETW branes of the bulk correspond to a deconfinement/confinement transition on the CFT side although further study needs to be made towards this correspondence.

\item   The BT results, which were shown to be equivalent to those obtained through the FMP method in absence of black holes, can be expressed in terms of entropies of BCFT$_2$'s with 2 nontrivial boundary conditions dual to ETW branes. Furthermore, the CDL result can be recast in the form of an entropy product of a CFT$_2$ thereby showing agreement with the expectations following from the analytic behaviour encountered in section \ref{sec:4}. 

\item  One of our  results is that the total action (or bounce) associated to the decay process carries an \emph{internal} entropic interpretation. In particular, for the BT and FMP cases, they can always be expressed as the difference of generalised entropies. Only the latter, however, provides the right setup for an island to emerge. In particular, the black hole mass, which in 2D is an explicit parameter of the theory, is, in a way, playing the role of time in the Hawking evaporation process, and is responsible for $S_{_{brane}}$ undergoing a phase transition, as explained in section \ref{sec:5}. 


\end{itemize}

Understanding the behaviour of the same physical process by means of different formalisms is equally important in all fields of research, and appears to be particularly promising within the context of quantum gravity. Recent progress towards understanding the behaviour of the partition function for QG within, either, the canonical or microcanonical ensemble, \cite{Marolf:2022jra, Marolf:2022ntb}, as well as the study of the phases of gauge theories and their holographic duals, \cite{Dias:2022eyq}, are certainly among the most promising directions deserving more investigation. Clearly our work leaves many open questions. We hope that our results  will be useful for addressing further questions related to vacuum transitions, early universe cosmology, and holography.

\section*{Acknowledgements}
We thank Ahmed Almheiri, Sebasti\'an C\'espedes,  Shanta de Alwis, Alexander Frenkel, I$\tilde{\text{n}}$aki Garc\'ia Etxebarria, Victor Gorbenko, Steven Gratton, Prem Kumar, Yolanda Lozano, Juan Maldacena, Francesco Muia, Carlos Nu$\tilde{\text{n}}$ez, Jorge Santos, Sakura Schafer-Nameki, Savdeep Sethi, Ronak Soni, Julian Sonner, Mark Van Raamsdonk, Aron Wall and Edward Witten for  helpful conversations. 
 The work of VP is partially supported by an STFC scholarship through DAMTP, and has been partially supported by an Angelo Della Riccia fellowship for the academic year 2020/21. The work of FQ has been partially supported by STFC consolidated grants ST/P000681/1, ST/T000694/1.

\bibliographystyle{plain}   
\bibliography{JM2,SWH,Banks:1983by,Giddings:1995gd,Maldacena:2010un,JM,RT,BB14,VanRaamsdonk:2016exw,Coleman:1980aw,Brown:1988kg,Fischler:1990pk,Hartle:1983ai,DeAlwis:2019rxg,Bachlechner:2016mtp,BBOC,Banihashemi:2022htw,LW,Farhi:1989yr,Susskind:2021yvs,Fujita:2011fp,BB-1,BB902,BB901,BB70,BB48,BB47,BB20,BB22,JM1,Penington:2019npb,JS,Antonini:2022xzo,Antonini:2022blk,VanRaamsdonk:2020tlr,MVRHFC,iic,BBTH,Geng:2021iyq,Geng:2021wcq,Geng2:2021wcq,Geng:2022slq,Langhoff:2021uct,Betzios:2019rds,Betzios:2021fnm,Bousso:2022gth,dadg,daeh,Chen:2020tes,Freivogel:2005qh,BBHEP,Fu:2019oyc,AP,ATC1,ATC,Jackiw:1984je,Teitelboim:1983ux,BB21,MVR,EW2,EW3,Susskind,EW1,Schwinger:1951nm,Banados:1992wn,Marolf:2022jra,Marolf:2022ntb,BB44,BBK,Gibbons:1976ue,B,BB71,BB72,CC,CHM,Raamsdonk:2020tin,BB3456,Witten:1998qj,BB-2,ATC11,BB24,ATC2,Anous:2019rqb,VanRaamsdonk:2021duo,AA,HS,Morvan:2022ybp,Klebanov:2007ws,Komargodski:2020mxzMVR1,BHCFT,BB29,Dias:2022eyq,BBAW2,BBAW}

\end{document}